\documentclass[english,11pt,oneside]{article}

\pdfoutput=1

\usepackage{amsmath}
\usepackage{amsfonts}
\usepackage{amssymb}
\usepackage{graphicx, rotating}
\usepackage{epstopdf}
\usepackage{epsfig}
\usepackage{latexsym}
\usepackage{multirow}
\usepackage{color}
\usepackage[dvipsnames]{xcolor}
\usepackage{slashed}
\usepackage[top=2.5 cm, bottom=2.5 cm, left=2.5 cm, right=2.5 cm]{geometry}
\usepackage{amstext}
\usepackage{array}  
\usepackage{hyperref}
\hypersetup{
colorlinks = true,
 linkcolor  = red, 
 linktocpage=true,
 citecolor={blue}
}
\usepackage{bbold}

\usepackage[utf8]{inputenc}
\usepackage{bm}
\usepackage{leftidx}
\usepackage[numbers,sort&compress]{natbib}

\usepackage{makecell}

\usepackage{authblk}
\usepackage{colortbl}
\usepackage{xcolor}

\usepackage{float}

\newcommand{\be}{\begin{equation}}
\newcommand{\ee}{\end{equation}}
\newcommand{\ba}{\begin{array}}
\newcommand{\ea}{\end{array}}
\newcommand{\eq}[1]{Eq.~(\ref{#1})}

\newcommand{\ab}{{\alpha\beta}}

\definecolor{LightGrey}{gray}{0.96}
\definecolor{Grey}{gray}{0.94}

\begin{document}

\title{\bf Effective comparison of neutrino-mass models\thanks{ULB-TH/21-16} }
\renewcommand\footnotemark{}

\author[1]{Rupert Coy}
\author[2]{Michele Frigerio}
\affil[1]{ \emph{Service de Physique Théorique, Université Libre de Bruxelles, Boulevard du Triomphe, CP225, 1050 Brussels, Belgium} \vspace{0.2cm}}
\affil[2]{ \emph{Laboratoire Charles Coulomb (L2C), University of Montpellier, CNRS, Montpellier, France}}
\date{\today}

\maketitle

\begin{abstract}
\noindent New physics in the lepton sector may account for neutrino masses, affect electroweak precision observables, induce charged-lepton flavour violation, and shift dipole moments.
The low-energy predictions of different models are most conveniently compared within the formalism of effective field theory.
To illustrate the benefits of this approach, we derive the Wilson coefficients for a set of representative models: the fermionic seesaw mechanisms (type I and III), the Zee model, and a minimal leptoquark model. In each case,
the Weinberg and the dipole operators have qualitatively different origins. In parallel, we present the model-independent constraints on the Wilson coefficients coming from various lepton observables. 
We then show that it becomes straightforward to understand 
the allowed parameter space for each model, and to discriminate between them.
The Zee and leptoquark models are suitable to address the muon $g-2$ anomaly. 
We also confront the models with the anomalies in the W-boson mass and semileptonic $B$-meson decays. 
\end{abstract}

\newpage
\tableofcontents


\section{Introduction}

Models for neutrino masses have a long and fascinating history 
\cite{Minkowski:1977sc,Yanagida:1979as,GellMann:1980vs,Mohapatra:1979ia,Cheng:1980qt,Magg:1980ut,Schechter:1980gr,Zee:1980ai,Lazarides:1980nt,Mohapatra:1980yp,Chikashige:1980ui,Gelmini:1980re,Babu:1988ki,Foot:1988aq,Ma:1998dn}, 
which started well before the experimental confirmation of neutrino oscillations.
The most compelling explanation for the smallness of neutrino masses amounts to assuming that they violate the accidental lepton-number symmetry of the Standard Model (SM) \cite{Weinberg:1979sa}. Then, the main question is what is the associated new physics scale.
In most realisations, this is larger than the electroweak scale. The only exceptions require SM-singlet, very weakly coupled, light particles, which we will not discuss here. 
A very large scale for lepton number violation is theoretically intriguing, e.g.~in connection with Grand Unification, or to efficiently realise baryogenesis via leptogenesis. 
Apart from neutrino oscillations, other experiments may not be sensitive enough to probe such a large new physics scale.
However, there are at least two independent arguments against such line of thought:
(i) the tininess of lepton number violation is technically natural even when new particles are not very heavy; (ii) concrete models for lepton-mass generation typically involve more than a single mass scale.
Therefore, natural neutrino mass models can predict a number of measurable effects in lepton-number conserving observables as well.

In this paper, we will not focus on the direct effects of new heavy fields at high energies, such as their impact on the cosmological history, or their 
production and phenomenology at colliders. 
These require a strongly model-dependent analysis, 
and experimental progress is expected to be slow, especially at the energy frontier.
We will rather focus on indirect new-physics effects in observables at the electroweak scale and below. These currently present some anomalies, and constraints will significantly improve in the near future.
We aim to argue that, rather than computing the low-energy predictions for each ultraviolet (UV) model, it is far more efficient to adopt the Effective Field Theory (EFT) language. This amounts to reducing each model to a set of Wilson Coefficients (WCs) for higher-dimensional SM operators.
Then the comparison of different models become straightforward, as each low-energy observable is in one-to-one correspondence with a certain combination of WCs. 
While this approach may appear self-evident for EFT practitioners, it turns out that (i) well-known and well-motivated models of neutrino masses have not been investigated systematically from the EFT perspective, and (ii) the EFT analysis identifies some structural properties of the models and of their phenomenology, which were previously overlooked.

In a previous paper \cite{Coy:2018bxr} we demonstrated the advantages of the EFT approach to the type-I seesaw model. Here we extend the analysis to a few other models of neutrino masses, and perform an EFT comparison of their phenomenology. We will consider both type-I
\cite{Minkowski:1977sc,Yanagida:1979as,GellMann:1980vs,Mohapatra:1979ia} and type-III  \cite{Foot:1988aq,Ma:1998dn} seesaw models, where neutrino masses are induced at tree level from the exchange of sterile neutrinos and weak-triplet fermions, respectively.\footnote{For an EFT treatment of type-II seesaw, see e.g.~the recent analysis \cite{Li:2022ipc} and references therein.}
Their EFTs are similar, yet  
their phenomenology is clearly distinguished by
one specific combination of WCs.
preliminary results on type-III were presented in a PhD thesis \cite{Coy:2019akn}.
In addition, we will consider two models which induce neutrino masses at one loop: the well-known Zee model \cite{Zee:1980ai}, involving a second Higgs doublet and a weak-singlet scalar; and a less studied, minimal model of leptoquarks (LQs) \cite{Cata:2019wbu,Zhang:2021dgl,Dedes:2021abc}, chosen to induce both neutrino masses \cite{Nieves:1981tv,Chua:1999si,Mahanta:1999xd} and dipole operators without chiral suppression \cite{Djouadi:1989md,Davidson:1993qk,Cheung:2001ip}.
We stress that the various EFT tools employed in our analysis could be easily applied to any other model of new physics, in the lepton sector and beyond. 

In section \ref{sec:structure}, we illustrate how a careful spurion analysis allows us to fully exploit the global symmetry structure of any given model, in order to derive the qualitative structure of the WCs.  This reduces the large set of couplings in the UV to a few combinations of parameters relevant in the infrared (IR).
In section \ref{sec:EFTcomp}, we summarise the general properties of the EFT matching and running procedure, which allows us to compute the set of low-energy WCs specific to each UV model. We then specialise to the four models introduced above, and explicitly derive their WCs at leading order. 
In section \ref{sec:SMEFTconstraints}, we consider in turn various low-energy lepton observables, and determine which combinations of WCs are constrained by each of them. The results collected in this section are, to a large extent, model-independent. 
In section \ref{sec:Pheno}, we apply these general constraints to the model-specific EFTs to characterise the phenomenology of the four models. We present plots which compactly illustrate the interplay of various constraints on the relevant parameter space of each model.
Finally, in section \ref{sec:Comp} we propose an effective procedure to compare the different models, and draw our conclusions.

\section{Models' structure}
\label{sec:structure}

In this section we define the field content and the interactions of the models that we wish to 
compare. We denote by $(R_c,R_w)_Y$ the $SU(3)_c \times SU(2)_w\times U(1)_Y$ field representations, see Table \ref{tablenewfields} for the list of new multiplets. Our conventions for the SM Lagrangian are the same as e.g.~in \cite{Jenkins:2013zja,deBlas:2017xtg}, 
in particular the charged lepton Yukawa coupling reads $\mathcal{L}_{SM} \supset - \overline{e_R} y_e H^\dagger l_L+h.c.$ (doublets on the right), 
the Higgs quartic coupling is normalised as $\lambda = m_h^2/(2v^2)$ where $v\simeq 246$ GeV is the electroweak scale, and
we define covariant derivatives with a plus sign, $D^\mu\equiv \partial^\mu + i g_a A^\mu_a T_a$.

We will assume that new states beyond the SM have masses $M$ significantly larger than $v$. 
Then the SMEFT Lagrangian involves higher-dimensional operators,  
\be
{\cal L}_{SMEFT} = \sum_i C^i Q_i ~,\qquad\quad c^i \equiv \left(\frac{v}{\sqrt{2}}\right)^{d_i-4} C^i ~,
\ee
where $d_i\equiv[Q_i]=5,6,\dots$, and we defined dimensionless WCs $c^i$. 
We adopt  the operator basis defined in \cite{Grzadkowski:2010es}. The operators of interest for our analysis are listed in Table \ref{tab:ops} for convenience.
It is useful to characterise the models' structure
in terms of small dimensionless parameters,
\be
\epsilon \equiv \dfrac{Y v}{\sqrt{2} M}~, 
\ee
where $Y$ are the couplings between the SM and the new states. 
The SMEFT below the scale $M$ will have WCs $c^i$ given by appropriate combinations of the $\epsilon$ parameters, determined by the symmetries of the model.

The structure of each model is most conveniently understood by a spurion analysis.
In particular, each SM fermion species $\psi$ is associated with a $U(3)_\psi$ family symmetry, with $\psi\to V_\psi \psi$. We will treat Yukawa couplings as well as WCs as spurions preserving the SM flavour symmetry $U(3)^5$, for example  $y_e \to V_e y_e V_l^\dag$.

\begin{table*}[tbp!]
\renewcommand{\arraystretch}{1.2}
\centering
\begin{tabular}{|c|c|c|c|}
\hline
Field & Spin & $(R_c,R_w)_Y$ & Model \\ \hline \hline
$N_R$ & $\frac{1}{2}$ & $(1,1)_0$ & seesaw type-I \\ \hline\hline
$\Sigma_R$ & $ \frac{1}{2}$ & $(1,3)_0$ & seesaw type-III \\ \hline\hline
$H_2$ & 0 & $(1,2)_{1/2}$ & Zee \\ \hline
$\delta$ & 0 & $(1,1)_1$ & Zee \\ \hline\hline
$D$ & 0 & $(3,2)_{1/6}$ & leptoquark \\ \hline
$S$ & 0 & $(3,1)_{-1/3}$ & leptoquark \\ \hline
\end{tabular}
\caption{ \sl \small Representations of the new, heavy fields in the four models under investigation.}
\label{tablenewfields} 
\end{table*}

\subsection{The seesaw of type I or III}
 
The first model we consider is the type-I seesaw \cite{Minkowski:1977sc,Yanagida:1979as,GellMann:1980vs,Mohapatra:1979ia}, defined as a SM extension by $n$ sterile neutrinos\footnote{At least two sterile neutrinos are required to explain the observed pattern of neutrino oscillations. Similarly, at least two weak-triplet chiral fermions are required in the type-III seesaw model.}, $N_R\sim (1,1)_0$. 
The Lagrangian density is given by ${\cal L}_{SM}$ plus
\begin{equation}
\mathcal{L}_N = i \overline{N_R} \slashed{\partial} N_R - \left( \frac{1}{2} \overline{N_R} M_N N_R^{\ c} + \overline{N_R} Y_N \tilde{H}^\dagger l_L + h.c. \right)~.
\end{equation}
The new interactions can be considered as spurions transforming under the flavour symmetry $U(n)_N \times U(3)_l$, according to
\be
M_N\to V_N M_N V_N^T ~,\qquad Y_N\to V_N Y_N V^\dag_l~.
\ee
It is convenient to define the dimensionless parameters 
\be
\epsilon_{ia} \equiv (M_N^{-1} Y_N)_{ia} \frac{v}{\sqrt{2}} 
~,\qquad \mu_{ij} \equiv (M_N)_{ij} \frac{\sqrt{2}}{v} ~.
\label{epsmu}\ee
Once the sterile neutrinos are integrated out, the possible $U(n)_N$-invariant combinations are, at the lowest orders,
\be 
{\cal O}(M^{-1}):~(\epsilon^T \mu \epsilon)_{ab}~,\qquad {\cal O}(M^{-2}):~(\epsilon^\dag\epsilon)_{ab}~,~~~[{\cal O}(M^{-1})]^2~.
\label{Ispu}\ee
It is understood, of course, that the hermitian conjugate combinations are allowed as well, and the last term stands for any possible product of two ${\cal O}(M^{-1})$ combinations.
Note we are interested only in combinations involving positive powers of $\epsilon$, as is evident from diagrammatic arguments
(combinations with negative powers of $\mu$ are possible at higher orders in $M^{-1}$).

The WCs of dim-5 and dim-6 operators are necessarily proportional to the combinations in \eq{Ispu}.
For example, the Weinberg and the dipole WCs transform as
\be 
c^W\to V_l^* c^W V_l^\dag~,\qquad\qquad c^{eB,eW}\to V_l c^{eB,eW} V_e^\dag~.
\label{WBW}\ee
Therefore, they can receive a tree-level contribution $c^W\propto (\epsilon^T \mu \epsilon)$, and a one-loop contribution 
$c^{eB,eW}\propto (\epsilon^\dag\epsilon\, y_e^\dag)$, respectively.

The second model we want to analyse is the type-III seesaw \cite{Foot:1988aq,Ma:1998dn}, where the SM is extended by $n$ weak-triplet chiral fermions, $\Sigma_R\sim (1,3)_0$.
The associated Lagrangian reads
\begin{equation}
\mathcal{L}_{\Sigma} =  i \overline{\Sigma_{R}^A} \slashed{D}^{AB} \Sigma_{R}^B - \left( \frac{1}{2} \overline{\Sigma_{R}^A} M_\Sigma \Sigma_{R}^{A\,c} + \overline{\Sigma_{R}^A} Y_\Sigma \tilde{H}^\dagger \sigma^A l_{L} + h.c. \right) ~,
\end{equation}
where $A,B=1,2,3$ are $SU(2)_w$ indices in the adjoint. The spurion analysis proceeds in strict analogy with the type-I seesaw case by simply replacing $M_N\to M_\Sigma$ and $Y_N\to Y_\Sigma$. 

We will distinguish the type-I and III spurions by adding an obvious subscript: $\epsilon_{N,\Sigma}$ and $\mu_{N,\Sigma}$.
The different $SU(2)_w$ structure will manifest as different numerical prefactors in the various WCs,
which will correspond to a significantly different phenomenology, as we will see in section \ref{ssec:tIandIII}.

\subsection{The Zee model}
\label{ssec:zee}

The third model, known as the Zee model \cite{Zee:1980ai}, amounts to adding to the SM a second scalar transforming like the SM Higgs, $H_2\sim (1,2)_{1/2}$, as well as a singly-charged, weak-singlet scalar, $\delta \sim (1,1)_1$. 
The SM Lagrangian is augmented by
\begin{align}
\mathcal{L}_{H_2,\delta} &=  (D_\mu \delta)^\dagger (D^\mu \delta) + (D_\mu H_2)^\dagger (D^\mu H_2) - V_{H_2,\delta}
\notag \\
&- \left(\overline{l_{L}^{\ c}} Y_\delta \,i\sigma_2 l_{L} \delta +  \overline{e_R} Y_2 H_2^\dagger l_L  
+ h.c. \right) + \dots  ~,
\end{align}
where the dots stand for $H_2$ Yukawa couplings to quarks, which we set to zero for simplicity. With this assumption, we remove constraints from hadronic observables, especially FCNC processes, as here we want to focus on lepton observables. On the other hand, large quark Yukawa couplings may 
contribute significantly to lepton observables at two loops: we will mention where our results depend on the assumption that 
$H_2$ is `quark-phobic'.  
Note that $Y_\delta$
is antisymmetric in flavour space, and its three independent entries can be taken real by rephasing the three $l_{L}$ fields. On the other hand, in the basis where $y_e$ is diagonal with real, positive eigenvalues, $Y_2$ is a generic, complex matrix. 

The scalar potential can be written as
\be 
V_{H_2,\delta} = M_\delta^2 \delta^\dag \delta + M_2^2 H_2^\dagger H_2 + \left[
M_{\delta 2} \tilde H^\dag  H_2 \delta^\dagger  + \lambda_2 (H^\dagger H) (H^\dagger H_2) 
+h.c.\right]+\dots ~,
\ee
where the dots stand for quartic couplings involving two or more heavy fields $H_2$ and/or $\delta$. These induce effective operators of dimension 8 or larger, which we neglect. Note that, with no loss of generality, we chose a basis for $H$ and $H_2$ such that 
$(M_{12}^2 H^\dag H_2+h.c.)$ vanishes.
This is the most convenient basis to decouple one Higgs doublet, under the hypothesis $M_2 \gg v/\sqrt{2}$.
Note that $M_{\delta 2}$ and $\lambda_2$ can be taken real by choosing the phases of $\delta$ and $H_2$.

Each relevant coupling of the Zee model can be treated as a spurion transforming under the SM lepton flavour symmetry, $U(3)_l\times U(3)_e$, 
as well as under the rephasing symmetry of the new scalars, $U(1)_{H_2}\times U(1)_\delta$, acting as $H_2 \to e^{i \phi_2} H_2$ and 
$\delta \to e^{i \phi_\delta} \delta$. 
It is convenient to define small, dimensionless parameters,
\be\ba{ll}
(\epsilon_\delta)_{ab} \equiv \dfrac{(Y_\delta)_{ab} v}{\sqrt{2}M_\delta}\to \left(V_l^*\epsilon_\delta V_l^\dag\right)_{ab} e^{-i\phi_\delta}~,\qquad&
\epsilon_\lambda \equiv \dfrac{\lambda_2 v}{\sqrt{2}M_2}\to \epsilon_\lambda \ e^{-i\phi_2} ~,\\
(\epsilon_2)_{ab} \equiv \dfrac{(Y_2)_{ab} v}{\sqrt{2}M_2}\to \left(V_e \epsilon_2 V_l^\dag\right)_{ab} e^{i\phi_2}~,\qquad\quad &
\mu_{Z} \equiv \dfrac{\sqrt{2}M_{\delta 2}}{v} \to \mu_{Z}\ e^{i(\phi_\delta-\phi_2)}~.
\ea\label{epsZee}\ee
Once the heavy scalars $H_2$ and $\delta$ are integrated out, the possible $U(1)_{H_2}\times U(1)_\delta$-invariant combinations are, at the lowest orders,
\be\ba{l}
{\cal O}(M^{-1}):~(\epsilon_\delta)_{ab}  \mu_{Z} (\epsilon_2)_{cd}~,\qquad (\epsilon_\delta)_{ab}  \mu_{Z} \epsilon^*_\lambda ~,\qquad \\
{\cal O}(M^{-2}):~(\epsilon_\delta)_{ab} (\epsilon^*_\delta)_{cd}~,\qquad (\epsilon_2)_{ab} (\epsilon^*_2)_{cd}~,\qquad (\epsilon_2)_{ab} \epsilon_\lambda ~,\qquad \epsilon_\lambda \epsilon_\lambda ^*~,\qquad [{\cal O}(M^{-1})]^2~,
\ea\ee
where we restricted ourselves to positive powers of $\epsilon$  
for obvious diagrammatic reasons. 

Note that the flavour indices of the $\epsilon$ may be contracted either among themselves or with the SM charged-lepton Yukawa coupling, $y_e$, 
as long as one respects $U(3)_l\times U(3)_e$ covariance. 
Each pair of contracted indices corresponds to a diagram with an internal $l_L$ or $e_R$ propagator.
Let us consider, as significant examples, the Weinberg and dipole WCs, transforming as in \eq{WBW}.

The Weinberg WC is symmetric in flavour space,  $c^W_{ab}=c^W_{ba}$, therefore it cannot be simply proportional to the antisymmetric
$(\epsilon_\delta)_{ab}$. A straightforward spurion analysis reveals that the minimal contributions to $c^W$ 
are proportional to 
\be
[(\epsilon_\delta y_e^\dag \epsilon_2)+(\dots)^T]\mu_Z~~{\rm(one~loop)} ~,\qquad 
[(\epsilon_\delta y_e^\dagger y_e)+(\dots)^T]\epsilon_\lambda ^*\mu_Z~~{\rm(two~loops)}~.
\label{Wzee}\ee
The former combination corresponds to the well-known one-loop contribution to neutrino masses in the Zee model, let us call it $m^{Z}_\nu$. 
The latter combination contributes to the Weinberg operator at two loops, and it also induces the dim-7 operator $(\overline{l_L^c} \tilde{H}^*)(\tilde{H}^\dagger l_L)(H^\dag H)$ 
at one loop. 
Thus, the spurion analysis has revealed a new, alternative mechanism which contributes to neutrino masses via $H_2$ and $\delta$.
The associated $\delta m^Z_\nu$ can be dominant over $m^Z_\nu$ when the  $\epsilon_2$ entries are smaller than the entries of 
$y_e\epsilon^*_\lambda$. 
However, such a matrix has vanishing diagonal entries in flavour space, $(\delta m^Z_\nu)_{aa}=0$. As a consequence, it is not sufficient alone to accommodate all current neutrino oscillation data.
Indeed, the entries $(m^Z_{\nu})_{aa}$ also vanish if $\epsilon_2$ is diagonal in the basis where $y_e$ is diagonal. This case is known as the `minimal' Zee model \cite{Zee:1980ai,Wolfenstein:1980sy}, which is presently excluded (see e.g. \cite{He:2003ih}). In contrast, a generic $\epsilon_2$ does allow for a good fit 
(see e.g. \cite{Herrero-Garcia:2017xdu}).

Next, let us consider the dim-6 dipole WCs $c^{eB,eW}$. 
The possible contributions to these WCs should transform  as in \eq{WBW}, therefore they have to be proportional to
\be\ba{ll}
\epsilon^\dag_\delta\epsilon_\delta y_e^\dag~,\quad 
\epsilon_2^\dag \epsilon_2 y_e^\dag~,\quad 
y_e^\dag \epsilon_2\epsilon_2^\dag~,& {\rm(one~loop)}\\
\epsilon_2^\dag \epsilon_\lambda^*~,\quad
(y_e^\dag\epsilon_2 y_e^\dag)\epsilon_\lambda~,\quad
(\epsilon_2^\dag y_e y_e^\dag)\epsilon_\lambda^*~,\quad
(y_e^\dag y_e \epsilon_2^\dag)\epsilon_\lambda^*~,
&  {\rm(two~loops)}\\
\ea\label{dipoleZee}\ee
where we dropped combinations requiring a higher number of loops or additional powers of $y_e$.
A few comments are in order to interpret the result in \eq{dipoleZee}:
\begin{itemize}
\item[(i)] The loop counting is done by closing all scalar lines on each other, except for the one $H$ field appearing in $Q_{eB,eW}$. These operators are induced at the new physics scale, $M\sim M_{2,\delta}$, but additional contributions to the electromagnetic dipole ${\cal O}_{e\gamma}$ may arise at the electroweak scale with a reduced number of loops, as $H$ lines can be replaced by the vev, $v$. Indeed one finds one-loop contributions to $c^{e\gamma}$ proportional to the terms in the second line of \eq{dipoleZee}.
\item[(ii)] Given the smallness of the charged-lepton Yukawa couplings, contributions without $y_e$ insertions can be dominant even if they require extra loops. Specifically, there is a two-loop diagram proportional to $\epsilon_2^\dag \epsilon_\lambda^*$,
corresponding to a Barr-Zee type diagram involving a $W$ loop \cite{Barr:1990vd}.
In this context, one should also consider an additional spurion combination, $\epsilon_2^\dag \epsilon_\lambda^* {\rm tr}(y_{u,d}^\dag y_{u,d})$.
It contributes to $c^{eB,eW}$ only at three loops. However, after electroweak symmetry breaking, it contributes to $c^{e\gamma}$ at two loops.
This corresponds to a Barr-Zee type diagram involving a  quark loop \cite{Barr:1990vd}.
\item[(iii)] Recall that we set to zero the $H_2$ Yukawa couplings to quarks. If they were added, additional spurion combinations would exist and give a potentially large, two-loop contribution to the leptonic dipoles, see e.g.~\cite{Davidson:2016utf}.
\item[(iv)] While the spurion analysis allows us to promptly identify the relevant loop diagrams, an actual computation is needed, in particular, to distinguish log-enhanced contributions from finite ones, as we will discuss in sections \ref{sec:EFTcomp} and \ref{sec:Pheno}.
\end{itemize}

\subsection{The minimal leptoquark model}

Let us now motivate the choice of our fourth and last model.
A scalar leptoquark (LQ) is the only single-field extension of the SM that may induce a leptonic dipole operator at one loop without chirality suppression, that is, without the insertion of a small lepton Yukawa coupling $y_e$,
see e.g.~\cite{Djouadi:1989md,Davidson:1993qk,Cheung:2001ip,Biggio:2014ela}.  
More precisely,  such a dipole is induced by either a weak singlet LQ $\sim (3,1)_{-1/3}$ or a weak doublet LQ $\sim (3,2)_{7/6}$, with the insertion of an up-quark Yukawa coupling, which can be of order one in the case of the top quark. Such an unsuppressed 
one-loop contribution to the dipole has the required size to address the discrepancy between the measurement of the muon magnetic dipole moment and the SM prediction, for LQ masses of order TeV. 
On the other hand, scalar LQs can induce the Weinberg operator at one loop, providing a self-contained model of neutrino masses, 
see e.g.~\cite{Nieves:1981tv,Chua:1999si,Mahanta:1999xd}.  Specifically, two multiplets are needed,  a weak doublet LQ $\sim (3,2)_{1/6}$ and either a singlet 
$(3,1)_{-1/3}$ or a triplet $(3,3)_{-1/3}$. In both cases, the two LQs have a cubic coupling to the SM Higgs, and the resulting Weinberg operator is proportional to the down-quark Yukawa coupling. 

In light of these considerations, the minimal LQ model that induces {\it both dipoles and neutrino masses at one loop and without $y_e$ suppression} is obtained by extending the SM with two scalar LQ multiplets, a weak doublet, $D \sim (3,2)_{1/6}$, and a weak singlet, $S \sim (3,1)_{-1/3}$.\footnote{
In a supersymmetric framework, $D$ and $S$ could be identified with the squarks $\tilde q_L$ and $\tilde d_R$. Here 
we do not impose any supersymmetric restriction on their couplings.} 
This combination of LQs has recently been studied in \cite{Cata:2019wbu,Zhang:2021dgl,Dedes:2021abc}, partly with similar motivations to ours.
The Lagrangian reads
\begin{align}
\mathcal{L}_{D,S} &= (D_\mu D)^\dagger (D^\mu D) + (D_\mu S)^\dagger (D^\mu S) - V_{D, S} \notag \\
&- \left( \overline{q_{L}^c} Y_L \,i\sigma_2l_{L} S^\dag + \overline{u_{R}^c} Y_R  e_{R} S^\dagger + \overline{d_{R}} Y_D D^T i\sigma_2 l_{L} + h.c. \right) ~.
\label{eq:LQmodel}
\end{align}
Note that we set to zero the $q_L q_L S$ and $u_R d_R S$ interactions, which are allowed by the SM gauge symmetry but, in conjunction with 
$q_L l_L S^\dag$ and $u_R e_R S^\dag$, would violate baryon number. This choice, corresponding to the baryon number assignments $B(S)=B(D)=+1/3$, allows both the Weinberg operator and the lepton dipole operators to be induced at one loop.
The scalar potential is given by
\begin{align}
V_{D,S} &= M_D^2 D^\dagger D + M_S^2 S^\dagger S+ (M_{DS} D^\dagger H S + h.c.) + \dots ~,
\end{align}
where the cubic coupling 
$DDS$ is again forbidden by baryon number conservation. The dots stand for quartic couplings, which all involve two or more heavy fields $D$ and $S$, and are therefore irrelevant for deriving the dim-5 and dim-6 WCs.
Lepton number is broken by the interplay of various couplings: as $Y_{L,R}\ne 0$ imply  $L(S)=+1$ and $Y_D\ne 0$ implies $L(D)=-1$, one finds that $M_{DS}\ne 0$ violates lepton number by two units.

Let us proceed as in the previous models, defining dimensionless parameters and analysing their transformation properties under the SM flavour symmetry and the LQ rephasing symmetry, $U(1)_D\times U(1)_S$, acting as $D \to e^{i \phi_D} D$ and 
$S \to e^{i \phi_S} S$:
\be\ba{ll}
(\epsilon_L)_{ab} \equiv \dfrac{(Y_L)_{ab} v}{\sqrt{2}M_S}\to \left(V_q^*\epsilon_L V_l^\dag\right)_{ab} e^{i\phi_S}~,\qquad\quad &
(\epsilon_R)_{ab} \equiv \dfrac{(Y_R)_{ab} v}{\sqrt{2}M_S}\to \left(V_u^*\epsilon_R V_e^\dag\right)_{ab} e^{i\phi_S}~,\qquad \\
(\epsilon_D)_{ab} \equiv \dfrac{(Y_D)_{ab} v}{\sqrt{2}M_D}\to \left(V_d  \epsilon_D V_l^\dag\right)_{ab} e^{-i\phi_D}~,\qquad\quad &
\mu_{DS} \equiv \dfrac{\sqrt{2}M_{DS}}{v} \to \mu_{DS}\ e^{i(\phi_D-\phi_S)}~.
\ea\ee
Note that $\mu_{DS}$ can be chosen real, while $\epsilon_{L,R,D}$ are arbitrary complex matrices.
The possible $U(1)_{D}\times U(1)_S$-invariant combinations are
\be\ba{l}
{\cal O}(M^{-1}):~(\epsilon_L)_{ab}  (\epsilon_D)_{cd} \mu_{DS}  ~,\qquad (\epsilon_R)_{ab} (\epsilon_D)_{cd} \mu_{DS} ~,\qquad \\
{\cal O}(M^{-2}):~(\epsilon_D)_{ab} (\epsilon^*_D)_{cd}~,\qquad (\epsilon_{L,R})_{ab} (\epsilon^*_{L,R})_{cd}~,\qquad 
(\epsilon_L)_{ab} (\epsilon^*_R)_{cd} ~, \qquad [{\cal O}(M^{-1})]^2~.
\ea\ee
Pairs of flavour indices corresponding to conjugate SM fermions 
can be contracted, either between the $\epsilon$ or with the SM Yukawa couplings, which transform as $y_e \to V_e y_e V_l^\dag$,
$y_u\to V_u y_u V_q^\dag$ and $y_d\to V_d y_d V_q^\dag$.

The minimal combinations that may contribute to the Weinberg WC $c^W$ are
\be
[(\epsilon_L^T y_d^\dag \epsilon_D)+(\dots)^T] \mu_{DS} ~,\qquad\quad 
[(y_e^T \epsilon_R^T y_u y_d^\dag\epsilon_D)+(\dots)^T] \mu_{DS} ~.
\ee
The latter is more Yukawa-suppressed, still it could dominate for $\epsilon_L$ much smaller than $\epsilon_R$.
The first combination does indeed contribute to $c^W$ at one loop.
By proceeding diagrammatically, it may seem that the second combination could contribute to $c^W$ at two loops, and it would also induce the dim-7 operator $(\overline{l_L^c}\tilde{H}^*)(\tilde{H}^\dag l_L)(H^\dag H)$ at one loop. 
However, a closer look reveals that these diagrams vanish due to $SU(2)_w$ contractions, essentially because $H^T i\sigma_2 H =0$. Thus, we are left only with the former, one-loop contribution to neutrino masses.

The minimal combinations contributing
to the dipole WCs $c^{eB,eW}$ at one loop are
\be 
\epsilon^\dag_L y_u^T \epsilon_R~,\qquad 
\epsilon_D^\dag \epsilon_D y_e^\dag~,\qquad 
\epsilon_L^\dag\epsilon_L y_e^\dag ~,\qquad 
y_e^\dag\epsilon_R^\dag\epsilon_R ~.
\label{dipoleLQ}\ee
The first combination is proportional to $y_u$ rather than $y_e$, that is to say, it avoids a chirality-flip on the external lepton line, which is a big suppression especially for electrons and muons. In addition, we will see that 
the one-loop diagram proportional to $y_u$ is the only one with a logarithmic enhancement.
Note that both $\epsilon_L$ and $\epsilon_R$ should be sizeable for a significant contribution to the dipole, while the product $\epsilon_D\mu_{DS}$ can be taken as small as needed to suitably suppress neutrino masses.

\begin{table}
\renewcommand{\arraystretch}{1.3}
\centering
\begin{tabular}{|c|c|} \hline
Name & Operator \\ \hline \hline
$Q_{W,ab}$ & $(\overline{l_{La}^c}  \tilde{H}^* ) (\tilde{H}^\dagger l_{Lb} )$ \\ \hline
\hline
$Q_{eB,ab}$ & $(\overline{l_{La}} \sigma_{\mu \nu} e_{Rb}) H B^{\mu \nu}$ \\ \hline
$Q_{eW,ab}$ & $(\overline{l_{La}} \sigma_{\mu \nu} e_{Rb}) \sigma^A H W^{A \mu \nu}$ \\ \hline
\hline
$Q_{Hl,ab}^{(1)}$ & $(\overline{l_{La}} \gamma_\mu l_{Lb})(H^\dagger i \overleftrightarrow{D^\mu} H)$ \\ \hline
$Q_{Hl,ab}^{(3)}$ & $(\overline{l_{La}} \gamma_\mu \sigma^A l_{Lb})(H^\dagger i \overleftrightarrow{D^\mu} \sigma^A H)$ \\ \hline 
$Q_{He,ab}$ & $(\overline{e_{Ra}} \gamma_\mu e_{Rb})(H^\dagger i \overleftrightarrow{D^\mu} H)$ \\ \hline
$Q_{eH,ab} $ & $(\overline{l_{La}} H e_{Rb}) (H^\dagger H)$ \\ \hline
\hline
$Q_H$ & $(H^\dagger H)^3$ \\ \hline
$Q_{HD}$ & $(H^\dagger D_\mu H)^* (H^\dagger D^\mu H)$ \\ \hline
$Q_{H\square}$ & $(H^\dagger H) \square (H^\dagger H)$ \\ \hline
\end{tabular}
~~~
\begin{tabular}{|c|c|} \hline
Name & Operator \\ \hline \hline
$Q_{ll,ab}$ & $(\overline{l_{La}} \gamma_\mu l_{Lb})(\overline{l_{Lc}} \gamma^\mu l_{Ld})$ \\ \hline
$Q_{le,ab}$ & $(\overline{l_{La}} \gamma_\mu l_{Lb})(\overline{e_{Rc}} \gamma^\mu e_{Rd})$ \\ \hline
$Q_{ee,ab}$ & $(\overline{e_{Ra}} \gamma_\mu e_{Rb})(\overline{e_{Rc}} \gamma^\mu e_{Rd})$ \\ \hline
\hline
$Q^{(1)}_{lq,abcd}$ & $(\overline{l_{La}} \gamma_\mu l_{Lb})(\overline{q_{Lc}} \gamma^\mu q_{Ld})$ \\ \hline
$Q^{(3)}_{lq,abcd}$ & $(\overline{l_{La}} \gamma_\mu \sigma^A l_{Lb})(\overline{q_{Lc}} \gamma^\mu \sigma^A q_{Ld})$ \\ \hline
$Q_{qe,ab cd}$ & $ (\overline{q_{La}} \gamma_\mu q_{Lb})(\overline{e_{Rc}} \gamma^\mu e_{Rd})$ \\ \hline
$Q_{lu,abcd}$ & $ (\overline{l_{La}} \gamma_\mu l_{Lb})(\overline{u_{Rc}} \gamma^\mu u_{Rd})$ \\ \hline
$Q_{ld,abcd}$ & $(\overline{l_{La}} \gamma_\mu l_{Lb})(\overline{d_{Rc}} \gamma^\mu d_{Rd})$ \\ \hline
$Q_{eu,abcd}$ & $ (\overline{e_{Ra}} \gamma_\mu e_{Rb})(\overline{u_{Rc}} \gamma^\mu u_{Rd})$ \\ \hline
$Q_{ed,ab cd}$ & $ (\overline{e_{Ra}} \gamma_\mu e_{Rb})(\overline{d_{Rc}} \gamma^\mu d_{Rd})$ \\ \hline
\hline
$Q^{(1)}_{lequ,ab cd}$ & $ (\overline{l_{La}} e_{Rb})\, \epsilon \, (\overline{q_{Lc}} u_{Rd})$ \\ \hline
$Q^{(3)}_{lequ,ab cd}$ & $ (\overline{l_{La}} \sigma_{\mu \nu} e_{Rb}) \, \epsilon \, (\overline{q_{Lc}} \sigma^{\mu \nu} u_{Rd})$ \\ \hline
$Q_{ledq,abcd}$ & $ (\overline{l_{La}} e_{Rb}) (\overline{d_{Rc}} q_{Ld})$ \\
\hline 
\end{tabular}
\caption{\small List of the SMEFT operators relevant for the analysis of the models presented in this paper. 
These are all the operators in the Warsaw basis which involve leptons, plus the three Warsaw-basis operators which involve only the Higgs field.}
\label{tab:ops}
\end{table}

\section{Models' effective description}
\label{sec:EFTcomp}

\subsection{Matching to the SMEFT}

At the new physics scale $M$, one can systematically match any given UV-complete model to the SMEFT, that is, an EFT involving operators built with SM fields only. Here we derive the set of WCs for the four models under investigation,
summarising our results in Tables~\ref{table-WCs1}-\ref{table-WCs4}, which enable a direct comparison between the models.

Let us recall, very schematically, the structure of the EFT `matching and running' procedure. Tree-level matching induces WCs $c \sim Y^2 (v/M)^{d-4}$, where $d$ is the operator dimension and $Y$ a typical coupling between SM and heavy states.
One-loop matching corresponds to an additional suppression by a factor $\alpha/(4\pi)$, where $\alpha=y^2/(4\pi)$  for  $y$ a typical SM cubic coupling, or $\alpha=\lambda/(4\pi)$ for $\lambda$ a SM quartic coupling. 
We will drop the obvious factors of $4\pi$ in the following. 
Matching at $n$ loops is needed to extract finite terms of order $\alpha^n$. 
Then WCs are evolved from the scale $M$ to lower scales, e.g. the electroweak scale $v$,
via Renormalisation Group Equations (RGEs). One-loop RGEs account for leading logarithmic corrections to WCs. These correspond to pieces of order 
$\alpha \log(M/v)$ in one-loop diagrams, as well as to the $\alpha^n\log^n(M/v)$ parts of $n$-loop diagrams. 
Two-loop RGEs take into account next-to-leading logarithms, i.e. terms of order $\alpha^{n}\log^{n-1}(M/v)$ 
coming from  $n$-loop diagrams, for $n \geq 2$. 
Such `matching and running' procedure should be repeated at any physical threshold, till one reaches the energy scale pertinent to the observable
of interest. In particular, in section \ref{SMtoL} we will deal with the matching at scale $v$ between the SMEFT and the LEFT, that is,
the lower-energy EFT with broken electroweak symmetry.

In our analysis, we will derive the leading non-vanishing contribution to each relevant WC, defined as the term with, first, the lowest power of $\alpha$ and, second, the highest power of logarithms. While a such term is theoretically dominant, in practice subleading terms can be larger, as soon as hierarchies exist among the various heavy masses $M_i$, the new couplings $Y_i$, and/or the SM couplings $y_i$. If not otherwise stated, we will assume that all new physics dimensionful parameters are of the same order, $M_i\sim M$, and all new physics couplings to the SM are also of the same order, $Y_i\sim Y$. 
Where this assumption is dropped, we will illustrate how the phenomenology can differ. On the other hand, as the hierarchy among the SM couplings $y_i$ is known, we will take it into account to determine the actual dominant contribution to each WC.

\definecolor{grayLLL}{gray}{0.90} 
\definecolor{grayLL}{gray}{0.70} 
\definecolor{grayL}{gray}{0.60}

\begin{table}[!t]
\renewcommand{\arraystretch}{2}
\centering
\hspace*{-.5cm}
\arrayrulecolor{black!70}
\begin{tabular}{
!{\color{black}\vline}>{\centering\arraybackslash}m{0.8cm}
!{\color{black}\vline}>{\centering\arraybackslash}m{3.4cm}
!{\color{black}\vline}>{\centering\arraybackslash}m{3.4cm}
!{\color{black}\vline}>{\centering\arraybackslash}m{3.4cm}
!{\color{black}\vline}>{\centering\arraybackslash}m{3.4cm}
!{\color{black}\vline}}
\hline
\textbf{WCs} & Seesaw I & Seesaw III & Zee & Leptoquarks \\ \hline \hline
$c^W_{ab}$ 
& $\frac12(\epsilon_N^T \mu_N \epsilon_N)_{ab}$ 
& $\frac12(\epsilon_\Sigma^T \mu_\Sigma \epsilon_\Sigma)_{ab}$ 
& \cellcolor{grayLL} $\makecell{-2 \mu_Z (\epsilon_\delta y_e^\dag \epsilon_2)_{ab}{\bf P}\\
-2\mu_Z(\epsilon_2^T y_e^* \epsilon_\delta^T)_{ab} {\bf P}} $ 
& \cellcolor{grayLL} $\makecell{3 \mu_{DS}(\epsilon_L^T y_d^\dag \epsilon_D)_{ab}{\bf P}\\
+3 \mu_{DS}(\epsilon_D^T y_d^* \epsilon_L)_{ab} {\bf P}} $  \\  
\hline \hline
$c^{eB}_{ab}$ 
& \cellcolor{grayLL} $- \frac{g_1}{24} (\epsilon_N^\dag \epsilon_N y_e^\dag)_{ab}{\bf P}$ 
& \cellcolor{grayLL} ${- \frac{g_1}{8} (\epsilon_\Sigma^\dag \epsilon_\Sigma y_e^\dag)_{ab}{\bf P}} $
& \cellcolor{grayLL} $\makecell{ - \frac{g_1}{3} (\epsilon_\delta^\dag \epsilon_\delta y_e^\dag)_{ab}{\bf P}\\ + \frac{5g_1}{48} (\epsilon_2^\dag \epsilon_2 y_e^\dag)_{ab}{\bf P}\\ + \frac{g_1}{24} (y_e^\dag \epsilon_2 \epsilon_2^\dag)_{ab} {\bf P}}$ 
& \cellcolor{grayLLL} $-\frac{5 g_1}{4} (\epsilon_L^\dag y_u^T \epsilon_R)_{ab}{\bf L}$  \\  
\hline
$c^{eW}_{ab}$ 
& \cellcolor{grayLL} $- \frac{5g_2}{24} (\epsilon_N^\dag \epsilon_N y_e^\dag)_{ab} {\bf P}$  
& \cellcolor{grayLL} ${- \frac{3g_2}{8} (\epsilon_\Sigma^\dag \epsilon_\Sigma y_e^\dag)_{ab} {\bf P}} $
& \cellcolor{grayLL} $\makecell{ - \frac{g_2}{6} (\epsilon_\delta^\dag \epsilon_\delta y_e^\dag)_{ab}{\bf P}\\ + \frac{g_2}{48} (\epsilon_2^\dag \epsilon_2 y_e^\dag)_{ab} {\bf P}}$ 
& \cellcolor{grayLLL} $\frac{3 g_2}{4} (\epsilon_L^\dag y_u^T \epsilon_R)_{ab} {\bf L} $ \\  
\hline \hline 
$c^{Hl(1)}_{ab}$ & $\frac{1}{4} (\epsilon_N^\dagger \epsilon_N)_{ab}$ & $ \frac{3}{4} (\epsilon^\dagger_\Sigma \epsilon_\Sigma)_{ab}$ & \cellcolor{grayLLL} $\makecell{ 
	\frac{2g_1^2}{3} (\epsilon_\delta^\dag \epsilon_\delta)_{ab}{\bf L}
\\- \frac{g_1^2}{3} (\epsilon_2^\dag \epsilon_2)_{ab} {\bf L}}$
& \cellcolor{grayLLL} $\makecell{
	- \frac{g_1^2}{3} (\epsilon_D^\dag \epsilon_D)_{ab} {\bf L}
\\- \frac{g_1^2}{6}( \epsilon_L^\dag \epsilon_L)_{ab}{\bf L}\\ 
- \frac{3}{2} (\epsilon_L^\dag y_u^T y_u^* \epsilon_L )_{ab}{\bf L}} $ \\  \hline 
$c^{Hl(3)}_{ab}$ 
& $-\frac{1}{4} (\epsilon_N^\dagger \epsilon_N)_{ab}$ 
& $ \frac{1}{4} (\epsilon^\dagger_\Sigma \epsilon_\Sigma)_{ab}$ 
& \cellcolor{grayLLL} $\makecell{\frac{2g_2^2}{3} (\epsilon_\delta^\dag \epsilon_\delta)_{ab} 
{\bf L}}  $  
& \cellcolor{grayLLL} $\makecell{\frac{1}{2} g_2^2 (\epsilon_L^\dag \epsilon_L)_{ab}{\bf L} \\
	 - \frac{3}{2} (\epsilon_L^\dag y_u^T y_u^* \epsilon_L)_{ab} {\bf L}}$ \\  \hline 
$c^{He}_{ab}$ 
& \cellcolor{grayLLL} $\makecell{\frac{1}{2} (y_e \epsilon_N^\dag \epsilon_N y_e^\dag)_{ab} {\bf L}
\\- \frac{g_1^2}{3}\text{tr}[\epsilon_N^\dag \epsilon_N]\delta_{ab} {\bf L}}$ 
& \cellcolor{grayLLL} $\makecell{\frac{3}{2} (y_e \epsilon_\Sigma^\dag \epsilon_\Sigma y_e^\dag)_{ab}{\bf L}\\ -g_1^2 \text{tr}[\epsilon_\Sigma^\dag \epsilon_\Sigma] \delta_{ab} {\bf L}} $ 
& \cellcolor{grayLLL} $\makecell{ 
	- \frac{g_1^2}{3} (\epsilon_2 \epsilon_2^\dag)_{ab}{\bf L} }$ 
& \cellcolor{grayLLL} $\makecell{ 3 (\epsilon_R^\dag y_u^* y_u^T \epsilon_R)_{ab}{\bf L}\\
- \frac{g_1^2}{6} ( \epsilon_R^\dag \epsilon_R)_{ab} {\bf L}}$ \\  \hline 
$c^{eH}_{ab}$ 
& \cellcolor{grayLLL} $\makecell{2\lambda (\epsilon_N^\dagger \epsilon_N y_e^\dag)_{ab} {\bf L}
\\ + \frac{g_2^2}{3}\text{tr}[\epsilon_N^\dagger \epsilon_N] (y_e^\dag)_{ab} {\bf L}
\\- 6 (c^{W\dag} c^W y_e)_{ab} {\bf L}
\\+ 8 \text{tr}[c^{W\dag} c^W] (y_e^\dag)_{ab} {\bf L}}$
& $(\epsilon^\dagger_\Sigma \epsilon_\Sigma y_e^\dag)_{ab} $   
& $\epsilon_\lambda^* (\epsilon_2^\dag)_{ab}$ 
& \cellcolor{grayLLL} $\makecell{6 ( \epsilon_L^\dag y_u^T y_u^* y_u^T \epsilon_R)_{ab}{\bf L}\\
 - 6 \lambda (\epsilon_L^\dag y_u^T \epsilon_R)_{ab} {\bf L}}$ \\  
\hline 
\end{tabular}
\caption{ \sl \small 
The leading-order contribution to the WCs of the SMEFT, in four neutrino-mass models. 
For this table we display the operators involving two leptons and no quarks. 
Light-grey cells correspond to WCs induced at one-loop leading-log, which are proportional to ${\bf L} \equiv \log(M/v)/(16\pi^2)$. 
Dark-grey cells correspond to WCs induced at one-loop finite order, which are proportional to ${\bf P} \equiv 1/(16\pi^2)$. 
Finally, when a given WC receives two contributions which involve the same  $\epsilon$'s, but different powers of $y_e$ or $y_d$, 
we dropped the subleading one.
}
\label{table-WCs1} 
\end{table}

\begin{table}[!p]
\renewcommand{\arraystretch}{2}
\centering
\hspace*{-.5cm}
\arrayrulecolor{black!70}
\begin{tabular}{
!{\color{black}\vline}>{\centering\arraybackslash}m{0.8cm}
!{\color{black}\vline}>{\centering\arraybackslash}m{3.4cm}
!{\color{black}\vline}>{\centering\arraybackslash}m{3.4cm}
!{\color{black}\vline}>{\centering\arraybackslash}m{3.4cm}
!{\color{black}\vline}>{\centering\arraybackslash}m{3.4cm}
!{\color{black}\vline}}
\hline
\textbf{WCs} & Seesaw I & Seesaw III & Zee & Leptoquarks \\ \hline \hline
$c^{H}$ 
& \cellcolor{grayLLL} $\makecell{\frac 43 \lambda g_2^2 \text{tr}[ \epsilon_N^\dag \epsilon_N ] {\bf L}
\\- 32 \lambda \text{tr}[c^{W\dag} c^W]{\bf L}}$
& \cellcolor{grayLLL} $\makecell{- \frac{4}{3} \lambda g_2^2 \text{tr}[\epsilon_\Sigma^\dag \epsilon_\Sigma]{\bf L} \\ - 32 \lambda \text{tr}[c^{W\dag} c^W]{\bf L}} $ 
& $\epsilon_\lambda^* \epsilon_\lambda$ 
& \cellcolor{grayLL} \textemdash \\  \hline %
$c^{H\square}$ 
& \cellcolor{grayLLL} $\makecell{\frac{g_1^2 + 3 g_2^2}{6}\text{tr}[\epsilon_N^\dag \epsilon_N] {\bf L} \\+ 2 \text{tr}[c^{W \dag} c^W]{\bf L}}$
& \cellcolor{grayLLL} $\makecell{\frac{g_1^2 - g_2^2}{2}\text{tr}[\epsilon_N^\dag \epsilon_N]{\bf L} \\+ 2 \text{tr}[c^{W \dag} c^W]{\bf L}}$ 
& \cellcolor{grayLL} \textemdash 
& \cellcolor{grayLL} \textemdash \\ \hline 
$c^{HD}$ 
& \cellcolor{grayLLL} $\makecell{\frac{2g_2^2}{3} \text{tr}[\epsilon_N^\dag \epsilon_N]{\bf L}\\+ 16 \text{tr}[c^{W \dag} c^W]{\bf L}}$
& \cellcolor{grayLLL} $\makecell{2g_2^2 \text{tr}[\epsilon_N^\dag \epsilon_N]{\bf L}\\+ 16 \text{tr}[c^{W \dag} c^W]{\bf L}}$ 
& \cellcolor{grayLL} \textemdash 
& \cellcolor{grayLL} \textemdash \\ \hline  
\end{tabular}
\caption{ \sl \small  
The leading-order contribution to the WCs of the SMEFT, in four neutrino-mass models. 
In this table we display the operators involving only the Higgs field. 
Conventions as in Table~\ref{table-WCs1}. We computed only WCs generated at tree level or one-loop leading-log: those induced at one-loop finite or higher order are shaded in dark grey and denoted by a dash (\textemdash).}
\label{table-WCs2} 
\end{table}

\begin{table}[!p]
\renewcommand{\arraystretch}{2}
\centering
\hspace*{-.5cm}
\arrayrulecolor{black!70}
\begin{tabular}{
!{\color{black}\vline}>{\centering\arraybackslash}m{0.8cm}
!{\color{black}\vline}>{\centering\arraybackslash}m{3.4cm}
!{\color{black}\vline}>{\centering\arraybackslash}m{3.4cm}
!{\color{black}\vline}>{\centering\arraybackslash}m{3.4cm}
!{\color{black}\vline}>{\centering\arraybackslash}m{3.4cm}
!{\color{black}\vline}}
\hline
\textbf{WCs} & Seesaw I & Seesaw III & Zee & Leptoquarks \\ \hline \hline
$c^{ll}_{abcd}$  
& \cellcolor{grayLLL} $\makecell{\frac{g_1^2-g_2^2}{24} (\epsilon^\dagger_N \epsilon_N)_{ab} \delta_{cd} {\bf L}
\\+ \frac{g_1^2-g_2^2}{24} \delta_{ab} (\epsilon_N^\dagger \epsilon_N)_{cd} {\bf L}
\\+ \frac{g_2^2}{12} (\epsilon_N^\dagger \epsilon_N)_{ad} \delta_{cb} {\bf L}
\\+ \frac{g_2^2}{12} \delta_{ad} (\epsilon_N^\dag \epsilon_N)_{cb} {\bf L}
\\+ 2(c^{W\dag})_{ac} (c^W)_{bd} {\bf L}}$ 
& \cellcolor{grayLLL} $\makecell{\frac{3g_1^2 + g_2^2}{24} (\epsilon_\Sigma^\dag \epsilon_\Sigma)_{ab} \delta_{cd}{\bf L} \\ + \frac{3g_1^2 + g_2^2}{24} \delta_{ab} (\epsilon_\Sigma^\dag \epsilon_\Sigma)_{cd}{\bf L} \\ - \frac{g_2^2}{12} (\epsilon_\Sigma^\dag \epsilon_\Sigma)_{ad} \delta_{cb}{\bf L} \\ - \frac{g_2^2}{12} \delta_{ad}(\epsilon_\Sigma^\dag \epsilon_\Sigma)_{cb} {\bf L}
	\\ + 2(c^{W\dag})_{ac} (c^W)_{bd}{\bf L}}$ 
& $(\epsilon_\delta^\dagger)_{ac} (\epsilon_\delta)_{db}$ 
& \cellcolor{grayLLL} $\makecell{ \frac{g_1^2}{6}(\epsilon_D^\dag \epsilon_D)_{ab} \delta_{cd} {\bf L} \\
+ \frac{g_1^2}{6} \delta_{ab} (\epsilon_D^\dag \epsilon_D)_{cd} {\bf L} \\
+ \frac{g_2^2}{2} (\epsilon_L^\dag \epsilon_L)_{ad} \delta_{bc} {\bf L} \\ 
+ \frac{g_2^2}{2} \delta_{ad} (\epsilon_L^\dag \epsilon_L)_{cb} {\bf L} \\ 
+ \frac{g_1^2 - 3 g_2^2}{12} (\epsilon_L^\dag \epsilon_L)_{ab} \delta_{cd} {\bf L} \\
+ \frac{g_1^2 - 3 g_2^2}{12} \delta_{ab} (\epsilon_L^\dag \epsilon_L)_{cd} {\bf L}}$ \\  \hline 
$c^{le}_{abcd}$ 
& \cellcolor{grayLLL} $\makecell{
	\frac{g_1^2}{6} (\epsilon_N^\dagger \epsilon_N)_{ab} \delta_{cd}{\bf L}}$ 
& \cellcolor{grayLLL} $\makecell{
	 \frac{g_1^2}{2} (\epsilon_\Sigma^\dag \epsilon_\Sigma)_{ab} \delta_{cd} {\bf L} } $ 
& $- \frac{1}{2}(\epsilon_2^\dag)_{ad} (\epsilon_2)_{cb}$ 
& \cellcolor{grayLLL} $\makecell{\frac{3}{2} (\epsilon_L^\dag y_u^T \epsilon_R)_{ad} (y_e)_{cb} {\bf L} \\
+ \frac{3}{2} (y_e^\dag)_{ad} (\epsilon_R^\dag y_u^* \epsilon_L)_{cb} {\bf L} \\
+ \frac{g_1^2}{3} (\epsilon_L^\dag \epsilon_L)_{ab} \delta_{cd} {\bf L} \\
+ \frac{2g_1^2}{3} (\epsilon_D^\dag \epsilon_D)_{ab} \delta_{cd} {\bf L} \\
+ \frac{2g_1^2}{3} \delta_{ab} (\epsilon_R^\dag \epsilon_R)_{cd} {\bf L}}$ \\  \hline 
$c^{ee}_{abcd}$ 
& \cellcolor{grayLL} \textemdash 
& \cellcolor{grayLL} \textemdash 
& \cellcolor{grayLLL} $\makecell{
 \frac{g_1^2}{3} ( \epsilon_2 \epsilon_2^\dag)_{ab} \delta_{cd} {\bf L} \\ 
+ \frac{g_1^2}{3} \delta_{ab} (\epsilon_2 \epsilon_2^\dag)_{cd} {\bf L} }$ 
& \cellcolor{grayLLL} $\makecell{ 
\frac{2g_1^2}{3} (\epsilon_R^\dag \epsilon_R)_{ab} \delta_{cd} {\bf L} \\
+ \frac{2g_1^2}{3} \delta_{ab} (\epsilon_R^\dag \epsilon_R)_{cd} {\bf L}}$ \\
\hline 
\end{tabular}
\caption{ \sl \small 
The leading-order contribution to the WCs of the SMEFT, in four neutrino-mass models. 
In this table we display the operators involving four leptons. 
Conventions as in Tables~\ref{table-WCs1} and \ref{table-WCs2}. 
}
\label{table-WCs3} 
\end{table}

\begin{table}[!p]
\renewcommand{\arraystretch}{2}
\centering
\hspace*{-0.5cm}
\arrayrulecolor{black!70}
\begin{tabular}{!{\color{black}\vline}>{\centering\arraybackslash}m{1cm}
!{\color{black}\vline}>{\centering\arraybackslash}m{3.5cm}
!{\color{black}\vline}>{\centering\arraybackslash}m{3.4cm}
!{\color{black}\vline}>{\centering\arraybackslash}m{3.4cm}
!{\color{black}\vline}>{\centering\arraybackslash}m{3.55cm}!{\color{black}\vline}}
\hline
\textbf{WCs} & Seesaw I & Seesaw III & Zee & Leptoquarks \\ \hline \hline
$c^{lq(1)}_{abcd}$ 
& \cellcolor{grayLLL} $\makecell{ 
- \frac{1}{4}(\epsilon_N^\dag \epsilon_N)_{ab} (y_u^\dag y_u)_{cd} {\bf L} \\ 
-\frac{g_1^2}{36}(\epsilon_N^\dag \epsilon_N)_{ab} \delta_{cd} {\bf L}}$ 
& \cellcolor{grayLLL} $\makecell{
- \frac{3}{4}(\epsilon_\Sigma^\dag \epsilon_\Sigma)_{ab} (y_u^\dag y_u)_{cd} {\bf L} \\ 
- \frac{g_1^2}{12} (\epsilon_\Sigma^\dag \epsilon_\Sigma)_{ab} \delta_{cd} {\bf L}}$ 
& \cellcolor{grayLLL} $ \makecell{ 
\frac{2g_1^2}{9} (\epsilon_\delta^\dag \epsilon_\delta)_{ab} \delta_{cd} {\bf L} \\ 
- \frac{g_1^2}{9} (\epsilon_2^\dag \epsilon_2 )_{ab} \delta_{cd} {\bf L} }$ 
& $\frac{1}{4} (\epsilon_L^\dagger)_{ac} (\epsilon_L)_{db}$ \\  \hline 
$c^{lq(3)}_{abcd}$ 
& \cellcolor{grayLLL} $\makecell{ 
-\frac{1}{4}(\epsilon_N^\dag \epsilon_N)_{ab} (y_u^\dag y_u)_{cd} {\bf L} \\
+\frac{g_2^2}{12}(\epsilon_N^\dag \epsilon_N)_{ab} \delta_{cd} {\bf L}}$ 
& \cellcolor{grayLLL} $\makecell{
\frac{1}{4}(\epsilon_\Sigma^\dag \epsilon_\Sigma)_{ab} (y_u^\dag y_u)_{cd} {\bf L} \\
- \frac{g_2^2}{12} (\epsilon_\Sigma^\dag \epsilon_\Sigma)_{ab} \delta_{cd} {\bf L} }$ 
& \cellcolor{grayLLL} $\frac{2g_2^2}{3} (\epsilon_\delta^\dag \epsilon_\delta)_{ab} \delta_{cd} {\bf L} $ 
& $-\frac{1}{4} (\epsilon_L^\dagger)_{ac} (\epsilon_L)_{db}$ \\  \hline 
$c^{qe}_{abcd}$ 
& \cellcolor{grayLL} \textemdash 
& \cellcolor{grayLL} \textemdash 
& \cellcolor{grayLLL} $- \frac{g_1^2}{9} \delta_{ab} (\epsilon_2 \epsilon_2^\dag)_{cd} {\bf L}$ 
& \cellcolor{grayLLL} $\makecell{ 
 \frac{1}{2} (\epsilon_L^* y_e^T)_{ac} (y_u^T \epsilon_R)_{bd} {\bf L} \\
+ \frac{1}{2} (y_u^\dag \epsilon_R^*)_{ac} (\epsilon_L y_e^\dag)_{bd} {\bf L} \\
+ \frac{1}{2} (y_u^\dag \epsilon_R^*)_{ac} (y_u^T \epsilon_R)_{wd} {\bf L} \\
- \frac{g_1^2}{3} (\epsilon_L^* \epsilon_L^T)_{ab} \delta_{cd} {\bf L} \\
- \frac{2g_1^2}{9} \delta_{ab} (\epsilon_R^\dag \epsilon_R)_{cd} {\bf L}}$ \\  \hline 
$c^{lu}_{abcd}$ 
& \cellcolor{grayLLL} $\makecell{ \frac{1}{2}(\epsilon_N^\dag \epsilon_N)_{ab} (y_u y_u^\dag)_{cd} {\bf L} \\
 - \frac{g_1^2}{9} (\epsilon_N^\dag \epsilon_N)_{ab}\delta_{cd} {\bf L} }$ 
& \cellcolor{grayLLL} $ \makecell{\frac{3}{2} (\epsilon_\Sigma^\dag \epsilon_\Sigma)_{ab} (y_u y_u^\dag)_{cd} {\bf L} \\ - \frac{g_1^2}{3} (\epsilon_\Sigma^\dag \epsilon_\Sigma)_{ab} \delta_{cd} {\bf L}} $ 
& \cellcolor{grayLLL} $ \makecell{ \frac{8g_1^2}{9} (\epsilon_\delta^\dag \epsilon_\delta)_{ab} \delta_{cd} {\bf L} \\ - \frac{4g_1^2}{9} (\epsilon_2^\dag \epsilon_2)_{ab} \delta_{cd} {\bf L} }$ 
& \cellcolor{grayLLL} $ \makecell{\frac{1}{2} ( \epsilon_L^\dag y_u^T )_{ac} (\epsilon_R y_e )_{db} {\bf L} \\
+\frac{1}{2} ( \epsilon_L^\dag y_u^T )_{ac} (y_u^* \epsilon_L )_{db} {\bf L} \\
+ \frac{1}{2} ( y_e^\dag \epsilon_R^\dag)_{ac} (y_u^* \epsilon_L )_{db} {\bf L} \\ 
- \frac{2g_1^2}{9} (\epsilon_L^\dag \epsilon_L)_{ab} \delta_{cd} {\bf L} \\
- \frac{4g_1^2}{9} (\epsilon_D^\dag \epsilon_D)_{ab} \delta_{cd} {\bf L} \\
- \frac{g_1^2}{3} \delta_{ab} (\epsilon_R^* \epsilon_R^T)_{cd} {\bf L}} $ \\  \hline 
$c^{ld}_{abcd}$ 
& \cellcolor{grayLLL} $\makecell{
	 \frac{g_1^2}{18} (\epsilon_N^\dag \epsilon_N)_{ab} \delta_{cd} {\bf L}}$ 
& \cellcolor{grayLLL} $\makecell{
	 \frac{g_1^2}{6}(\epsilon_\Sigma^\dag \epsilon_\Sigma)_{ab} \delta_{cd}{\bf L}}$ 
& \cellcolor{grayLLL} $ \makecell{ - \frac{4g_1^2}{9} (\epsilon_\delta^\dag \epsilon_\delta)_{ab} \delta_{cd} {\bf L} \\ + \frac{2g_1^2}{9} (\epsilon_2^\dag \epsilon_2)_{ab} \delta_{cd} {\bf L}}$ 
& $- \frac{1}{2} (\epsilon_D^\dag)_{ad} (\epsilon_D)_{cb} $ \\  \hline 
$c^{eu}_{abcd}$ & \cellcolor{grayLL} \textemdash
& \cellcolor{grayLL} \textemdash 
& \cellcolor{grayLLL} $- \frac{4g_1^2}{9} (\epsilon_2 \epsilon_2^\dag)_{ab} \delta_{cd} {\bf L}$ & $\frac{1}{2} (\epsilon_R^\dag)_{ac} (\epsilon_R)_{db} $ \\  \hline 
$c^{ed}_{abcd}$ 
& \cellcolor{grayLL} \textemdash 
& \cellcolor{grayLL} \textemdash  
& \cellcolor{grayLLL} $ \frac{2g_1^2}{9} (\epsilon_2 \epsilon_2^\dag)_{ab} \delta_{cd} {\bf L}$ 
& \cellcolor{grayLLL} $ \makecell{
	 \frac{2g_1^2}{3} \delta_{ab} (\epsilon_D \epsilon_D^\dag)_{cd} {\bf L} \\
+ \frac{4g_1^2}{9} (\epsilon_R^\dag \epsilon_R)_{ab} \delta_{cd} {\bf L}}$ \\  
\hline\hline 
$c^{lequ(1)}_{abcd}$ & \cellcolor{grayLL} \textemdash 
& \cellcolor{grayLL} \textemdash 
& \cellcolor{grayLLL} $2 \text{tr}[y_e^\dag \epsilon_2] (\epsilon_2^\dag)_{ab} (y_u^\dag)_{cd} {\bf L}$ & $- \frac{1}{2} (\epsilon_L^\dagger)_{ac} (\epsilon_R)_{db}$ \\  \hline 
$c^{lequ(3)}_{abcd}$ & \cellcolor{grayLL} \textemdash 
& \cellcolor{grayLL} \textemdash 
& \cellcolor{grayLL} \textemdash 
& $ \frac{1}{8} (\epsilon_L^\dagger)_{ac} (\epsilon_R)_{db}$ \\  \hline 
$c^{ledq}_{abcd}$ & \cellcolor{grayLL} \textemdash 
& \cellcolor{grayLL} \textemdash
& \cellcolor{grayLLL} $2 \text{tr}[y_e^\dag \epsilon_2] (\epsilon_2^\dag)_{ab} (y_d)_{cd} {\bf L}$ & \cellcolor{grayLLL} $\makecell{- 3 (\epsilon_L^\dag y_u^T \epsilon_R)_{ab} (y_d)_{cd} {\bf L} \\
- (\epsilon_D^\dag y_d)_{ad} (\epsilon_D y_e^\dag)_{cb} {\bf L} \\
+ (\epsilon_L^\dag y_d^T)_{ac} (\epsilon_L y_e^\dag)_{db} {\bf L} \\
+ (\epsilon_L^\dag y_d^T)_{ac} (y_u^T \epsilon_R)_{db} {\bf L}}$ \\  \hline 
\end{tabular}
\caption{ \sl \small 
The leading-order contribution to the WCs of the SMEFT, in four neutrino-mass models. 
In this table we display the operators involving two leptons and two quarks. 
Conventions as in Tables~\ref{table-WCs1} and \ref{table-WCs2}. 
}
\label{table-WCs4} 
\end{table}

Let us begin by displaying the tree-level EFT Lagrangians (up to dim-6 operators) for the four models under consideration.
They can be easily obtained by employing the equations of motion for the new, heavy fields, in order to express them as a function of the SM, light fields, and thus remove them from the initial Lagrangian.
Integrating out the sterile neutrinos $N_R$ of the type-I seesaw, one obtains \cite{Broncano:2002rw}
\begin{equation}
\mathcal{L}_\text{I}^\text{tree} = \frac{1}{\sqrt{2}v} (\epsilon_N^T \mu_N \epsilon_N)_{ab} Q_{\substack{W\\ab}} + \frac{1}{2v^2} (\epsilon_N^\dag \epsilon_N)_{ab} \left(Q_{\substack{Hl(1)\\ab}} - Q_{\substack{Hl(3)\\ab}}\right)~.
\label{type1EFT}
\end{equation}
Integrating out the triplet chiral fermions $\Sigma_R$ of the type-III seesaw, 
\begin{equation}
\mathcal{L}_\text{III}^\text{tree} = \frac{1}{\sqrt{2}v} (\epsilon_\Sigma^T \mu_\Sigma \epsilon_\Sigma)_{ab} Q_{\substack{W\\ab}} + \frac{1}{2v^2} (\epsilon_\Sigma^\dag \epsilon_\Sigma)_{ab} \left(3Q_{\substack{Hl(1)\\ab}} + Q_{\substack{Hl(3)\\ab}} \right) + \frac{2}{v^2} (\epsilon_\Sigma^\dag \epsilon_\Sigma y_e^\dag)_{ab} Q_{\substack{eH\\ab}} ~.
\label{type3EFT}
\end{equation}
Integrating out the scalars $H_2$ and $\delta$ of the Zee model,
\begin{equation}
\mathcal{L}_\text{Zee}^\text{tree} = \frac{2 \epsilon_\lambda^*}{v^2} (\epsilon_2^\dag)_{ab} Q_{\substack{eH\\ab}} + \frac{2}{v^2} (\epsilon_\delta^\dag)_{ac} (\epsilon_\delta)_{db} Q_{\substack{ll\\abcd}} - \frac{1}{v^2} (\epsilon_2)_{ad} (\epsilon_2^\dag)_{cb} Q_{\substack{le\\abcd}} + \frac{2 |\epsilon_\lambda|^2}{v^2} Q_H ~.
\label{zeeEFT}
\end{equation}
Integrating out the leptoquarks $D$ and $S$,
\begin{align}
\mathcal{L}_\text{LQ}^\text{tree} &= \frac{1}{2v^2} (\epsilon_L^\dag)_{ac} (\epsilon_L)_{db} \left(Q_{\substack{lq(1)\\abcd}} - Q_{\substack{lq(3)\\abcd}}\right) + \frac{1}{4v^2} (\epsilon_L^\dag)_{ac} (\epsilon_R)_{db} \left(Q_{\substack{lequ(3)\\abcd}} - 4Q_{\substack{lequ(1)\\abcd}} \right) \notag \\
&+ \frac{1}{v^2} (\epsilon_R^\dag)_{ac}(\epsilon_R)_{db} Q_{\substack{eu\\abcd}} - \frac{1}{v^2} (\epsilon_D^\dag)_{ad}(\epsilon_D)_{cb} Q_{\substack{ld\\abcd}} ~.
\label{lqEFT}
\end{align}
The operators induced at tree level may mix into other operators via one-loop diagrams.
The full set of one-loop RGEs for the SMEFT WCs are listed in \cite{Jenkins:2013zja,Jenkins:2013wua,Alonso:2013hga}. 
We employed them to identify the WCs generated via operator mixing, and to extract the corresponding one-loop leading-log contributions
in the four models.

In the Zee and the LQ models, the dim-5 Weinberg WC arises from finite one-loop diagrams, which we evaluated explicitly.
Similarly, the dim-6 lepton dipole WCs are one-loop finite in the Zee model as well as in the type-I and III seesaw models.
Since these WCs are crucial for phenomenology, we determined them by an explicit one-loop matching computation. 
We used Package-X \cite{Patel:2015tea} to cross-check some of our computations.

The leading-order contributions to the WCs in the four models are displayed in Tables~\ref{table-WCs1} (two leptons, no quarks), 
\ref{table-WCs2} (Higgs only), \ref{table-WCs3} (four leptons) and \ref{table-WCs4} (two leptons, two quarks). 
More precisely, the twenty-seven WCs listed across the tables are those generated either at tree level or at one-loop leading-log level, 
in at least one of the four models, with one exception: we did not display WCs of operators involving quarks but not leptons.
This is because in our subsequent phenomenological analysis we focus predominantly on leptonic observables, which are more constraining. 
In the case of the Weinberg and the dipole operators (Table~\ref{table-WCs1}) we also reported the one-loop finite piece of the WCs,
when it is the leading contribution. One should be aware that any linear combination of WCs, which vanishes according to our tables, is generally non-zero at the next subleading order (one more loop, or one less log).

The derivation of the EFTs of the four models under consideration has been partially completed in the literature. 
The tree-level computation has been performed for all cases (see e.g. \cite{deBlas:2017xtg} for a general tree-level analysis). 
Our results \cite{Coy:2018bxr} for the one-loop leading-log WCs in the type-I seesaw model was confirmed by the recent analysis \cite{Zhang:2021jdf}, which actually performs the full one-loop matching for this model. 
Our result \cite{Coy:2018bxr} for the one-loop finite WCs of the dipole operators in the type-I seesaw agrees with \cite{Zhang:2021tsq}.
In the case of the type-III seesaw, our results for the one-loop WCs are new. 
For the Zee model, the dim-6 WCs are generated either by the singlet scalar $\delta$, or by the second Higgs doublet $H_2$.
The one-loop matching of the SM extended by $\delta$ was performed in \cite{Bilenky:1993bt}: we agree with their results at leading-log order and with their one-loop finite dipole WCs. 
We here provided the complete set of WCs for a quark-phobic $H_2$ at one-loop leading-log order.
The WCs of bosonic operators in a two-Higgs-doublet model have been matched at one-loop in \cite{Henning:2014wua}.
The one-loop matching of the SM extended by $S$ was performed in \cite{Gherardi:2020det} and the matching of the SM augmented by both the $S$ and $D$ was completed by \cite{Dedes:2021abc}. 
We find general agreement, except for the one-loop finite contribution of the LQ $D$ to $c^{eB}$  in \cite{Dedes:2021abc}, see the discussion around Eq.~\eqref{cegLQ1}.
We also agree with the literature for the one-loop finite Weinberg operator WC in the Zee and LQ models.

\subsection{Matching at the electroweak scale}\label{SMtoL}

The SMEFT WCs displayed in Tables \ref{table-WCs1}-\ref{table-WCs4} are pertinent before electroweak symmetry breaking. 
In order to make contact with experiments performed at energies well below the electroweak 
scale $v$, one has to match onto the low-energy EFT (LEFT),
by integrating out electroweak-scale states, namely the $h$, $Z$ and $W$ bosons and the top quark, and then run down to the scale relevant to each experiment.

Just as for the SMEFT, we are interested in the leading contribution to each LEFT WC.
The expansion in the number of couplings and logarithms proceeds analogously to the SMEFT case detailed above. The typical logarithm is  
$\log(v/m_f)$ for an experiment whose relevant energy scale is the light-fermion mass $m_f$. 
The tree-level matching of the SMEFT onto the LEFT is relatively straightforward, and it is outlined e.g.~in Ref.~\cite{Jenkins:2017jig}, wherein the LEFT basis is also defined. The one-loop RGEs for the LEFT have been compiled in \cite{Jenkins:2017dyc}. 
For compactness we do not list all the LEFT WCs which arise at leading order in the four models under investigation. 
In section \ref{sec:SMEFTconstraints} we will rather provide specific combinations of LEFT WCs, when relevant for a given observable.

Let us discuss some general, relevant features of the LEFT, the effective theory of leptons and light quarks 
with gauge symmetry $SU(3)_c \times U(1)_{em}$.
Since QCD and QED conserve lepton and baryon numbers, flavour and CP, the violations of these symmetries reside entirely in the WCs determined by the matching at electroweak scale. 
While different scalar and tensor-current four-fermion operators mix under the RGEs, neither mix with vector-current four-fermion operators: this is relevant e.g. for the $\mu\to e$ conversion rate, 
see \eq{mueconversion}. 
We note that no other LEFT operator mixes into the operator 
$\mathcal{O}^{V,LL}_{\nu e,abba}\equiv (\bar\nu_{La}\gamma_\mu\nu_{Lb})(\bar e_{Lb}\gamma^\mu e_{La})$, 
which interferes with the SM amplitude for $\ell_a \to \ell_b \nu_a \bar{\nu}_b$. 
As will be discussed in section \ref{EWPT}, this process is used to measure $G_F$, which leads to several stringent bounds on the parameter space of our four models.

The electromagnetic dipole operator, $\mathcal{O}^{e\gamma}_{ab} \equiv (v/\sqrt{2}) \overline{\ell_a} \sigma_{\mu \nu} P_R \ell_b F^{\mu \nu}$, deserves special treatment. 
We adopt the normalisation $\mathcal{L}_{LEFT} \supset (2/v^2) c^{e\gamma}_{ab} \mathcal{O}^{e\gamma}_{ab} + h.c.$, consistent with our SMEFT conventions introduced at the start of section \ref{sec:structure}. 
Firstly, $c^{e\gamma}$ receives a UV contribution from the matching of the SMEFT dipole operators,
\begin{equation}
c^{e\gamma (UV)}_{ab} = c_w c^{eB}_{ab} - s_w c^{eW}_{ab} \, .
\label{cegMatchTree}
\end{equation}
Here $c^{eB,eW}$ are evaluated at the scale $v$, therefore they may include contributions both from one-loop diagrams at the new physics scale $M$, and from mixing with other SMEFT operators as one runs from $M$ to $v$.

Secondly, matching at the scale $v$ may provide a further correction to $c^{e\gamma}$ of a comparable size. 
In particular, SMEFT operators that do not mix into the dipole may still provide a finite one-loop contribution to the dipole, implying a significant constraint on their WC. 
Indeed, the SMEFT operators $Q_{Hl}^{(1)}$, $Q_{Hl}^{(3)}$ and $Q_{He}$ modify the couplings of leptons to the gauge bosons, $W$ and $Z$, see e.g.~Eqs.~\eqref{zlV}-\eqref{znu} for the shift in the $Z$ couplings. 
As a consequence, the SM one-loop electroweak contribution to the dipole, $c^{e\gamma(SM)}$, is shifted by \cite{Dekens:2019ept}
\begin{equation}
c^{e\gamma (IR)}_{ab} = \frac{e}{48\pi^2} \left[ (4 s_w^2 - 2) c^{Hl(1)} y_e^\dag + (3 + 4s_w^2) c^{Hl(3)} y_e^\dag + 4 s_w^2 y_e^\dag c^{He} \right]_{ab} \, .
\label{cegMatchLoop}
\end{equation}
Since $c^{Hl(1)}$, $c^{Hl(3)}$ and $c^{He}$ can be generated at tree-level at scale $M$, $c^{e\gamma(IR)}$ can be of the same order as $c^{e\gamma(UV)}$.
There are also contributions to $c^{e\gamma \text{ (IR)}}$ proportional to $(c^{W\dag} c^W y_e^\dag)$ from two insertions of the Weinberg operator, however these are completely negligible as neutrino masses are tiny. 
Next, the running from the electroweak scale down to a light-fermion mass scale $m_f$ generates a contribution to the dipole WC from mixing of other LEFT operators \cite{Jenkins:2017dyc},
\begin{equation}
c^{e\gamma(IR,log)}_{ab} = \frac{e}{8\pi^2}\left[ - 4\ c^{T,RR}_{\substack{ed \\ abcd}} (y_d)_{dc} 
+ 8\ c^{T,RR}_{\substack{eu \\ abcd}} (y_u)_{dc} + c^{S,RR}_{\substack{ee\\adcb}} (y_e)_{dc} \right] \log \frac{v}{m_f} \, .
\label{cegRGE}
\end{equation}
The tree-level matching of the SMEFT onto these LEFT operators reads 
\be
c^{T,RR}_{ed}=0~,\quad
c^{T,RR}_{\substack{eu\\abcd}} = - c^{lequ(3)}_{abcd}~,\quad
c^{S,RR}_{\substack{ee\\abcd}} = -\frac{v^2}{2m_h^2}\left[c^{eH}_{ab} (y_e^\dag)_{cd} + (y_e^\dag)_{ab} c^{eH}_{cd}\right]\, .
\ee
The size of $c^{e\gamma(IR,log)}$ may be comparable to $c^{e\gamma(UV)}$ and/or $c^{e\gamma(IR)}$ if at least one of these SMEFT WCs is 
induced at tree level. 
In summary, our full EFT result for $c^{e\gamma}$ is obtained by adding the contributions in Eqs.~(\ref{cegMatchTree}), 
(\ref{cegMatchLoop}), and (\ref{cegRGE}).

There may be also a non-perturbative contribution to $c^{e\gamma}$, at scales $\mu \lesssim 2$ GeV, from operators involving leptons and 
light quarks, see e.g.~\cite{Dekens:2018pbu}. 
This term will not be relevant for our phenomenological discussion, as we will not consider new physics coupled to light 
quarks.\footnote{In principle, this contribution could be significant in the LQ model, if generic couplings to light quarks were considered.} 
Note that, in some new physics models, some two-loop contributions to $c^{e\gamma}$ may also be relevant, even sometimes dominant: in particular,
we will include them when we address the phenomenology of the Zee model, in section \ref{ZeePheno}. 
Finally, let us also remark that, at the charged-lepton mass scale, besides the $c^{e\gamma}$ contribution, there may be LEFT four-lepton operators that give a one-loop finite contribution to dipole transitions
\cite{Aebischer:2021uvt}, as we will see in section \ref{radiativedecays}.
 
Except for the electromagnetic dipole operator, we will neglect one-loop finite contributions as well as two-loop corrections, as they provide only subleading contributions to the WCs of the other relevant LEFT operators.

\section{Experimental constraints on leptonic operators}
\label{sec:SMEFTconstraints}

In this section, we derive general bounds on combinations of WCs.
We focus mainly on leptonic observables, for the following reasons:
(i) new physics which induces neutrino masses necessarily modifies the lepton sector, therefore one expects effects in other leptonic observables;
(ii) constraints from lepton experiments are outstanding, e.g. the precision measurements of  $Z$-boson couplings to leptons \cite{ALEPH:2005ab},  and the stringent bounds on lepton flavour violation (LFV), most notably in $\mu \to e$ transitions \cite{TheMEG:2016wtm,Bellgardt:1987du,Dohmen:1993mp,Bertl:2006up};
(iii) in the near future one expects significantly better sensitivities, as well as tests of current anomalies, e.g. in the muon anomalous magnetic moment, $a_\mu\equiv (g-2)_\mu/2$ \cite{Bennett:2006fi,Aoyama:2020ynm,Abi:2021gix}.\footnote{The existence of a discrepancy between the SM prediction of $a_\mu$ and its experimental determination has recently been questioned by a new lattice QCD calculation of the hadronic vacuum polarisation contribution to $a_\mu$ \cite{Borsanyi:2020mff}.  
However, this result is intriguingly in tension with the calculation of that same contribution using dispersion relations, as reviewed in \cite{Aoyama:2020ynm}.}

One can place a bound on the combination of WCs corresponding to each given observable.
In principle, these combinations may depend on the full set of SMEFT WCs. 
However, we will take into account only the subset of SMEFT WCs listed in Tables \ref{table-WCs1}-\ref{table-WCs4} (and the corresponding subset of LEFT WCs which they match onto), which are those relevant for our models.
In fact, many lepton observables  only depend on this subset of WCs, and in this case our expressions apply to any UV model: this occurs for all flavour-violating observables,
as well as for the Fermi constant $G_F$, the $G_F$-universality tests, and the electric dipole moments (EDMs).
The other flavour-conserving observables may be affected, in a generic UV model, by the redefinition of the gauge-boson wavefunctions, the gauge couplings and the weak mixing angle, due to the SMEFT operators $Q_{HB}, Q_{HW}, Q_{HWB}$: since we neglect those, our expressions will not be fully general for $m_Z$, $s_w^2$,  the rate for $Z \to \ell_a^+ \ell_a^-$, and the magnetic dipole moments $a_{\ell}$, but they will still apply to our set of  models.
A compact summary of most bounds is provided in Table \ref{tab:WCbounds}.
For a few observables, the relevant combination of WCs is too lengthy to fit in the table, so we will display it 
in the main text. 
The constraints are all on dim-6 WCs, since the coefficient of the Weinberg operator is already required to be very small by neutrino masses, $c^W \sim (\sqrt{2}/v) m_\nu \lesssim 10^{-12}$. We do not consider operators with dimension larger than six. 

Our analysis has a significant overlap with previous articles discussing constraints on dim-6 WCs,
see in particular Refs.~\cite{Crivellin:2013hpa,Falkowski:2014tna,Berthier:2015oma,Feruglio:2015gka,Falkowski:2015krw,Falkowski:2017pss,Frigerio:2018uwx,Calibbi:2021pyh}.
We do improve some bounds, e.g. the contribution to  the magnetic dipole moment in \eq{magmomentEW} seems to have been overlooked in previous literature,
and we provide a perhaps more complete compilation of constraints in a coherent EFT framework. 
The model-independent bounds on WCs obtained in this section can be applied straightforwardly to a given UV model. We will specialise them to the WCs of the four neutrino-mass models, 
derived in section \ref{sec:EFTcomp}, in order to study their phenomenology in section \ref{sec:Pheno}.

\subsection{Corrections to $m_Z$ and $G_F$ and their implications}
\label{EWPT}

Three precisely-measured parameters, the mass of the $Z$ boson $m_Z$, the Fermi constant $G_F$, and the electromagnetic coupling $\alpha$, are the inputs to compute the SM predictions for other electroweak observables.
Therefore, the EFT operators which shift these parameters modify, in turn, the predictions for the derived observables. 
A contribution to the  $Z$-boson mass, i.e. a $Z_\mu Z^\mu$ term, arises from $c^{HD}$,
\begin{equation}\label{mZshift}
m_Z^2 \simeq m^2_{Z,0} (1 + c^{HD} ) \,,
\end{equation}
where $m_Z$ is the physical mass extracted from experiments, while $m^2_{Z,0}=(g_1^2+g_2^2)v^2/4$ is the usual combination of SM parameters.
We neglect other SMEFT operators shifting $m_Z$, namely $Q_{HB}$, $Q_{HW}$ and $Q_{HWB}$, as they are not generated in our models at tree-level, nor at one-loop leading-log order. 
The Fermi constant, extracted from the measurement of the width of muon decay, $\mu \to \nu_\mu e \bar{\nu}_e$, is given by
\begin{equation}
G_F \simeq G_{F,0} \left( 1 - c^G \right) ~,\qquad c^G \equiv c^{ll}_{e\mu \mu e} + c^{ll}_{\mu ee\mu} - 2 c^{Hl(3)}_{ee} - 2 c^{Hl(3)}_{\mu \mu} ~,
\label{gfShift}
\end{equation}
where $G_{F,0} \equiv 1/(\sqrt{2}v^2)$.

The EFT shift of the electromagnetic coupling $\alpha$ depends on the experiment chosen to define it.
For example, if one starts from the measurement of the electron anomalous magnetic moment, $a_e$ \cite{Hanneke:2008tm}, there is a 
(small) shift in $\alpha$ proportional to $y_e {\rm Re\,}c^{e\gamma}_{ee}$. 
Then, one can compare with the value of $\alpha$ extracted from atomic-frequency measurements. There is currently a significant 
discrepancy between the Caesium \cite{Parker:2018vye} and Rubidium \cite{Morel:2020dww} determinations.
Even including large systematics to accommodate this discrepancy, the new physics contribution cannot exceed roughly 
$|\Delta a_e| \lesssim 10^{-12}$,
which corresponds to an extremely stringent constraint, $|{\rm Re\,}c^{e\gamma}_{ee}| \lesssim 3 \cdot 10^{-8}$. 
Since all other observables that we will consider are measured with much less precision, from now on we can safely neglect the EFT shift in $\alpha$.

Let us come to observables derived from $m_Z$, $G_F$ and $\alpha$. The (tree-level) SM predictions for the  $W$-boson mass and the weak mixing angle can be written as
\be 
m^2_W =  \frac{m^2_Z}{2}\left[1+\left(1-\frac{2\sqrt{2}\pi\alpha}{G_F m_Z^2}\right)^{1/2}\right]  ~,\qquad
\sin^2 2\theta_w = \frac{2\sqrt{2}\pi\alpha}{G_F m_Z^2}~.
\label{mWsw}\ee 
On the other hand, the EFT prediction $m_{W,0}$ is obtained from an analogous expression, but with 
$m_Z\to m_{Z,0}$ and $G_F\to G_{F,0}$.
One obtains
\begin{equation}
m_{W,0} \simeq m_W \left[ 1 + \frac{s_w^2 c^G - c_w^2 c^{HD}}{2(1-2s_w^2)} \right]\, ,
\label{mW0}\end{equation}
where $m_W = 80.3505 \pm 0.0077$ \cite{deBlas:2022hdk}
is the 
SM prediction (including radiative corrections).
Note there is no dim-6 operator in the SMEFT which can induce a mass term $W_\mu W^\mu$. 
One should therefore directly match $m_{W,0}$ (without a further shift) to the 
measured value. After the recent $m_W$ measurement performed by the CDF experiment \cite{CDF:2022hxs}, a conservative global average of experiments gives $m_{W,exp} = 80.413 \pm 0.015$ GeV \cite{deBlas:2022hdk}, in $3.7\sigma$ tension with the SM. 
We conservatively allow for $m_{W,exp} - 4 \sigma \leq m_{W,0} \leq m_{W,exp} + 2 \sigma$, in order not to exclude new physics models which negligibly affect $m_W$. 
The resulting bound is reported in Table \ref{tab:WCbounds}.

Similarly, the EFT prediction for the weak mixing angle is given by
\begin{equation}
s_{w,0}^2 \simeq s_w^2 \left[ 1 + \frac{c_w^2 (c^{HD} - c^G)}{1-2s_w^2} \right] \,.
\label{sw0}
\end{equation}
In the following, we shall employ $s^2_{w,0}$ in the expression of observables depending on the weak mixing angle.
The mixing angle can be extracted from the measurement of different observables. In general, each such observable will receive a specific correction from the EFT WCs, so one can write
schematically $s^2_{w,exp} = s^2_{w,0} + \Delta s^2_{w,0}$. As an example, let us consider the effective leptonic weak mixing angle.
It is customary to introduce an effective angle for each SM fermion species, by an appropriate combination of $Z$ couplings to fermions, $\bar{s}_f^2 \equiv (g^V_{f}/g^A_{f} - 1)/4 Q_f $,
where $g^{V,A}_f$ are defined by \eq{Zcouplings} and $Q_f$ is the electric charge. 
LEP measured the effective leptonic weak mixing angle, from various asymmetries  in $Z$ decays: $\bar{s}_{\ell,exp}^2= 0.23153 \pm 0.00004$ \cite{ALEPH:2005ab,pdg}.
We can compute this quantity by (i) inserting $s^2_{w,0}$ in the SM expression for $g^{V,A}_\ell$ and (ii) 
adding the corrections to the $Z$ couplings to charged leptons, given in Eqs.~\eqref{zlV} and \eqref{zlA}.\footnote{
Ours is just a rough estimate, as the LEP measurement actually involves a combination of various (both quark and lepton) asymmetries. It is difficult to consistently model all EFT corrections 
to the experimental value, still we expect that \eq{swexp} captures the correct size of the leading EFT correction in our models.}
Note we are taking into account only the subset of SMEFT operators listed in Table \ref{tab:ops}. 
We thus obtain
\begin{equation}
\bar{s}_{\ell,exp}^2 \simeq s_{w,0}^2 - 2 s_{w,0}^2 (c^{Hl(1)} + c^{Hl(3)}) - (1 - 2 s_{w,0}^2) c^{He} \, ,
\label{swexp}
\end{equation}  
where $c^{Hl(1)}\equiv \sum_a c^{Hl(1)}_{aa}/3$ and similarly for the other WCs, as the LEP measurement assumes lepton universality.
Now, we can use \eq{sw0} to rewrite $s^2_{w,0}$ as a function of $s^2_w$, the latter being the SM prediction at tree level. 
Finally, in order to include SM radiative corrections, we replace $s^2_w \to \bar s^2_\ell$, where $\bar s^2_\ell = 0.23148\pm 0.00012$ is the full SM prediction from \cite{deBlas:2016ojx}.
By comparing $\bar{s}_{\ell,exp}^2$ with $\bar s^2_\ell$  we obtain a bound on a combination of WCs, reported
in Table \ref{tab:WCbounds}.

Just as the muon decay is modified by the EFT operators, so too are the decays $\tau \to \nu_\tau \ell_a  \bar{\nu}_a$ for $a = e,\mu$. 
This leads to a shift in the corresponding Fermi constant, $G_F^{a\tau}$, analogous to that of Eq.~\eqref{gfShift}. 
Then the ratio of two effective Fermi constants at leading order is given by
\begin{equation}
\frac{G^{a b}_F}{G^{ac}_F} -1 \simeq c^{ll}_{acca} + c^{ll}_{caac} - c^{ll}_{abba} - c^{ll}_{baab} + 2 c^{Hl(3)}_{bb} - 2 c^{Hl(3)}_{cc} \, .
\label{gfratio}
\end{equation}
Precise experimental determinations of these ratios provide bounds on the associated combination of WCs, for different values of $a,b,c$, see Table \ref{tab:WCbounds}. 
We considered $2\sigma$ limits, except for the constraint from $G_F^{\mu \tau}/G_F$, where there is a $\sim 2\sigma$ discrepancy between data and the SM prediction: in that case we allowed for a $3\sigma$ deviation in the direction of the SM, so as not to rule out vanishing WCs.

The effects of the EFT shift in $m_Z$ and $G_F$ 
propagates to other observables, as will be discussed in the following when relevant.

\subsection{Higgs boson decays}

The Higgs boson couplings to charged leptons can be written as 
\begin{equation}
\mathcal{L}_h = - \frac{1}{\sqrt{2}} \overline{\ell_a} h (Y_e)_{ab} P_L \ell_b + h.c. \, .
\end{equation}
There are two EFT effects: firstly, $c^{eH}$ modifies the Higgs coupling to charged leptons and the charged lepton mass matrix; secondly, $c^{H\square}$ and $c^{HD}$ modify the Higgs kinetic term.
The effective Higgs-charged lepton coupling is consequently modified to
\begin{equation}
(Y_e)_{ab} \simeq \frac{\sqrt{2} m_a}{v} \left( 1 + 2 c^{H\square} - \frac{1}{2} c^{HD} \right) \delta_{ab} 
- 2 \left(c^{eH} \right)^\dag_{ab} \, ,
\end{equation}
at $\mathcal{O}(c)$ in the EFT. 
The rate of LFV Higgs decays is ($a\ne b$)
\begin{equation}
\Gamma(h \to \ell_a \ell_b) \equiv \Gamma(h \to \ell_a^+ \ell_b^-) + \Gamma(h \to \ell_a^- \ell_b^+) \simeq \frac{m_h}{4\pi} \left( | c^{eH}_{ab}|^2 + | c^{eH}_{ba}|^2 \right) \, .
\end{equation}
Comparing with current bounds \cite{Aad:2019ojw,Sirunyan:2021ovv} and future limits expected of the HL-LHC \cite{Banerjee:2016foh} lead to the constraints given in Table \ref{tab:WCbounds}. 
The constraints from flavour-conserving Higgs decays are too weak to set any useful bound on the associated WCs.

\subsection{$Z$ boson decays}
\label{zdecays}

The $Z$ boson couplings to fermions can be parametrised by
\begin{equation}
\mathcal{L}_Z = - \frac{e}{2s_w c_w} Z^\mu \overline{f_a} \gamma_\mu \left( g^V_{f,ab} - g^A_{f,ab} \gamma_5 \right) f_b \, .
\label{Zcouplings}\end{equation}
In the SM, $g^V_{f,ab} = [T_3 (f_L) - 2 s_w^2 Q(f)]\delta_{ab}$ and $g^A_{f,ab} = T_3(f_L)\delta_{ab}$ at tree level. 
These effective couplings are modified by EFT operators in two ways. 
Firstly, $c^{Hl(1,3)}$ and $c^{He}$ generate explicit corrections in the Lagrangian, namely 
\begin{align}
\delta g^V_{\ell,ab} &= - c^{Hl(1)}_{ab} - c^{Hl(3)}_{ab} - c^{He}_{ab}  \, , \label{zlV} \\
\delta g^A_{\ell,ab} &= - c^{Hl(1)}_{ab} - c^{Hl(3)}_{ab} + c^{He}_{ab}  \, , \label{zlA} \\
\delta g^V_{\nu,ab} &= \delta g^A_{\nu,ab} = c^{Hl(3)}_{ab} - c^{Hl(1)}_{ab} \, \label{znu} .
\end{align} 
Secondly, in the SM expression for $g^V_{f,ab}$ one should shift $s^2_w$ to $s^2_{w,0}$, given by \eq{sw0}. 
The rate for flavour-violating $Z$ decays to charged leptons is then
\begin{align}
&\Gamma(Z \to \ell_a \ell_b) \equiv \Gamma(Z \to \ell_a^+ \ell_b^-) + \Gamma(Z \to \ell_a^- \ell_b^+) \notag \\
&\simeq \frac{\sqrt{2}m_Z^3 G_F}{3\pi} 
\left( | c^{Hl(1)}_{ab} + c^{Hl(3)}_{ab}|^2 + | c^{He}_{ab}|^2 + \frac{1}{2}\left| s_w c^{eB}_{ab} + c_w c^{eW}_{ab} \right|^2 + \frac{1}{2}\left| s_w c^{eB}_{ba} + c_w c^{eW}_{ba} \right|^2 \right) \, .
\end{align}
Comparing with bounds from the LHC leads to the constraints in Table \ref{tab:WCbounds}. 
In the flavour-violating decays considered above, the shifts of $m_Z$ and $G_F$
enter in the partial width only at $\mathcal{O}(c^3)$, and thus we neglect them. 

In contrast, flavour-conserving $Z$-decay rates are modified at $\mathcal{O}(c)$, both by corrections to $g^{V,A}_\ell$ and by the shift 
of an overall factor $m_Z^2 G_F \to m^2_{Z,0} G_{F,0}$ (such factor stands for the combination of SM couplings which enters the decay rate,
while the additional power of $m_Z$ in the decay rate is due to kinematics, therefore it is not shifted).  
Combining the various EFT modifications leads to
\begin{align}
\Gamma(Z \to \ell_a^+ \ell_a^-) \simeq 
\Gamma(Z \to \ell_a^+ \ell_a^-)^\text{SM} &\left[1 + \frac{1-2s_w^2-4s_w^4}{(1 - 2s_w^2)(1 - 4 s_w^2 + 8 s_w^4)} (c^G - c^{HD}) \right. \notag \\
& \left. + \frac{4 - 8 s_w^2}{1-4s_w^2 + 8 s_w^4} (c^{Hl(1)}_{aa} + c^{Hl(3)}_{aa}) - \frac{8 s_w^2}{1 - 4s_w^2 + 8 s_w^4} c^{He}_{aa} \right] \, .
\end{align}
From this expression we obtain the limits listed in Table \ref{tab:WCbounds}.

The effective number of neutrinos is defined by \cite{ALEPH:2005ab}
\begin{equation}
N_\nu \equiv \sum \limits_{ab=e,\mu,\tau} \frac{\Gamma(Z \to \nu_a \overline{\nu_b})}
{\Gamma(Z \to \nu_c \overline{\nu_c})^\text{SM}} \
\frac{\Gamma(Z \to \ell_c^+ \ell_c^-)^\text{SM}}{\Gamma(Z \to \ell_c^+ \ell_c^-)}  ~ ,
\label{nnu}\end{equation}
where $c$ stands for a single flavour (no sum). Since in \cite{ALEPH:2005ab} the decays into charged leptons are assumed to be flavour universal, we define $\Gamma(Z \to \ell_c^+ \ell_c^-)$ as the average over the three flavours. 
The ratio of $Z$ decay widths in \eq{nnu} is affected by the shift in the $Z$ couplings to leptons, given in Eqs.~\eqref{zlV}-\eqref{znu}, and by
the shift of $s^2_w$ to $s^2_{w,0}$, in the SM expression for $g^V_{\ell,cc}$. 
Then, the prediction for $N_\nu$ at $\mathcal{O}(c)$ is given by
\begin{align}
N_\nu & = 3 - \frac{12 s_w^2 c_w^2 (1 - 4 s_w^2)}{(1-2s_w^2)(1-4s_w^2 + 8 s_w^4)} (c^G - c^{HD}) \notag \\
&+ \frac{8}{1-4s_w^2 + 8 s_w^4} \sum \limits_a \left[ s_w^2 c^{He}_{aa} - (1 - 3 s_w^2 + 4 s_w^4) c^{Hl(1)}_{aa} - s_w^2 (1 - 4s_w^2) c^{Hl(3)}_{aa} \right] \, .
\end{align}
We note that the value of $N_\nu$ measured at LEP had a long-standing $2\sigma$ tension with the SM \cite{ALEPH:2005ab}. 
However, an updated calculation of the Bhabha scattering cross-section has recently reduced the disagreement to only $0.5\sigma$ 
\cite{Janot:2019oyi}, leading to the constraint in Table \ref{tab:WCbounds}.

We focused on $Z$-boson decays while neglecting $W$-boson decays, since the former are measured with better precision by a factor of a few compared to the latter \cite{ALEPH:2005ab,pdg}.

\subsection{Lepton decays into three charged leptons}\label{eto3e}

Stringent constraints on LFV come from the experimental bounds on the decays of a charged lepton into three charged leptons. 
There are three types of such decays, $\ell_a^- \to \ell_b^- \ell_b^+ \ell_b^-$, $\ell_a^- \to \ell_b^- \ell_b^+ \ell_c^-$, and $\ell_a^- \to \ell_b^- \ell_c^+ \ell_b^-$, where $a \neq b \neq c$.
These processes are induced at tree level in the SMEFT by four-lepton operators, as well as by two-lepton operators involving Higgs fields or a gauge field-strength, see Refs.~\cite{Kuno:1999jp,Crivellin:2013hpa,Frigerio:2018uwx}.

The process $\ell_a^- \to \ell_b^- \ell_c^+ \ell_b^-$ violates flavour by two units, therefore it is induced by four-lepton operators only, 
with rate
\begin{equation}
\Gamma(\tau^- \to \ell_b^- \ell_c^+ \ell_b^- ) = \frac{m_\tau^5}{384\pi^3 v^4} \left( |c_{b\tau bc}^{le}|^2 + |c_{bcb \tau}^{le}|^2 + 2 |c^{ll}_{b \tau bc} + c^{ll}_{bcb \tau}|^2 + 2|c^{ee}_{b \tau bc} + c^{ee}_{bcb \tau}|^2 \right) \, .
\label{taubcb}
\end{equation}
The corresponding bounds are reported in Table \ref{tab:WCbounds}. 
The rates for $\ell_a^- \to \ell_b^- \ell_b^+ \ell_b^-$ and $\ell_a^- \to \ell_b^- \ell_b^+ \ell_c^-$ similarly involve four-lepton WCs, as well as several other WCs. 
The corresponding complete, lengthy combinations of WCs can be found for instance in Refs.~\cite{Kuno:1999jp,Crivellin:2013hpa,Frigerio:2018uwx}. 
In the context of each specific model, we will present the simplified expression for these rates, and discuss the associated restrictions on the parameter space.
Current (future expected) $90\%$ C.L. experimental limits for these decays are \cite{Bellgardt:1987du,Hayasaka:2010np,Blondel:2013ia,Kou:2018nap}
	\begin{align}
		&BR(\mu^- \to e^-e^+e^-) < 1 \times 10^{-12} (1 \times 10^{-16}) \, , &\\
		&BR(\tau^- \to e^-e^+e^-) < 27 (0.43) \times 10^{-9} \, , &
		&BR(\tau^- \to \mu^-\mu^+\mu^-) < 21 (0.33) \times 10^{-9} \, , &\\
		&BR(\tau^- \to e^-e^+ \mu^-) < 18 (0.3) \times 10^{-9} \, ,& 
		&BR(\tau^- \to \mu^-\mu^+ e^-) < 27 (0.45) \times 10^{-9} \, .&
	\end{align}

\begin{table}[!htbp]
\renewcommand{\arraystretch}{1.4}
\hspace*{-1.3cm}
\centering
\begin{tabular}{|c|c|c|}
\hline
Observable & Bound & C.L. \\ \hline \hline
$m_W$ & $ 0.23 c^G - 0.77 c^{HD} \in 
[-0.6,13] \times 10^{-4}$ \cite{deBlas:2022hdk} & \ref{EWPT} \\ \hline
$s_w^2$ & $c^{HD} - c^G - 1.40 c^{Hl(1+3)} - 1.62 c^{He} \in [-6.1, 9.1] \times 10^{-4}$ \cite{ALEPH:2005ab,pdg} & $2\sigma$ \\ \hline
$G_F^{\mu \tau}/G_F^{e\tau}$ & $c^{ll}_{\tau ee \tau} + c^{ll}_{e \tau \tau e} - c^{ll}_{\tau \mu \mu \tau} - c^{ll}_{\mu \tau\tau \mu} + 2 c^{Hl(3)}_{\mu\mu} - 2 c^{Hl(3)}_{ee} \in [-1.0, 4.6] \times 10^{-3}$ \cite{Pich:2013lsa} & $2\sigma$ \\ \hline
$G_F^{e\tau}/G_F$ & $c^{ll}_{e \mu \mu e} + c^{ll}_{ \mu ee \mu} - c^{ll}_{e \tau \tau e} - c^{ll}_{\tau ee \tau} + 2 c^{Hl(3)}_{\tau\tau} - 2 c^{Hl(3)}_{ \mu \mu}  \in [-1.9, 4.1]\times 10^{-3}$ \cite{Pich:2013lsa} & $2\sigma$ \\ \hline
$G_F^{\mu\tau}/G_F$ & $c^{ll}_{ \mu ee  \mu} + c^{ll}_{e \mu \mu e} - c^{ll}_{ \mu \tau\tau  \mu} - c^{ll}_{\tau \mu \mu \tau} + 2 c^{Hl(3)}_{\tau\tau} - 2 c^{Hl(3)}_{ee} \in [-1.5, 6.0]\times 10^{-3}$ \cite{Pich:2013lsa} & \ref{EWPT} 
\\ \hline \hline
$h \to e \mu$ & $\sqrt{|c^{eH}_{e\mu}|^2 + |c^{eH}_{\mu e}|^2} \lesssim 1.6(0.8) \times 10^{-4}$ \cite{Aad:2019ojw} (\cite{Banerjee:2016foh}) & $95\%$ \\ \hline
$h \to e \tau$ & $\sqrt{|c^{eH}_{e\tau}|^2 + |c^{eH}_{\tau e}|^2} \lesssim 9.6(3.6) \times 10^{-4} $ \cite{Sirunyan:2021ovv} (\cite{Banerjee:2016foh}) & $95\%$ \\ \hline
$h \to \mu \tau$ & $\sqrt{|c^{eH}_{\mu\tau}|^2 + |c^{eH}_{\tau \mu}|^2} \lesssim 8.0(3.6) \times 10^{-4} $ \cite{Sirunyan:2021ovv} (\cite{Banerjee:2016foh}) & $95\%$ 
\\ \hline \hline
$Z \to e \mu$ & $\sqrt{| c^{Hl(1)}_{e\mu} + c^{Hl(3)}_{e\mu}|^2 + |c^{He}_{e\mu}|^2 + \frac{1}{2}\left| s_w c^{eB}_{e\mu} + c_w c^{eW}_{e\mu} \right|^2 + \frac{1}{2}\left| s_w c^{eB}_{\mu e} + c_w c^{eW}_{\mu e} \right|^2} \lesssim 1.2 \times 10^{-3}$ \cite{Aad:2014bca} & $95\%$ \\ \hline
$Z \to e \tau$ & $\sqrt{| c^{Hl(1)}_{e\tau} + c^{Hl(3)}_{e\tau} |^2 + |c^{He}_{e\tau}|^2 + \frac{1}{2}\left| s_w c^{eB}_{e\tau} + c_w c^{eW}_{e\tau} \right|^2 + \frac{1}{2}\left| s_w c^{eB}_{\tau e} + c_w c^{eW}_{\tau e} \right|^2} \lesssim 3.1 \times 10^{-3}$  \cite{Aad:2021jpw} & $95\%$ \\ \hline
$Z \to \mu \tau$ & $\sqrt{| c^{Hl(1)}_{\mu\tau} + c^{Hl(3)}_{\mu\tau} |^2 + |c^{He}_{\mu\tau}|^2 + \frac{1}{2}\left| s_w c^{eB}_{\mu\tau} + c_w c^{eW}_{\mu\tau} \right|^2 + \frac{1}{2}\left| s_w c^{eB}_{\tau \mu} + c_w c^{eW}_{\tau \mu} \right|^2} \lesssim 3.5 \times 10^{-3}$  \cite{Aad:2021jpw} & $95\%$ \\ \hline
$Z \to e^+ e^-$ & $1.19 (c^G - c^{HD}) + 4.27 (c^{Hl(1)}_{ee} + c^{Hl(3)}_{ee}) - 3.68 c^{He}_{ee} \in [-4.2 , 2.0] \times 10^{-3}$ \cite{ALEPH:2005ab,pdg} & $2\sigma$ \\ \hline
$Z \to \mu^+ \mu^-$ & $1.19 (c^G - c^{HD}) + 4.27 (c^{Hl(1)}_{\mu \mu} + c^{Hl(3)}_{\mu \mu}) - 3.68 c^{He}_{\mu \mu} \in [-4.7, 4.3] \times 10^{-3}$  \cite{ALEPH:2005ab,pdg} & $2\sigma$ \\ \hline
$Z \to \tau^+ \tau^-$ & $1.19 (c^G - c^{HD}) + 4.27 (c^{Hl(1)}_{\tau \tau} + c^{Hl(3)}_{\tau \tau}) - 3.68 c^{He}_{\tau \tau} \in [-2.2, 8.2] \times 10^{-3}$ \cite{ALEPH:2005ab,pdg} & $2\sigma$ \\ \hline
$N_\nu$ & $0.58 (c^{HD} - c^G) + 11.1 c^{He} - 24.8 c^{Hl(1)} - 0.82 c^{Hl(3)} \in [-0.019, 0.011]$ \cite{ALEPH:2005ab,pdg,Janot:2019oyi} & $2\sigma$ 
\\ \hline \hline
$\tau^- \to e^- \mu^+ e^-$ 
& $|c_{e \tau e \mu}^{le}|^2 + |c_{e \mu e \tau}^{le}|^2 + 2 |c^{ll}_{e \tau e \mu} + c^{ll}_{e \mu e \tau}|^2 
+ 2|c^{ee}_{e \tau e \mu} + c^{ee}_{e \mu e \tau}|^2 \lesssim 8.4 (0.2)\times 10^{-8}$ \cite{Hayasaka:2010np} (\cite{Kou:2018nap})  & $90\%$
\\ \hline
$\tau^- \to \mu^- e^+ \mu^-$ 
& $|c_{\mu \tau \mu e}^{le}|^2 + |c_{\mu e \mu \tau}^{le}|^2 + 2 |c^{ll}_{\mu \tau \mu e} + c^{ll}_{\mu e \mu \tau}|^2 
+ 2 |c^{ee}_{\mu \tau \mu e} + c^{ee}_{\mu e \mu \tau}|^2 \lesssim 9.5(0.2)\times 10^{-8}$ \cite{Hayasaka:2010np} (\cite{Kou:2018nap}) & $90\%$
\\ \hline \hline
$\mu \to e \gamma$ & $\sqrt{ |c^{e\gamma,obs}_{e\mu}|^2 + |c^{e\gamma,obs}_{\mu e}|^2 } \lesssim 6.4(2.4) \times 10^{-12}$ 
\cite{TheMEG:2016wtm} (\cite{MEGII:2018kmf}) & $90\%$ \\ \hline
$\tau \to e \gamma$ & $\sqrt{ |c^{e\gamma,obs}_{e\tau}|^2 + |c^{e\gamma,obs}_{\tau e}|^2 } \lesssim 7.1(2.1) \times 10^{-8}$ 
\cite{Aubert:2009ag} (\cite{Kou:2018nap}) & $90\%$ \\ \hline
$\tau \to \mu \gamma$ & $\sqrt{ |c^{e\gamma,obs}_{\mu\tau}|^2 + |c^{e\gamma,obs}_{\tau \mu}|^2 } \lesssim 8.2(1.2) \times 10^{-8}$ 
\cite{Aubert:2009ag} (\cite{Kou:2018nap}) & $90\%$ 
\\ \hline\hline
$a_e$ & $|\text{Re}\ c^{e\gamma,obs}_{ee} | \lesssim 3 \times 10^{-8} $ \cite{Hanneke:2008tm,Parker:2018vye,Morel:2020dww} 
& \ref{dipolemoments}
\\ \hline
$a_\mu$ & $\text{Re}[ c^{e\gamma,obs}_{\mu \mu} + 4.3 \times 10^{-7} (c^G - c^{HD}) ] \in [-0.5, 4.6 ] \times 10^{-7}$ \cite{Aoyama:2020ynm,Abi:2021gix} 
& \ref{dipolemoments}
\\ \hline
$d_e$ & $\left| \text{Im}\ c^{e\gamma,obs}_{ee}\right| \lesssim 1.5 \times 10^{-14}$ \cite{Andreev:2018ayy} & $90\%$
\\ \hline
$d_\mu$ & $\left| \text{Im}\ c^{e\gamma,obs}_{\mu\mu} \right| \lesssim 2.5 \times 10^{-4}$ \cite{Bennett:2008dy} & $95\%$
\\ \hline
\end{tabular}
\caption{
Experimental constraints on the leptonic WCs. 
The expected future bounds are in parentheses. 
The definitions of $c^G$ and $c^{e\gamma,obs}$ are given in Eqs.~\eqref{gfShift} and \eqref{cegobs}. 
The WCs $c^{Hl(1,3)}$ and $c^{He}$ with no subscript are averaged over flavour, as defined below Eq.~\eqref{swexp}. 
Bounds on a few additional observables ($\ell_a^- \to \ell_b^- \ell_b^+ \ell_b^-$, $\ell_a^- \to \ell_b^- \ell_b^+ \ell_c^-$ and $\mu \to e$ conversion in nuclei) depend on more
lengthy combinations of WCs, and are discussed in the text (sections \ref{eto3e} and \ref{mutoe}, respectively). 
In the last column we provide the confidence level for each bound or, if more specification is needed, the section where the bound is derived. 
}
\label{tab:WCbounds} 
\end{table}

\subsection{Radiative charged-lepton decays}
\label{radiativedecays}
Some of the strongest bounds on new physics in the lepton sector come from radiative charged-lepton decays. 
The rate of $\ell_a \to \ell_b \gamma$ decays is given by
\begin{equation}
BR(\ell_a \to \ell_b \gamma) \simeq \frac{m_a^3}{2\pi v^2 \Gamma_a} \left( |c^{e\gamma,obs}_{ab}|^2 + |c^{e\gamma,obs}_{ba}|^2 \right) \,,
\end{equation}
where $\Gamma_a$ is the total width of the charged lepton $\ell_a$.
Here, $c^{e\gamma,obs}$ is the effective combination of WCs that enters into dipole observables. 
It is given by
\begin{align}
c^{e\gamma,obs}_{ab} &= c^{e\gamma}_{ab} - \frac{e}{16\pi^2} c^{V,LR}_{\substack{ee\\acdb}} (y_e^\dag)_{cd} \, .
\label{cegobs}
\end{align}
The first term, the value of $c^{e\gamma}$ at the scale of the decaying lepton mass, is given by the sum of the contributions from Eqs.~\eqref{cegMatchTree}, \eqref{cegMatchLoop} and \eqref{cegRGE}. 
The second term corresponds (see e.g. \cite{Aebischer:2021uvt}) to the one-loop contribution to dipole observables of a LEFT four-lepton operator, 
$\mathcal{O}^{V,LR}_{ee,abcd} \equiv (2/v^2) ( \overline{\ell}_a \gamma_\mu P_L \ell_b) ( \overline{\ell}_c \gamma^\mu P_R \ell_d)$, whose matching to the SMEFT reads
\begin{equation}
c^{V,LR}_{\substack{ee\\abcd}} = c^{le}_{abcd} + 2 s_w^2 \left( c^{Hl(1)} + c^{Hl(3)} \right)_{ab} \delta_{cd} - (1 - 2s_w^2) \delta_{ab} c^{He}_{cd} \,.
\end{equation}
Comparing with experiments leads to the extremely stringent bounds listed in Table \ref{tab:WCbounds}, which are expected to be strengthened by a factor of a few in the MEG upgrade (for $\mu \to e \gamma$) \cite{MEGII:2018kmf} and at Belle-II (for $\tau \to e \gamma, \mu \gamma$) \cite{Kou:2018nap}.

\subsection{Magnetic and electric dipole moments}
\label{dipolemoments}

The EFT modifications to the magnetic and electric dipole moments take two forms. 
Firstly, there is the direct shift from the EM dipole operator, given by
\begin{align}
\Delta a_a^\text{(EM)} &\simeq \frac{4\sqrt{2} m_a}{e v} \text{Re}[c^{e\gamma,obs}_{aa}] \, , \label{magmoment} \\
\Delta d_a^\text{(EM)} &\simeq -\frac{2\sqrt{2}}{v} \text{Im}[c^{e\gamma,obs}_{aa}] \label{electricmoment} \, ,
\end{align}
with $c^{e\gamma,obs}$ as in Eq.~\eqref{cegobs}. 
Secondly, 
as the $W$ and $Z$ couplings to leptons are corrected by the shift in the weak mixing angle, given in \eq{sw0}, the SM one-loop electroweak contribution to $a_a$ is correspondingly 
shifted by	
\begin{equation}
\Delta a_a^\text{EW} \simeq \frac{\alpha m_a^2}{24\pi m_W^2} \frac{3 - s_w^2 - 4 s_w^4}{s_w^2(1-2s_w^2)} (c^G - c^{HD}) \, . 
\label{magmomentEW}
\end{equation}
Although the prefactor of the WCs in Eq.~\eqref{magmomentEW} is much smaller than the prefactor of $c^{e\gamma,obs}$ in Eq.~\eqref{magmoment}, 
one should keep in mind that $c^{e\gamma,obs}$ is necessarily loop-suppressed 
as well as Yukawa-suppressed.\footnote{The required Yukawa coupling may not be small in some specific models, for example it is $y_t \simeq 1$ 
in the LQ model that we consider, see Table \ref{table-WCs1}.} 
Thus, the two corrections in Eqs.~\eqref{magmoment} and \eqref{magmomentEW} could be of the same order. 
To the best of our knowledge, the correction to the magnetic dipole moment in Eq.~\eqref{magmomentEW} has been overlooked previously in the literature, and it can potentially be a significant effect.
In principle, the shift in $s_w$ also affects the SM prediction for the electric dipole moments, however 
the experimental upper limits are many orders of magnitude above the SM theoretical prediction.

We use experimental data on the electron and muon dipole moments to place bounds on $c^{e\gamma,obs}_{ee}$ and $c^{e\gamma,obs}_{\mu \mu}$, 
as well as 
$(c^G-c^{HD})$, listed in Table \ref{tab:WCbounds}. 
Since $a_\mu^{exp}$ is currently $4.2 \sigma$ larger than the SM prediction \cite{Aoyama:2020ynm,Abi:2021gix}, 
with $a_\mu^\text{exp} - a_\mu^{SM} = (2.51 \pm 0.59) \times 10^{-9}$, we very conservatively allow for $a_\mu^{exp}-5\sigma \le a_\mu^{SM} +\Delta a_\mu \le a_\mu^{exp}+2\sigma$, in order not to exclude new physics models with a negligible $\Delta a_\mu$. 
As already detailed in section \ref{EWPT}, one can compare $a_e^{exp}$ with 
atomic-frequency measurements of $\alpha$ \cite{Parker:2018vye,Morel:2020dww} to set a bound on $c^{e\gamma,obs}_{ee}$, which is also reported in Table \ref{tab:WCbounds}.

\subsection{$\mu \to e$ conversion in nuclei}\label{mutoe}

Beyond charged-lepton decays to three leptons and radiative decays, the $\mu \to e$ conversion in nuclei is a third powerful low-energy test of charged-LFV.
The conversion rate is sensitive to several low-energy EFT coefficients \cite{Cirigliano:2009bz}, \begin{equation}
\begin{array}{rcll}
\Gamma_N & =& \dfrac{m_\mu^5}{v^4} \Bigg|  \dfrac{v D_N c^{e\gamma*}_{\mu e}}{\sqrt{2} m_\mu} 
 + 4 \displaystyle\sum_{i=p,n} & \Bigg[ 
\displaystyle\sum_{q=u,d,s}\left( \frac{m_i}{m_q}  c^{S,L}_{\substack{eq \\ e \mu}} f^q_{Si} S^i_N
+ c^{V,R}_{\substack{eq \\ e\mu}} f_{Vi}^q V^i_N \right) \\
&&& + \displaystyle\sum_{Q=c,b,t} \frac{2m_i}{27 m_Q} c^{S,L}_{\substack{eQ \\ e\mu}} \Big(1 - \sum_{q=u,d,s}f_{Si}^q \Big) S^i_N  
 \Bigg] \Bigg|^2  \\
& + & \dfrac{m_\mu^5}{v^4} \Bigg|  \dfrac{v D_N c^{e\gamma}_{e\mu}}{\sqrt{2} m_\mu} 
 + 4 \displaystyle\sum_{i=p,n} & \Bigg[ 
\displaystyle\sum_{q=u,d,s}\left(\dfrac{m_i}{m_q} c^{S,R}_{\substack{eq \\ e\mu}} f^q_{Si}S^i_N + c^{V,L}_{\substack{eq \\ e\mu}} f_{Vi}^q V^i_N \right) \\
&&& + \displaystyle\sum_{Q=c,b,t} \dfrac{2m_i}{27 m_Q} c^{S,R}_{\substack{eQ \\ e\mu}} \Big(1 - \sum_{q=u,d,s}f_{Si}^q\Big) S^i_N  
\Bigg] \Bigg|^2 \, .
\end{array}
\label{mueconversion}
\end{equation}
The nucleus-dependent form factors $D_N$, $S^i_N$, and $V^i_N$ are given in Table 1 of \cite{Kitano:2002mt}, the nucleon form factors $f^q_{Si}$ and $f^q_{Vi}$ can be found e.g. in \cite{Crivellin:2017rmk}, 
and $c^{S/V , X}_{\substack{e\psi \\ ab}}$ are the WCs of the operators
\begin{equation}
\mathcal{O}^{S , X}_{\substack{e\psi \\ ab}} \equiv \frac{2}{v^2} (\overline{\ell_a}  P_X \ell_b) (\overline{\psi} \psi) \, ,
\quad\quad
\mathcal{O}^{V , X}_{\substack{e\psi \\ ab}} \equiv \frac{2}{v^2} (\overline{\ell_a} \gamma_\nu P_X \ell_b)  (\overline{\psi} \gamma^\nu \psi) \, ,
\end{equation}
for $\psi=q,Q$. The evolution of the LEFT WCs from the nucleon mass scale up to the electroweak scale is analysed in \cite{Crivellin:2017rmk}. 
The matching of the SMEFT operators onto the LEFT four-fermion operators 
can be found e.g.~in \cite{Jenkins:2017jig}. 
Note that at scale $m_Q$ ($Q=c,b,t$), the operators involving heavy quarks match at one loop onto $(\overline{\ell^a} P_X \ell_b)G^A_{\mu\nu}G^{A\mu\nu}$, which couple the leptons to the gluon content of the nucleons: this effect is accounted for by the second and fourth lines of \eq{mueconversion}.

The best bounds on $\mu \to e$ conversion come from experiments with titanium \cite{Dohmen:1993mp} and gold \cite{Bertl:2006up}, with $BR(\mu \text{Au} \to e \text{Au}) < 7 \times 10^{-13}$ and $BR(\mu \text{Ti} \to e \text{Ti}) < 4.3 \times 10^{-12}$ at $90\%$ C.L., where $BR(\mu N \to e N) \equiv \Gamma_N/\Gamma_N^\text{capt}$ with the nucleus capture rates given in e.g.~Table 8 of \cite{Kitano:2002mt}.
The strongest expected future limits are from conversions in aluminium \cite{Kuno:2013mha} and titanium \cite{Barlow:2011zza,Knoepfel:2013ouy},
$BR(\mu \text{Ti} \to e \text{Ti}) < 10^{-18}$ and $BR(\mu \text{Al} \to e \text{Al}) < 10^{-16}$ at $90\%$ C.L.. 
It is not very informative to translate experimental results into a bound on the lengthy combination of WCs given in Eq.~\eqref{mueconversion}. 
We will rather present such bound in the context of each of our four models, where the conversion rate reduces to a more compact expression.
In principle, comparing the conversion rates on different nuclei, it could be possible to disentangle the various WCs contributing to the rate \cite{Davidson:2018kud}.
While \eq{mueconversion} provides the leading, spin-independent contribution to the conversion rate, spin-dependent contributions may also be exploited, in order to provide constraints on a larger set of WCs 
\cite{Cirigliano:2017azj,Davidson:2017nrp}.

\section{Models' phenomenology}
\label{sec:Pheno}

Having derived the leading order WCs of the four neutrinos mass models we consider (summarised in Tables \ref{table-WCs1}, \ref{table-WCs2}, \ref{table-WCs3} and \ref{table-WCs4}), as well as the general bounds on the WCs of the SMEFT (see Table \ref{tab:WCbounds}), we can now proceed to study the phenomenology. 
For each model we will demonstrate the correlations between different observables and identify which experimental bounds are the most constraining.

\subsection{Type-I and III seesaw}
\label{ssec:tIandIII}

\begin{figure}[tb]
\hspace{-1.0cm}
\mbox{
\includegraphics[width=0.52\textwidth]{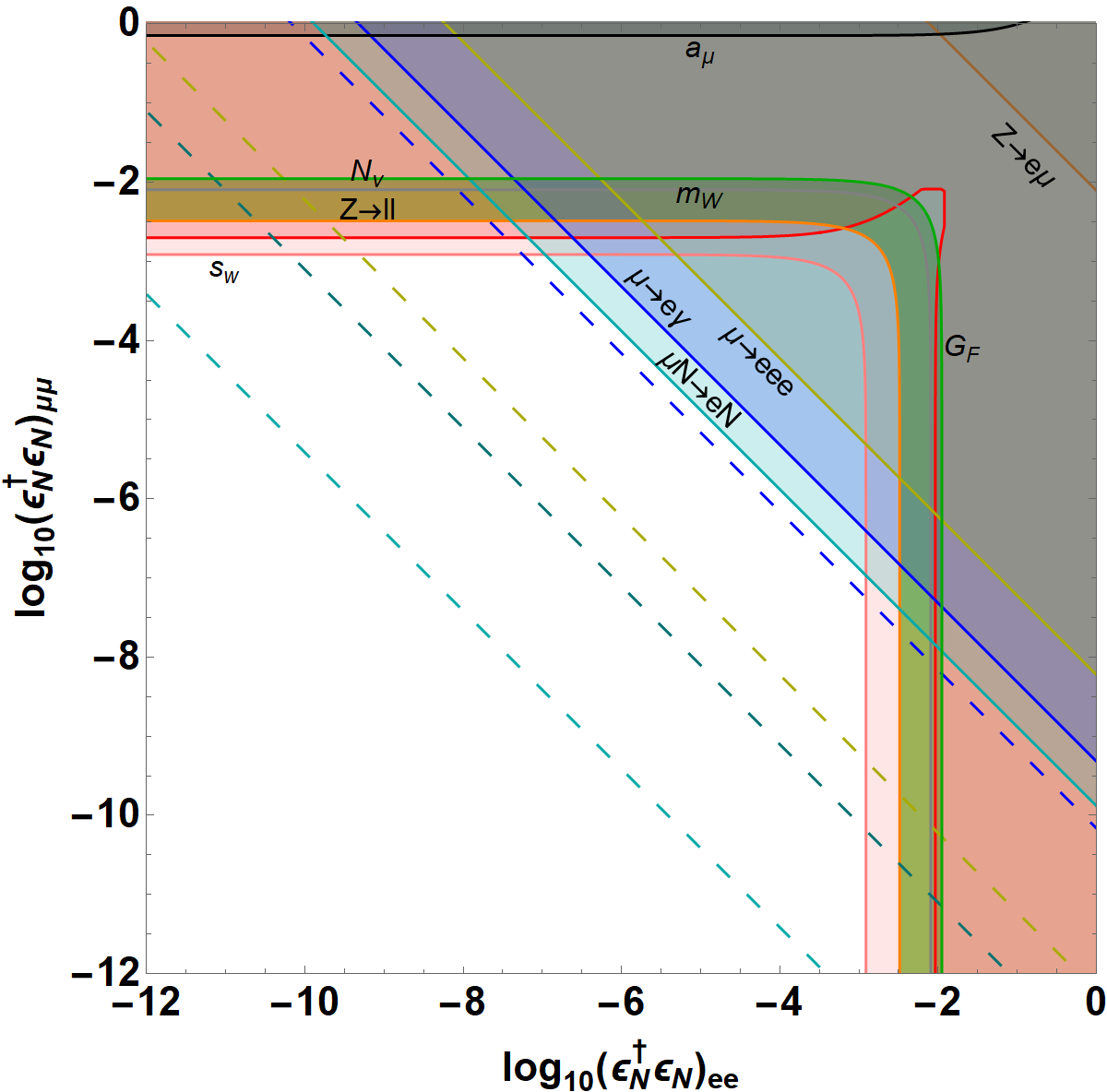}
\quad
\includegraphics[width=0.52\textwidth]{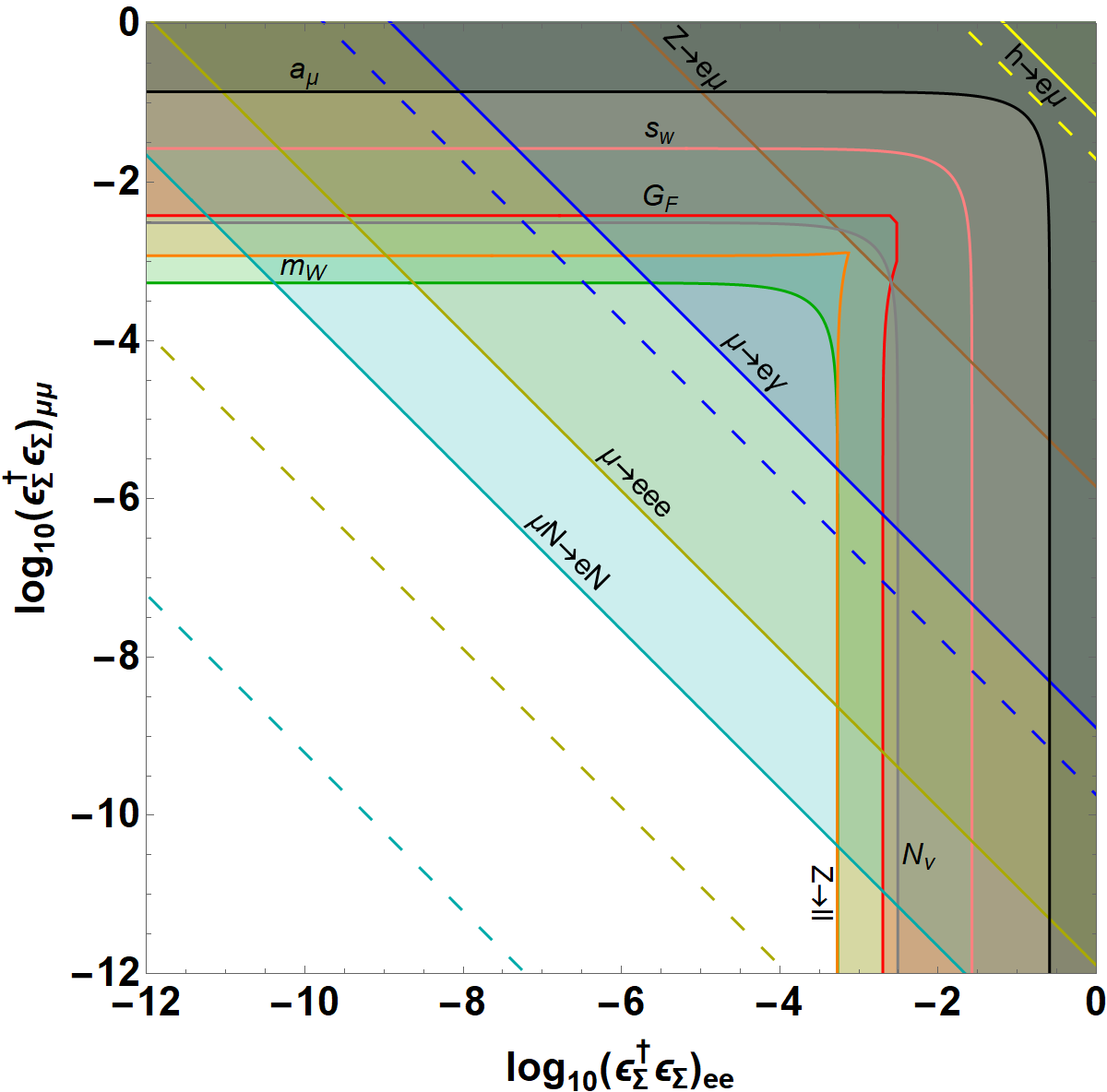}
}
\caption{Bounds on the seesaw model of type-I (left panel) and type-III (right panel), in the $e-\mu$ sector. Constraints are shown as a function of 
$({\epsilon}^\dag\epsilon)_{ee}$ and $({\epsilon}^\dag\epsilon)_{\mu\mu}$. 
We have set $(\epsilon^\dag\epsilon)_{\tau\tau}=0$ and, in RGE-induced WCs proportional to $\log M_i$, we have taken  $M_i=10\,{\rm TeV}$.
Constraints are from $s_w$ (pink), $G_F$ universality (red), $Z\to \ell_a^+\ell_a^-$ 
(orange), $m_W$ (green), $N_\nu$ (grey), $a_\mu$ (black), $h \to e\mu$ (yellow), $Z\to e\mu$ (brown), 
$\mu \to e \gamma$ (blue), $\mu \to eee$ (mustard), 
and $\mu \to e$ conversion in nuclei (cyan). 
Note that $\tau$ flavour is conserved, given the above choice of parameters. 
Current bounds are denoted by solid lines and shading, (a few)  expected future bounds are indicated by dashed lines. 
For $m_W$ in the type-I seesaw, we shaded in light green the $2\sigma$ region preferred by the current anomaly. 
}
\label{fig:SeesawEMU}
\end{figure}

\begin{figure}[tb]
\hspace{-1.0cm}
\mbox{
\includegraphics[width=0.52\textwidth]{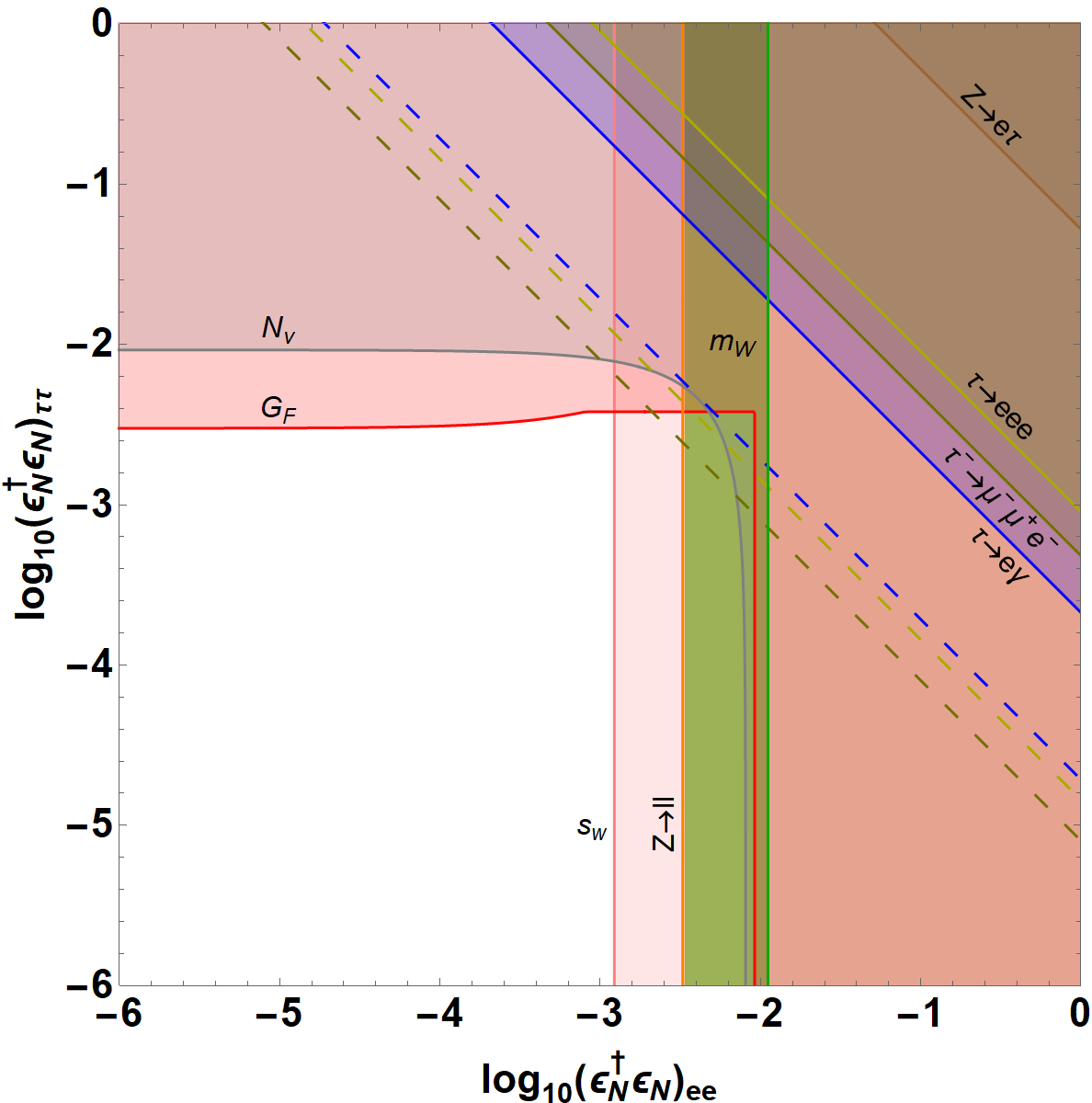}
\quad
\includegraphics[width=0.52\textwidth]{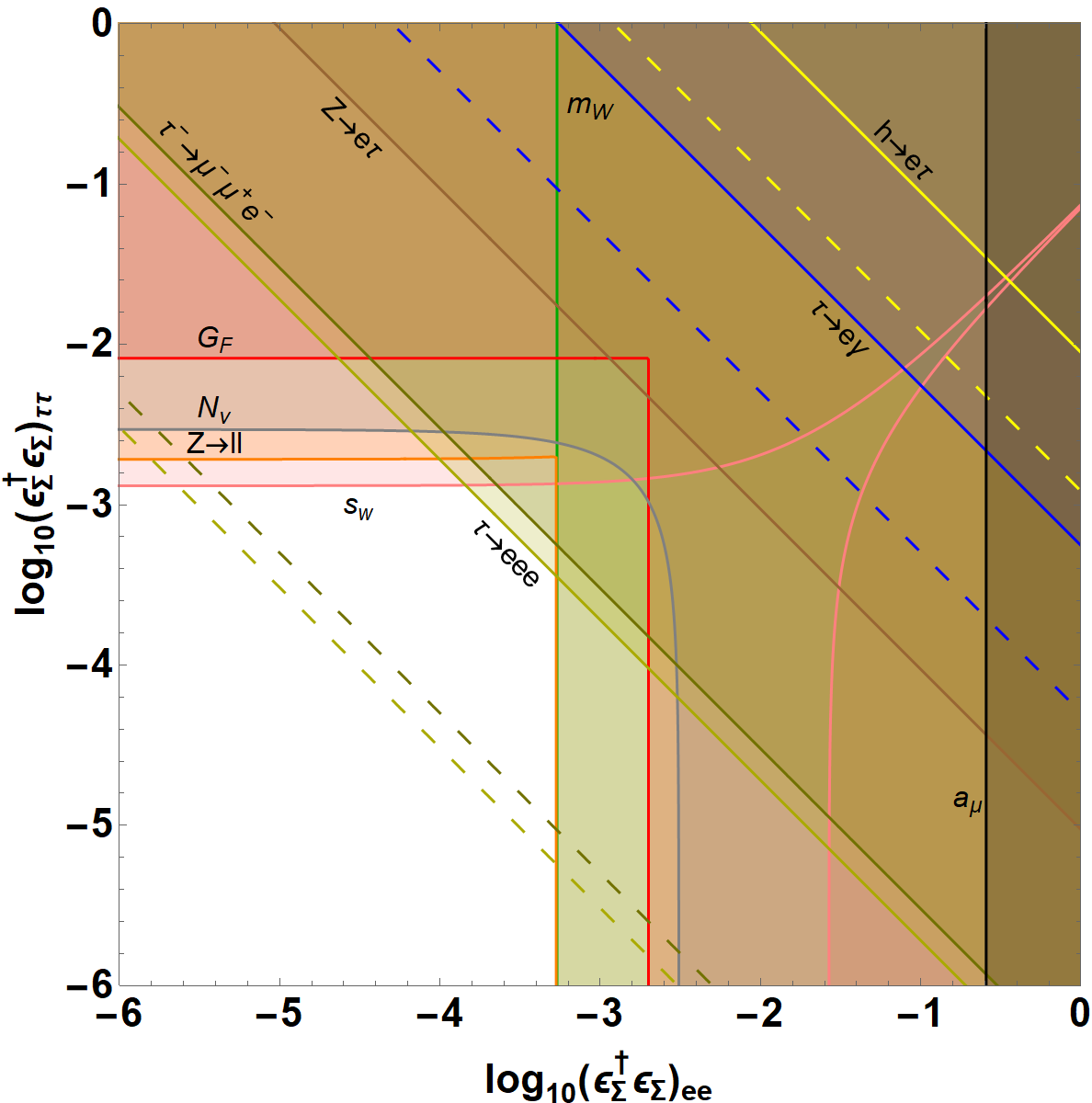}
}
\caption{Bounds on the seesaw model of type-I (left panel) and type-III (right panel), in the $e-\tau$ sector.  Constraints are shown as a function of 
$({\epsilon}^\dag\epsilon)_{ee}$ and $({\epsilon}^\dag\epsilon)_{\tau\tau}$. 
We have set $(\epsilon^\dag\epsilon)_{\mu\mu}=0$ and, in RGE-induced WCs proportional to $\log M_i$, we have taken  $M_i=10\,{\rm TeV}$.
Constraints are from $s_w$ (pink), $G_F$ universality (red), $Z\to \ell_a^+ \ell_a^-$ (orange), $m_W$ (green), $N_\nu$ (grey), $a_\mu$ (black), 
$h \to e\tau$ (yellow), $Z\to e\tau$ (brown), 
$\tau \to e \gamma$ (blue), $\tau \to eee$ (mustard), and $\tau \to \mu^+\mu^- e$ (dark mustard). 
Note that $\mu$ flavour is conserved, given the above choice of parameters.
Current bounds are denoted by solid lines and shading, (a few) expected future bounds are indicated by dashed lines. For $m_W$ in the type-I seesaw, we shaded in light green the $2\sigma$ region preferred by the current anomaly.
}
\label{fig:SeesawETAU}
\end{figure}

\begin{figure}[tb]
\hspace{-1.0cm}
\mbox{
\includegraphics[width=0.52\textwidth]{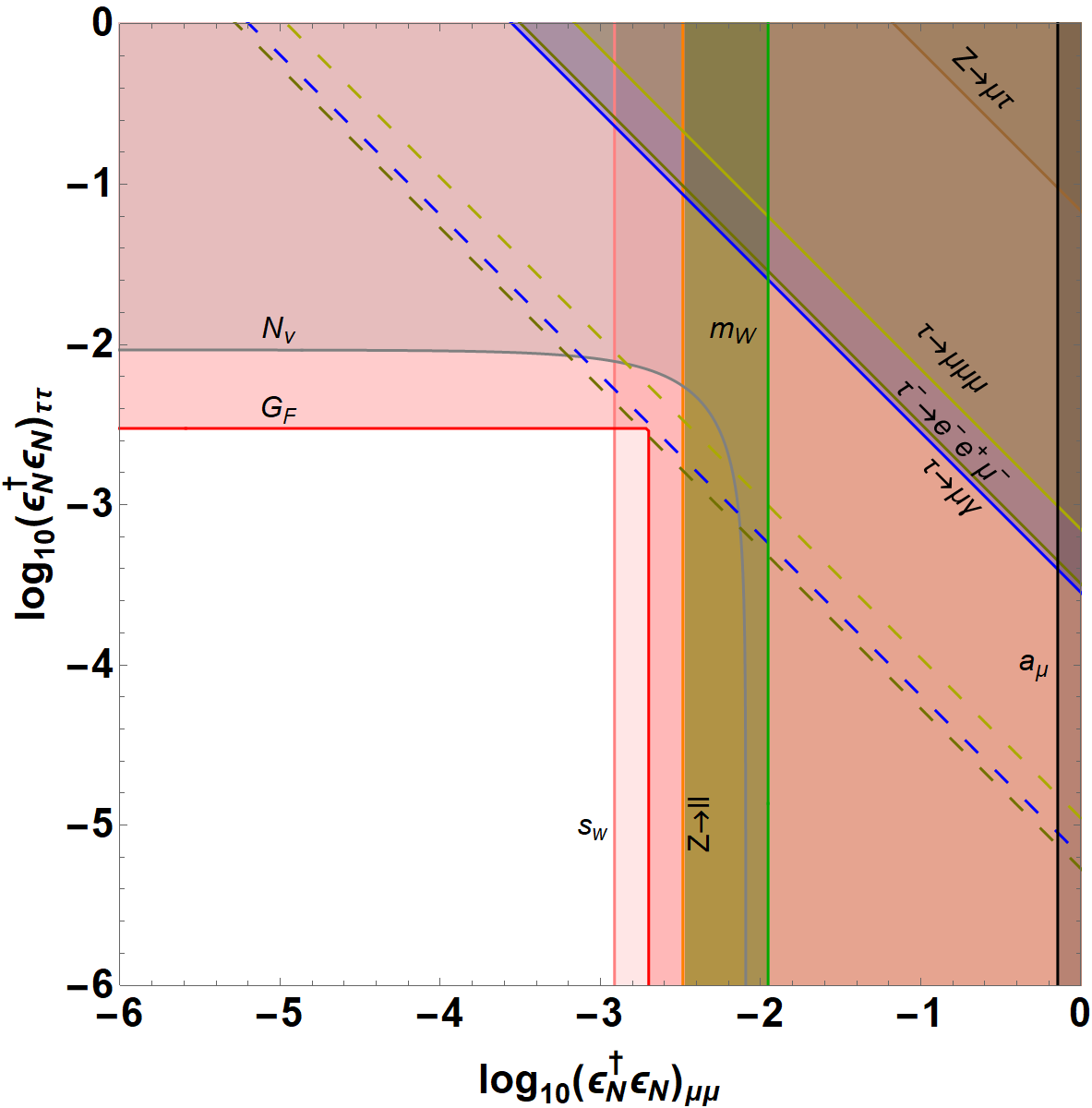}
\quad
\includegraphics[width=0.52\textwidth]{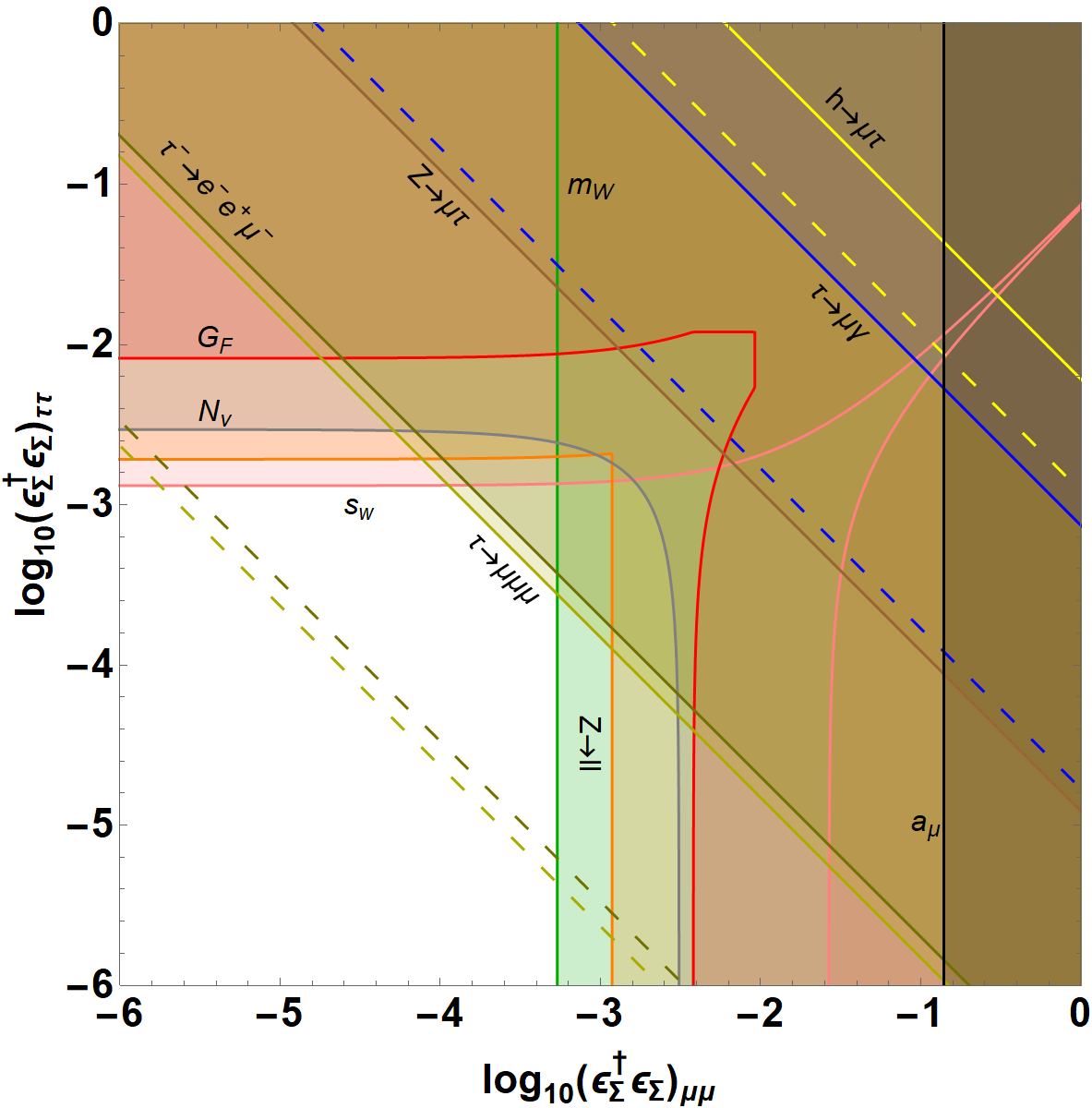}
}
\caption{Bounds on the seesaw model of type-I (left panel) and type-III (right panel), in the $\mu-\tau$ sector.  Constraints are shown as a function of 
$({\epsilon}^\dag\epsilon)_{\mu\mu}$ and $({\epsilon}^\dag\epsilon)_{\tau\tau}$. 
We have set $(\epsilon^\dag\epsilon)_{ee}=0$ and, in RGE-induced WCs proportional to $\log M_i$, we have taken  $M_i=10\,{\rm TeV}$.
Constraints are from $s_w$ (pink), $G_F$ universality (red), $Z\to \ell_a^+ \ell_a^-$ (orange), $m_W$ (green), $N_\nu$ (grey), $a_\mu$ (black), 
$h \to \mu\tau$ (yellow), $Z\to \mu\tau$ (brown), 
$\tau \to \mu \gamma$ (blue), $\tau \to \mu\mu\mu$ (mustard), and $\tau \to (e^+e^-)\mu$ (dark mustard).
Note that $e$ flavour is conserved, given the above choice of parameters. 
Current bounds are denoted by solid lines and shading, (a few) expected future bounds are indicated by dashed lines. 
For $m_W$ in the type-I seesaw, we shaded in light green the $2\sigma$ region preferred by the current anomaly.
}
\label{fig:SeesawMUTAU}
\end{figure}

The type-I and III seesaw mechanisms 
have identical flavour structures, their only difference being the $SU(2)_w$ representation of the new fermions. 
We have previously studied \cite{Coy:2018bxr} the phenomenology of the type-I seesaw EFT in detail.
Here we will recall a few salient features of the seesaw parameter space, holding for both type-I and III, then we will detail the qualitative differences between the two. Finally we will show in Figs.~\ref{fig:SeesawEMU}-\ref{fig:SeesawMUTAU} a comprehensive compilation of the experimental constraints in the case of type-I (left panel) and type-III (right panel).

Let us call $n$ the number of sterile neutrinos $N_R$ (triplets $\Sigma_R$) in the type-I (III) seesaw.
In both models, the light neutrino mass matrix is proportional to $c^W = (\epsilon^T \mu \epsilon)/2$, where $\epsilon$ and $\mu$ are defined in \eq{epsmu}, and we drop the subscript $N$ or $\Sigma$ whenever the discussion applies to both types of seesaw.
One needs $n\ge 2$ to accommodate neutrino oscillation data.
The combination of parameters defined by $c^W$ is highly constrained by the tininess of neutrino masses, therefore one can safely neglect 
the $(c^{W\dag} c^W)$ terms in the dim-6 WCs, as they are proportional to $m_\nu^2$.
On the other hand, the dim-6 WCs contain terms proportional to $(\epsilon^\dag \epsilon)$, or its log-enhanced version, $(\epsilon^\dag \log[\mu] \epsilon)$,
as shown in Tables \ref{table-WCs1}-\ref{table-WCs4}. For $n\ge 2$ these combinations of parameters are independent from $c^W$, they preserve lepton number and may lead to a number of observable effects.

The flavour structure of the matrix $(\epsilon^\dag \epsilon)$ is highly constrained in  the limit $c^W \to 0$. 
It can be shown \cite{Coy:2018bxr} that for $n=2$ or $3$, the matrix is exactly factorised, 
$(\epsilon^\dag \epsilon)_{ab} = \lambda_a \lambda_b$, where $a,b= e,\mu,\tau$ and $\lambda_{a,b}$ can be taken real and positive with no loss of generality.
This implies that flavour-violating entries are determined by flavour-conserving ones, $(\epsilon^\dag \epsilon)_{ab} = \sqrt{(\epsilon^\dag \epsilon)_{aa}(\epsilon^\dag \epsilon)_{bb}}$.
For $n>3$, 
the equality is relaxed to a Cauchy-Schwarz inequality, $|(\epsilon^\dag \epsilon)_{ab}| \leq \sqrt{(\epsilon^\dag \epsilon)_{aa}(\epsilon^\dag \epsilon)_{bb}}$, therefore flavour-violating entries are still bounded, and may actually be set to zero while (some) flavour-violating entries remain large  \cite{Coy:2018bxr}.
The analogous relations hold for the matrix $(\epsilon^\dag \log[\mu] \epsilon)$ as well.
In view of these general relations, it will be convenient to display the experimental constraints in the planes  
$(\epsilon^\dag \epsilon)_{aa}$ - $(\epsilon^\dag \epsilon)_{bb}$, for $ab=e\mu,e\tau,\mu\tau$, see 
Figs.~\ref{fig:SeesawEMU},\ref{fig:SeesawETAU},\ref{fig:SeesawMUTAU} respectively. 
The (maximal value of the) off-diagonal parameter $(\epsilon^\dag \epsilon)_{ab}$ is uniquely determined by the diagonal ones for $n=2,3$
($n>3$). In this way the constraints from flavour-conserving observables are correlated with those from LFV observables.

The different $SU(2)_w$ structure of the type-I and type-III seesaw is responsible for the discrepant values of the WCs in their respective EFTs. 
The key difference concerns the linear combination $c^{Hl(1)} + c^{Hl(3)}$, which modifies $Z$-boson couplings to charged leptons.
In type-I it arises only at one loop, via RGE effects, while it is non-zero at tree level in type III:
\begin{equation}
\ba{rcl}
(c^{Hl(1)}_{ab} + c^{Hl(3)}_{ab})_{\rm type-I}& \simeq &
- \dfrac{1}{16\pi^2} \left[\dfrac{g_1^2 + 17g_2^2}{12}(\epsilon_N^\dag \epsilon_N)_{ab} + \dfrac{g_1^2 - g_2^2}{6} \text{tr}[\epsilon_N^\dag \epsilon_N] \delta_{ab} \right] \log\dfrac{M_N}{v}~,\\
(c^{Hl(1)}_{ab} + c^{Hl(3)}_{ab})_{\rm type-III} &\simeq& (\epsilon_\Sigma^\dag \epsilon_\Sigma)_{ab}~.
\ea
\end{equation}
As a consequence, several type-III rates are enhanced relatively to type-I, by about four orders of magnitude, namely LFV $Z$ decays, lepton-to-three-lepton decays and $\mu \to e$ conversion in nuclei, because at leading order these processes are induced by the very same 
$Z \bar\ell_a \ell_b$ coupling. 
A second difference concerns the operator $Q_{eH}$, induced at tree level in the type-III and at one-loop leading-log in the type-I.
However, this affects only the constraint from LFV Higgs decays, which is not significant in either case.
Among the lepton observables we consider, only dipoles (viz. dipole moments and radiative charged-lepton decays) are loop-suppressed in type-III seesaw. In contrast, all LFV processes are loop-suppressed in the type-I case.

The constraints on the seesaw parameter space are derived by inserting the seesaw WCs, listed in 
Tables~\ref{table-WCs1}-\ref{table-WCs4}, into the general expressions for the observables presented in section~\ref{sec:SMEFTconstraints}.
For most observables, the combinations of WCs subject to constraints are summarised in table~\ref{tab:WCbounds}. Here we display only the few additional constraints which did not fit in that table.

Particularising the $\mu\to e$ conversion rate of \eq{mueconversion} to the case of type-I seesaw, the leading contributions come from the WCs
\begin{align}
	c^{V,L}_{\substack{eu \\ e\mu}} &= \left( \frac{1}{2} - \frac{4}{3} s_w^2 \right) \left(c^{Hl(1)}_{e\mu} + c^{Hl(3)}_{e\mu} \right)  + \frac{1}{2}(c^{lq(1)}_{e\mu uu} - c^{lq(3)}_{e\mu uu} + c^{lu}_{e\mu uu}) \, , \\
	c^{V,L}_{\substack{ed \\ e\mu}} &= \left( - \frac{1}{2} + \frac{2}{3} s_w^2 \right) \left(c^{Hl(1)}_{e\mu} + c^{Hl(3)}_{e\mu} \right)   + \frac{1}{2}(c^{lq(1)}_{e\mu dd} + c^{lq(3)}_{e\mu dd} + c^{ld}_{e\mu dd}) \, .
\end{align}
Note we included the contributions of all relevant WCs generated at one-loop leading-log order, except for $c^{He}_{e\mu}$ and $c^{eH}_{e\mu}$: 
the former is relatively suppressed by two powers of $y_e$, and the latter 
by a factor of $m_{p,n}/v$, see Eq. \eqref{mueconversion}. 
In the type-III seesaw, the leading order WCs read simply
\begin{align}
	&c^{V,L}_{\substack{eu \\ e\mu}} = \left( \frac{1}{2} - \frac{4}{3} s_w^2 \right) \left(c^{Hl(1)}_{e\mu} + c^{Hl(3)}_{e\mu} \right) \, ,&
	&c^{V,L}_{\substack{ed \\ e\mu}} = \left( -\frac{1}{2} + \frac{2}{3} s_w^2 \right) \left(c^{Hl(1)}_{e\mu} + c^{Hl(3)}_{e\mu} \right)\, .&
\end{align}
Here we retained only the vector-current WCs induced at tree level (the scalar-current contribution is suppressed by a factor of $m_{p,n}/v$), 
as we are just interested in deriving the most stringent constraint on  $({\epsilon_\Sigma}^\dag\epsilon_\Sigma)_{e\mu}$.

Next, let us consider the upper bound on $\ell_a \to \ell_b \ell_b \ell_b$ decays \cite{Bellgardt:1987du,Hayasaka:2010np,Blondel:2013ia,Kou:2018nap}. In type-I seesaw the rate is given by
\begin{align}
\Gamma(\ell_a \to \ell_b \ell_b \ell_b) & \simeq \frac{m_a^5}{384 \pi^3 v^4} \Big( 2 \left| c^{ll}_{abbb} + c^{ll}_{bbab} - (1- 2s_w^2) (c^{Hl(1)}_{ab } + c^{Hl(3)}_{ab }) \right|^2
 \nonumber\\
&+ \left|c^{le}_{abbb} + 2s_w^2 (c^{Hl(1)}_{ab} + c^{Hl(3)}_{ab}) \right|^2 \Big) ~,
\label{lto3l}\end{align}
where we dropped the doubly Yukawa-suppressed $c^{He}_{ab}$ and $c^{eH}_{ab}$ contributions. 
The rate in \eq{lto3l} also applies to the type-III seesaw, however one can drop the $c^{ll}$ and $c^{le}$ pieces as they are relatively 
loop-suppressed with respect to $(c^{Hl(1)}_{ba}+c^{Hl(3)}_{ba})$, which is induced at tree level. 
Bounds on the seesaw parameters are obtained by comparing with the experimental constraints listed in section \ref{eto3e}.
Finally, the rates for the $\tau \to \ell_b^+\ell_b^-\ell_c$ decays \cite{Hayasaka:2010np,Kou:2018nap} are, for the type-I and III seesaw models,
\begin{align}
\Gamma(\tau \to \ell_b^+\ell_b^-\ell_c) &\simeq \frac{m_\tau^5}{384 \pi^3 v^4} \Big( |c^{ll}_{c\tau bb} + c^{ll}_{bb c\tau} + c^{ll}_{cbb\tau} + c^{ll}_{b\tau cb} - (1 - 2s_w^2) (c^{Hl(1)}_{c\tau} 
 + c^{Hl(3)}_{c\tau})|^2 +\nonumber\\
& +|c^{le}_{c\tau bb} + 2 s_w^2 (c^{Hl(1)}_{c\tau} + c^{Hl(3)}_{c\tau})|^2 \Big) \, . 
 \end{align}
As before, while in type-I the WCs in these combinations are all of the same order,  in type-III $c^{ll}$ and $c^{le}$ are relatively loop suppressed.

All the relevant constraints are illustrated in Figs.~\ref{fig:SeesawEMU}, \ref{fig:SeesawETAU}, \ref{fig:SeesawMUTAU},
where we set to zero, in turn, the $\tau,\mu,e$ row and column of the matrix $(\epsilon^\dag\epsilon)_{ab}$.
For the LFV parameters we fixed $(\epsilon^\dag \epsilon)_{ab} = \sqrt{(\epsilon^\dag \epsilon)_{aa}(\epsilon^\dag \epsilon)_{bb}}$,
which is the case for $n=2,3$, and is an upper bound on the LFV entry for $n>3$: in the latter case the constraints from LFV observables can be relaxed.
For WCs induced at one-loop leading-log order, depending on $\epsilon^\dag\log(M/v)\epsilon$, we assumed for concreteness that the heavy fermions are degenerate, with mass matrix $M=(10{\rm~TeV})\mathbb{1}_n$. 
For the observables generated at leading log (that is, the LFV observables in type-I seesaw), the bounds strengthen logarithmically as $M_N$ grows above $10{\rm~TeV}$ (barring cancellations).

From Figs.~\ref{fig:SeesawEMU}, \ref{fig:SeesawETAU} and \ref{fig:SeesawMUTAU}, we see that bounds from flavour-conserving processes constrain the type-I and type-III seesaw to a similar degree. 
In the type-I, the most stringent bounds come  from $s_w$, $(\epsilon_N^\dag \epsilon_N)_{ee}$ and  $(\epsilon_N^\dag \epsilon_N)_{\mu \mu} \lesssim 1.2 \times 10^{-3}$, or from $G_F$ universality, $(\epsilon_N^\dag \epsilon_N)_{\tau \tau} \lesssim 2.5 \times 10^{-3}$. 
These exclude a possible resolution of the current $m_W$ anomaly, given by the green bands in the left panels. 
In the type-III, $s_w$ sets the best bound in the $\tau$ sector with $(\epsilon_\Sigma^\dag \epsilon_\Sigma)_{\tau \tau} \lesssim  1.3 \times 10^{-3}$, while $Z \to \ell_a^+ \ell_a^-$ gives $(\epsilon_\Sigma^\dag \epsilon_\Sigma)_{ee} \lesssim 5.3 \times 10^{-4}$ and $m_W$ gives $(\epsilon_\Sigma^\dag \epsilon_\Sigma)_{\mu \mu} \lesssim 5.4 \times 10^{-4}$. 
Note that, for Yukawa couplings $Y \sim 1$, a constraint $(\epsilon^\dag \epsilon) \lesssim 10^{-3}$ corresponds to a mass scale $M \gtrsim 5$ TeV. 
The various electroweak precision observables give comparable limits because the shift in the Fermi constant, $c^G$, defined in 
Eq.~\eqref{gfShift}, is generated in both models at tree level with the same magnitude (and opposite sign), and this shift directly or indirectly affects the relevant flavour-conserving processes, 
as explained in section~\ref{sec:SMEFTconstraints}. 
One exception is the $\tau$ channel of type-I seesaw, 
where the only relevant constraint comes from $G_F$ universality because new physics coupled to $\tau$ does not interfere with the SM muon decay, 
$\mu \to e \nu_\mu \bar{\nu}_e$, hence it is not constrained by $\tau$-independent flavour-conserving processes. 
On the other hand, the $\tau$ channel in the type-III model is constrained 
by the significant corrections to the $Z$ couplings to leptons, which occur at tree-level in the type-III but not the type-I: these indirectly affect the effective number of neutrinos, $N_\nu$, 
as well as $s_w$ (see section \ref{EWPT}), which provides the strongest constraint. 
Finally, note that both type-I and III seesaw predict a negative shift in $a_\mu$, thus the bound comes from the lower limit on the 
combination of WCs in Table \ref{tab:WCbounds}: 
since the shift is opposite to the direction of the anomaly, the bound is an order of magnitude stronger than if the shift 
in $a_\mu$ were positive.

In contrast to the comparable bounds from flavour-conserving observables, most LFV bounds on $(\epsilon^\dag\epsilon)_{ab}$ are about four 
orders of magnitude stronger in type-III than in type-I, 
due to the loop suppression of all LFV processes in type-I, as explained above. 
The exception is given by radiative charged-lepton decays, which of course occur at loop level in both models:
in this case the bounds are slightly stronger in type-I, because the value of $c^{e\gamma,obs}$ is a factor of $3/(8s_w^2) \simeq 1.6$ larger.
In the $e\mu$ sector (Fig.~\ref{fig:SeesawEMU}),  the single strongest bound comes from $\mu \to e $ conversion in nuclei.  
Assuming order-one Yukawa couplings, the current (future) limits probe scales as large as $M\simeq 60\, (1900)$ TeV in the type-I 
and 0.45 (11) PeV in the type-III. In the two cases, we took into account the different dependence on the nuclear form factors $V_N^p$ and $V_N^n$.
In the sectors $e\tau$ (Fig.~\ref{fig:SeesawETAU}) and $\mu\tau$ (Fig.~\ref{fig:SeesawMUTAU}),
in the case of type-I the flavour-conserving processes set better constraints than the $\tau \to e$ and $\tau \to \mu$ transitions, even including expected future bounds.
In contrast, in the type-III model the LFV processes set a slightly better constraint, but only in the region where 
$({\epsilon_\Sigma}^\dag\epsilon_\Sigma)_{ee,\mu\mu} \sim ({\epsilon_\Sigma}^\dag\epsilon_\Sigma)_{\tau\tau}$, that indeed corresponds to maximal flavour violation.
Thus, future LFV searches in $\tau$ decays will explore relevant parameter space of the type-III model.

We refer to \cite{Coy:2018bxr} for the comparison of our results with the literature on the type-I seesaw model.
Coming to type-III, we compared our results (see also \cite{Coy:2019akn}) with previous phenomenological studies of the type-III seesaw \cite{Abada:2007ux,Kamenik:2009cb,Biggio:2019eeo}. We generally find agreement where there is an intersection, with a few exceptions: 
our result for the LFV Higgs-boson partial decay widths is a factor $4/9$ smaller than in \cite{Biggio:2019eeo}. 
The definition of the effective number of neutrinos $N_\nu$ in \cite{Abada:2007ux,Biggio:2019eeo} 
disagrees with \eq{nnu}, which is the LEP definition \citep{ALEPH:2005ab}, thus leading to a different bound. 
The \cite{Biggio:2019eeo} result for $\ell_a \to 3\ell_b$ is a factor $4s_w^4/(1-4s_w^2 + 6s_w^4) \simeq 0.42$ smaller than ours, which 
agrees with \cite{Abada:2007ux}.

\subsection{Zee model}\label{ZeePheno}

The Zee model has a richer flavour structure than a single-type seesaw model, since it involves two Yukawa coupling matrices besides the SM ones, parameterised by the matrices $\epsilon_2$ and $\epsilon_\delta$ defined in \eq{epsZee}. Recall that $\epsilon_2$ is a generic complex matrix in flavour space, 
in particular its flavour-conserving and flavour-violating components are independent and CP violation can be large.
In contrast, $\epsilon_\delta$ is antisymmetric and can be chosen real, therefore both flavour-conserving and flavour-violating $\delta$-mediated processes are controlled by only three independent parameters.

\begin{figure}[tb]
	\hspace{-1.0cm}
	\mbox{
		\includegraphics[width=0.52\textwidth]{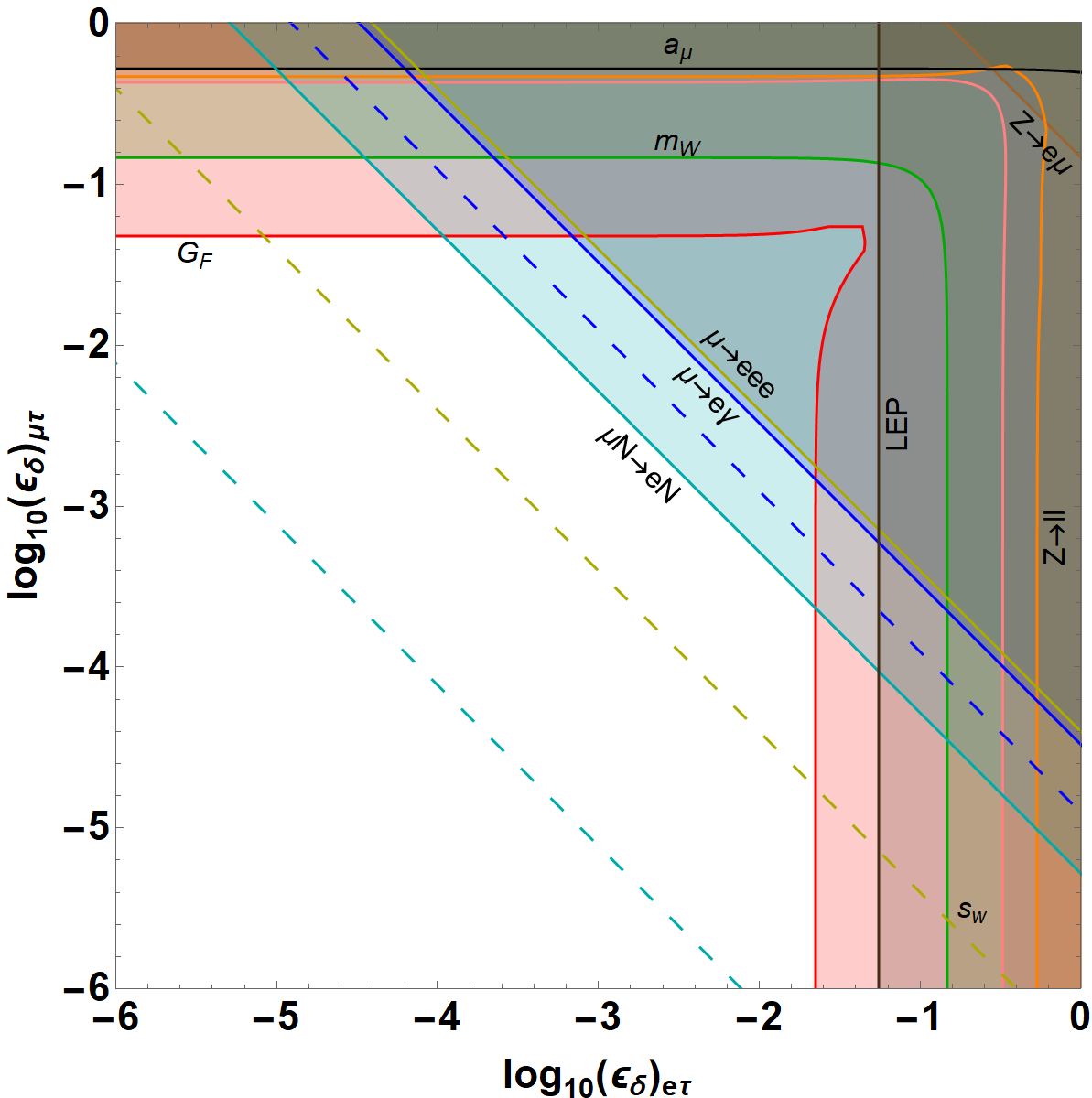}
		\quad
		\includegraphics[width=0.52\textwidth]{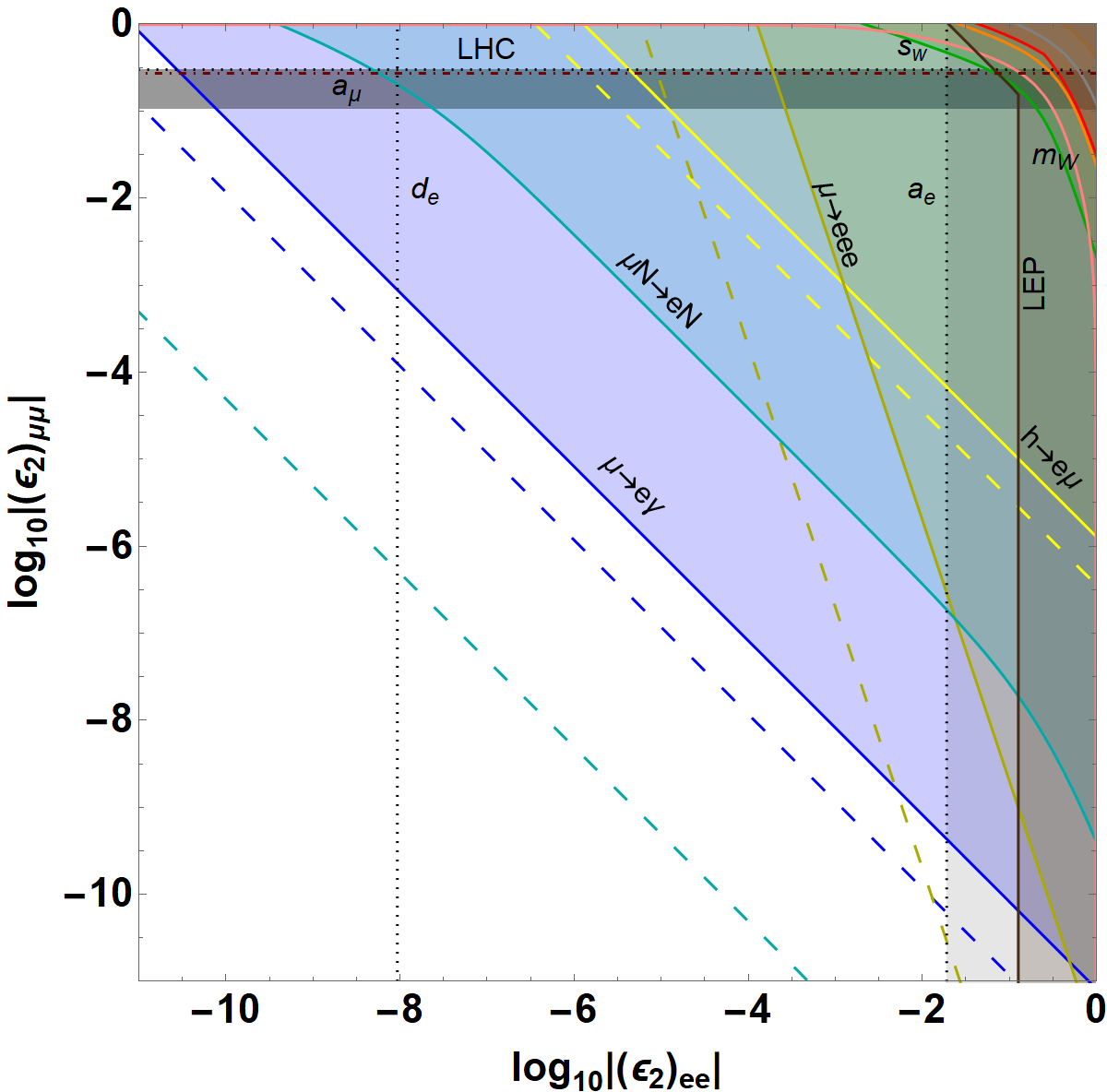}
	}
	\caption{
	Bounds on the Zee model, as a function of $(\epsilon_\delta)_{e\tau}$ and $(\epsilon_\delta)_{\mu\tau}$ (left panel),
	and as a function of $|(\epsilon_2)_{ee}|$ and $|(\epsilon_2)_{\mu \mu}|$, assuming $|(\epsilon_2)_{e\mu}|=|(\epsilon_2)_{\mu e}| = \sqrt{|(\epsilon_2)_{\mu\mu}(\epsilon_2)_{ee}|}$ 
	(right panel).
	For each panel, all other entries of $\epsilon_\delta$ and $\epsilon_2$ are set to zero. 
	With this choice of parameters, LFV is permitted only in the $e-\mu$ sector, in both panels. 
	In RGE-induced WCs proportional to $\log M_\delta$ ($\log M_2$) we have taken  $M_\delta(M_2)=10\,{\rm TeV}$.
	Some bounds in the right panel also depend on the choice of $\epsilon_\lambda$, and we set $\epsilon_\lambda=0.1$. 
	The colour scheme for the constraints is the same as in Fig.~\ref{fig:SeesawEMU}, as indicated by the labels. 
	Additionally, the dark brown bounds are from $e^+ e^- \to e^+ e^-, \tau^+ \tau^-$ at LEP, 
	while the dark red dot-dashed bound in the right panel is from a LHC search. 
	In the right panel, the $a_a$ ($d_a$) bounds are dotted, as they constrain only the real (imaginary) part of 
	$(\epsilon_2)_{aa}$. Therefore, only the weaker of the two bounds applies to $|(\epsilon_2)_{aa}|$ (light grey shading). 
	For $a_\mu$, we also shaded in dark grey the $2\sigma$ region preferred by the current anomaly. 
	}
	\label{fig:ZeeEMu}
\end{figure}

\begin{figure}[tb]
	\hspace{-1.0cm}
	\mbox{
		\includegraphics[width=0.52\textwidth]{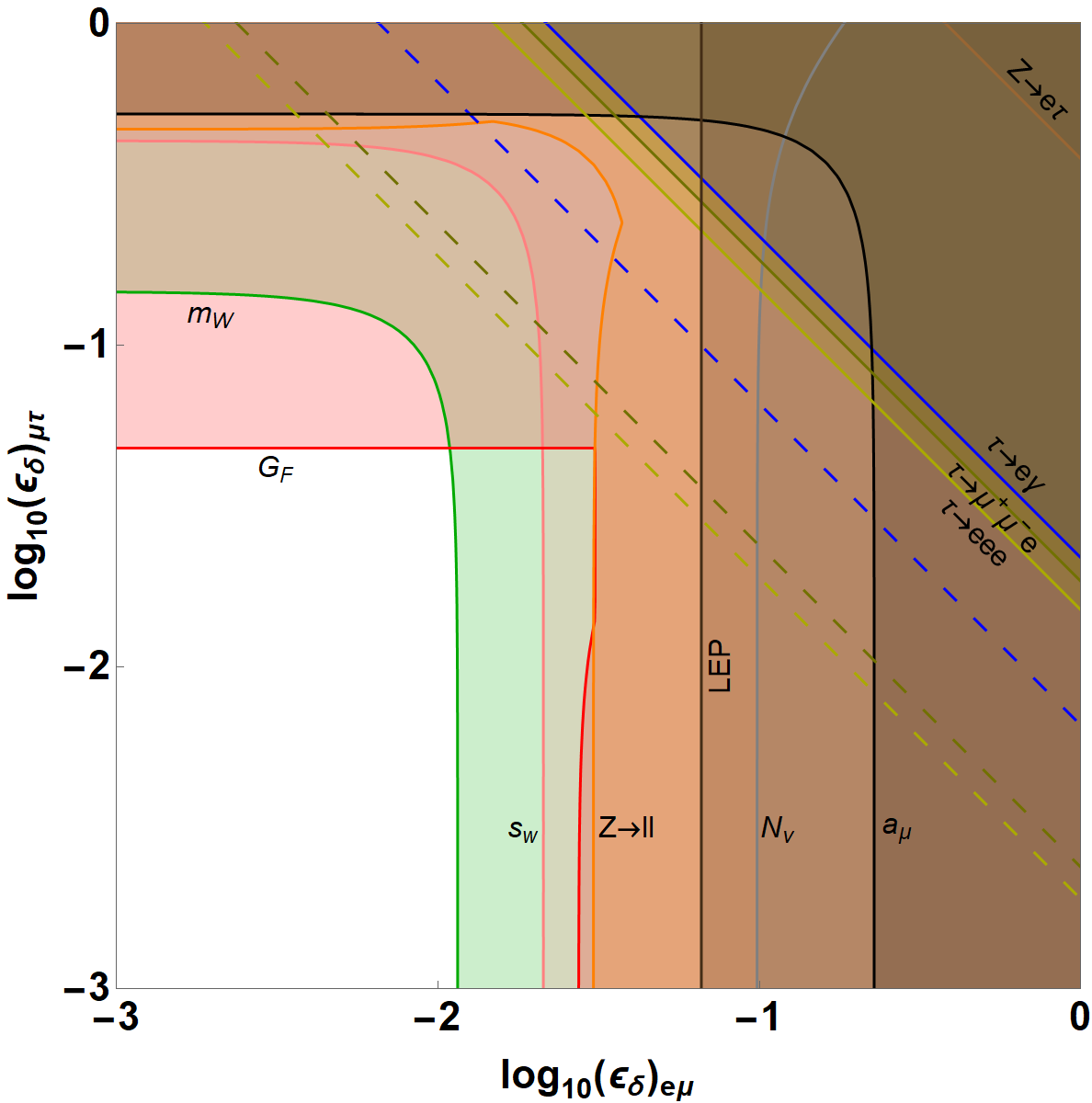}
		\quad
		\includegraphics[width=0.52\textwidth]{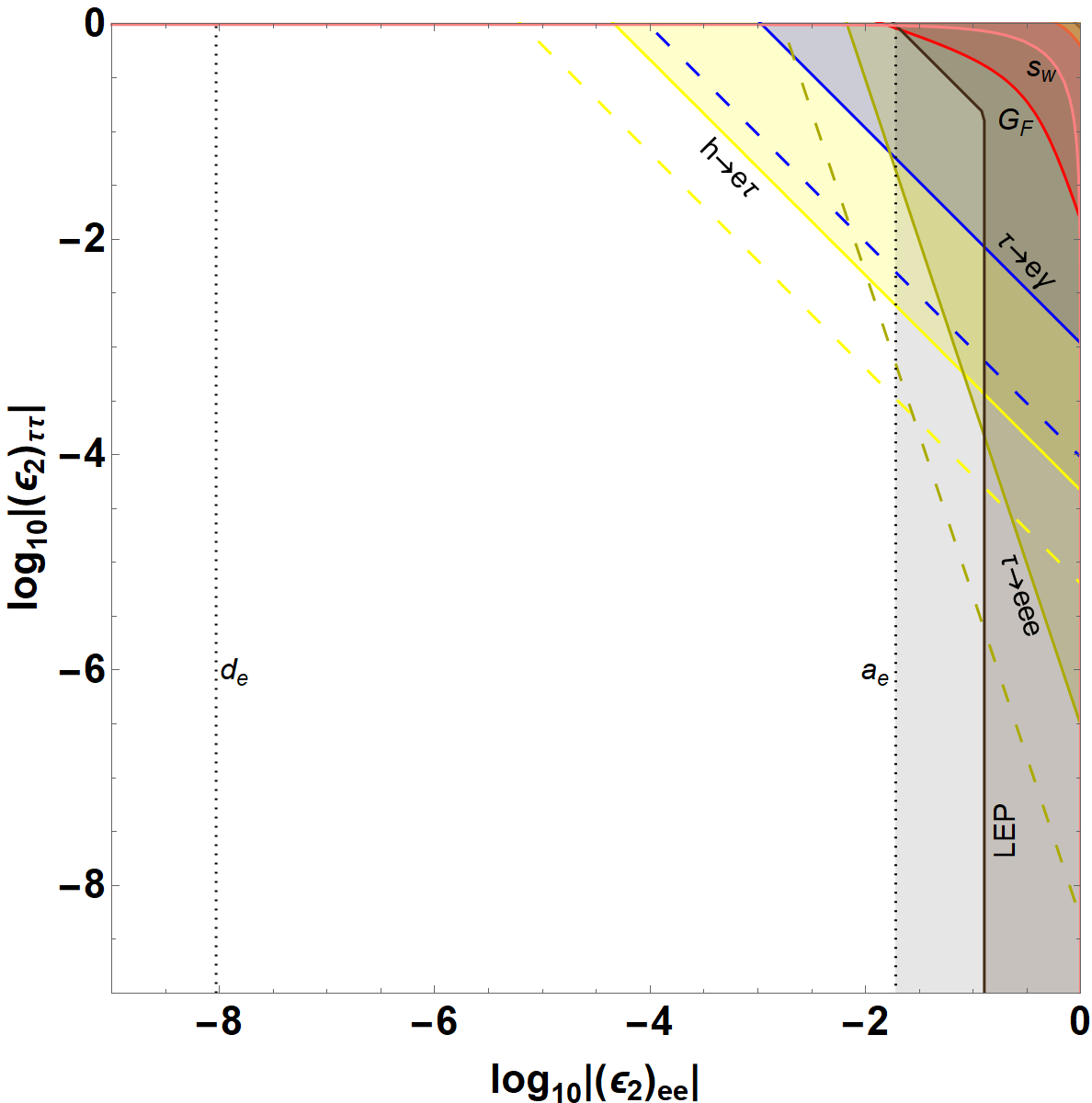}
	}
	\caption{Bounds on the Zee model, as a function of $(\epsilon_\delta)_{e\mu}$ and $(\epsilon_\delta)_{\mu\tau}$ (left panel),
	 as well as $|(\epsilon_2)_{ee}|$ and $|(\epsilon_2)_{\tau \tau}|$, assuming $|(\epsilon_2)_{e\tau}|=|(\epsilon_2)_{\tau e}| = \sqrt{|(\epsilon_2)_{ee}(\epsilon_2)_{\tau\tau}|}$ (right panel).
	 For each panel, all other entries of $\epsilon_\delta$ and $\epsilon_2$ are set to zero. 
	 With this choice of parameters, LFV is permitted only in the $e-\tau$ sector, in both panels. 
	 In RGE-induced WCs proportional to $\log M_\delta$ ($\log M_2$) we have taken $M_\delta(M_2)=10\,{\rm TeV}$.
	 Some bounds in the right panel also depend on the choice of $\epsilon_\lambda$, and we set $\epsilon_\lambda=0.1$. 
	 The colour scheme for the constraints is the same as in Fig.~\ref{fig:SeesawETAU}, as indicated by the labels.
	 Additionally, the dark brown bounds are from $e^+ e^- \to e^+ e^-, \mu^+ \mu^-$ at LEP.
	 In the right panel, the $a_e$ ($d_e$) bounds are dotted, as they constrain only the real (imaginary) part of 
	 $(\epsilon_2)_{ee}$. Therefore, only the weaker of the two bounds applies to $|(\epsilon_2)_{ee}|$ (light grey shading).
	}
	\label{fig:ZeeETau}
\end{figure}

\begin{figure}[tb]
	\hspace{-1.0cm}
	\mbox{
		\includegraphics[width=0.52\textwidth]{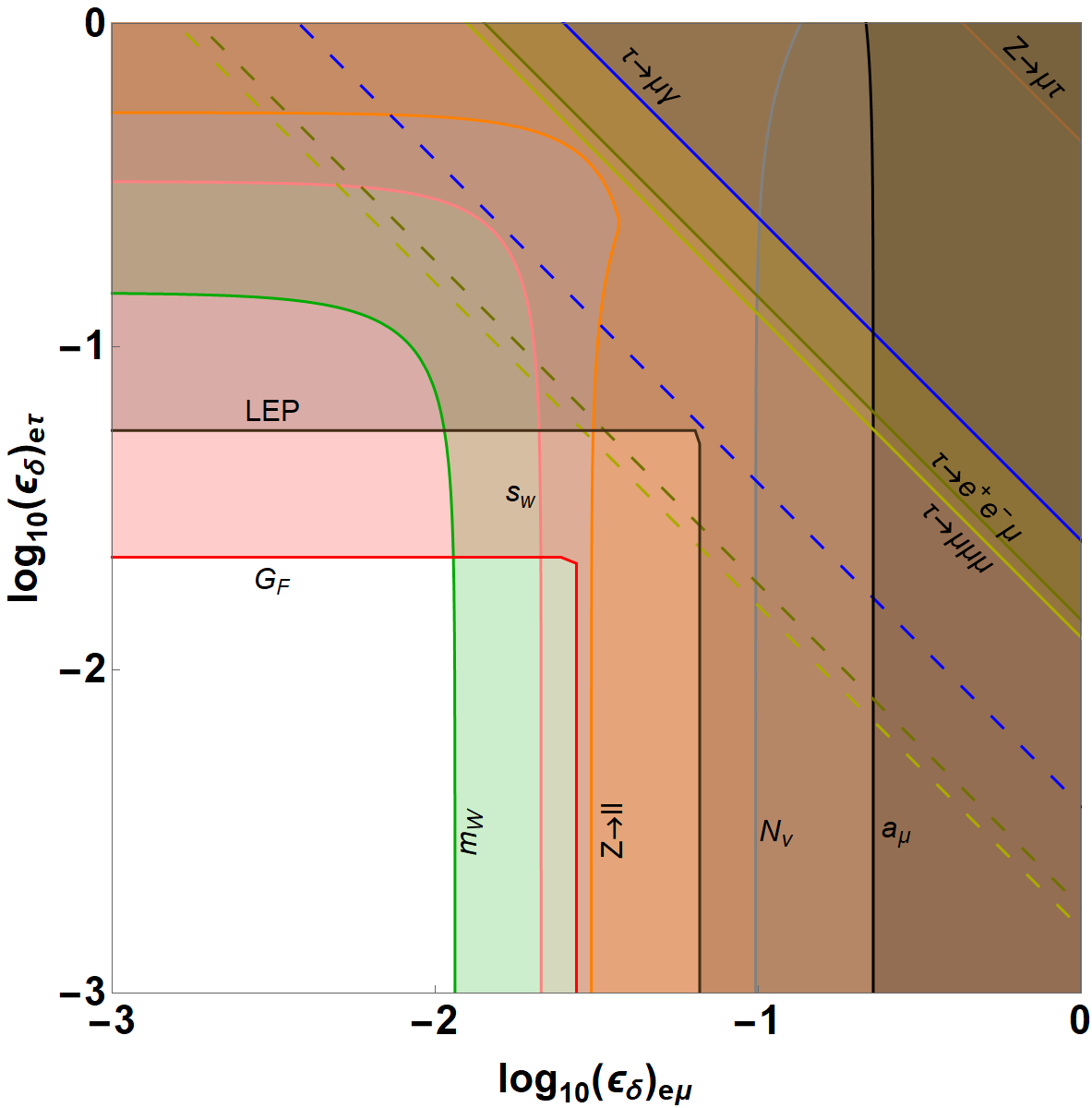}
		\quad
		\includegraphics[width=0.52\textwidth]{ZeeMuMuTauTauMW.png}
	}
	\caption{
	 Bounds on the Zee model, as a function of $(\epsilon_\delta)_{e\mu}$ and $(\epsilon_\delta)_{e\tau}$ (left panel),
	 as well as $|(\epsilon_2)_{\mu\mu}|$ and $|(\epsilon_2)_{\tau \tau}|$, assuming $|(\epsilon_2)_{\mu\tau}|=|(\epsilon_2)_{\tau\mu}| = \sqrt{|(\epsilon_2)_{\mu\mu}(\epsilon_2)_{\tau\tau}|}$ (right panel).
	 For each panel, all other entries of $\epsilon_\delta$ and $\epsilon_2$ are set to zero. 
	 With this choice of parameters, LFV is permitted only in the $\mu-\tau$ sector, in both panels. 
	 In RGE-induced WCs proportional to $\log M_\delta$ ($\log M_2$) we have taken  $M_\delta(M_2)=10\,{\rm TeV}$.
	 Some bounds in the right panel also depend on the choice of $\epsilon_\lambda$, and we set $\epsilon_\lambda=0.1$. 
	 The colour scheme for the constraints is the same as in Fig.~\ref{fig:SeesawMUTAU}, as indicated by the labels. 
	 Additionally, the dark brown bounds in the left panel are from $e^+ e^- \to \mu^+ \mu^-, \tau^+ \tau^-$ at LEP, 
	 while the dark red dot-dashed bound in the right panel is from a LHC search. 
	 In the right panel, the muon magnetic dipole moment bound is dotted, as it constrains only the real part of 
	 $(\epsilon_2)_{\mu\mu}$.
	 We also shaded in dark grey the $2\sigma$ region 
	 preferred by the current $a_\mu$ anomaly.
	}
	\label{fig:ZeeMuTau}
\end{figure}

We again use the tininess of neutrino masses as a starting point for our analysis. 
With the neutrino mass matrix $m_\nu \propto c^W \propto \mu_Z[\epsilon_\delta y_e^\dag \epsilon_2 + (\ldots)^T ]$, 
there are various limits leading to $m_\nu \to 0$. 
One possibility is that one single coupling is vanishingly small, (i) $\mu_Z \to 0$, (ii) $\epsilon_2 \to 0$, or (iii) $\epsilon_\delta \to 0$. 
In each case, lepton number is conserved, given appropriate lepton number assignments for the scalar fields $H_2$ and $\delta$. 
Alternatively, it is possible to keep all couplings sizeable, and still arrange the various flavour entries to achieve $[\epsilon_\delta y_e^\dag \epsilon_2 + (\ldots)^T] \to 0$, however this requires an apparently unnatural tuning. 
Note also (see Tables \ref{table-WCs1}-\ref{table-WCs4}) that, in contrast with $c^W$, the dim-6 WCs never depend on $\mu_Z$, 
nor on the product of $\epsilon_2$ and $\epsilon_\delta$.
In view of these considerations, we will analyse in detail the limits (ii) and (iii), corresponding to fully decoupled $H_2$ or $\delta$, respectively.
This allows us to comprehensively study the phenomenological impact of $\epsilon_\delta$ and $\epsilon_2$ separately, 
which will be illustrated in the left and right panels of Figs.~\ref{fig:ZeeEMu}-\ref{fig:ZeeMuTau}, respectively. 
Neutrino masses can be suppressed even when both matrices are sizeable at the same time: in this case it is straightforward to combine the constraints on the two limiting cases that we will present below, barring unlikely cancellations.

Before turning to experimental bounds, we comment on the dipole operator in the Zee model.
As already mentioned in section \ref{ssec:zee}, in the presence of a second Higgs doublet there are two-loop contributions from Barr-Zee diagrams \cite{Barr:1990vd} which may dominate over one-loop contributions since the additional loop-suppression can be compensated for by less chiral suppression (fewer charged-lepton Yukawa couplings, replaced by gauge or heavy-quark Yukawa couplings). 
Adding the two largest pieces, involving a top loop and a $W$ loop respectively, the Barr-Zee contribution to the EM dipole WC is approximately given by \cite{Harnik:2012pb,Davidson:2016utf} 
\begin{equation}
	c^{e\gamma,BZ}_{ab} \simeq \frac{e^3}{128\pi^4} \left[ \frac{8}{3} y_t^2 - \left( 6 + \log \frac{M_2}{v} \right) \right] \epsilon_\lambda^* (\epsilon_2^\dag)_{ab} \, ,
	\label{cegBZ}
\end{equation}
where the first term comes from the top loop and the second from the $W$ loop. 
This contribution should be compared with the one-loop contributions to the dipole WCs given in table \ref{table-WCs1}. 
The latter are suppressed by a charged-lepton Yukawa, however one should notice that (i) in the case of the third family, the $y_\tau$ suppression is comparable to the additional Barr-Zee loop suppression;
(ii) in the limit $\epsilon_\lambda \to 0$ the Barr-Zee contribution vanishes, while both $\epsilon_2$ and $\epsilon_\delta$ induce one-loop contributions; (iii) in the limit $\epsilon_2 \to 0$, only the one-loop contribution from $\epsilon_\delta$ survives.

The experimental bounds are derived inserting the Zee model WCs into the combinations of WCs constrained 
in Table \ref{tab:WCbounds}.
As usual, let us discuss the few additional constraints that did not fit in that table, starting with $\mu\to e$ conversion in nuclei.
In the Zee model, there is a tree-level contribution to 
$c^{eH}$, which matches onto the two-quark-two-lepton scalar-current WCs of \eq{mueconversion}, according to
\begin{align}
&	c^{S,L}_{\substack{eq\\e\mu}} = - \frac{v m_q}{2 \sqrt{2} m_h^2} c^{eH*}_{\mu e}\, ,&
&	c^{S,R}_{\substack{eq\\e\mu}} = - \frac{v m_q}{2 \sqrt{2} m_h^2} c^{eH}_{e\mu}\, ,&
\end{align}
for $q$ running over both light and heavy quarks. 
In contrast, the vector-current  WCs of \eq{mueconversion} are generated at one loop in the Zee model, according to
\begin{align}
&	c^{V,R}_{\substack{eu\\e\mu}} = \left(\frac{1}{2}- \frac{4}{3} s_w^2 \right) c^{He}_{e\mu} + \frac{1}{2} \left(c^{eu}_{e\mu uu} + c^{qe}_{uue\mu} \right) \, ,\qquad
c^{V,R}_{\substack{ed\\e\mu}} = \left( - \frac{1}{2} + \frac{2}{3} s_w^2 \right) c^{He}_{e\mu} + \frac{1}{2} \left(c^{ed}_{e\mu dd} + c^{qe}_{dde\mu} \right) \, ,& \notag \\
&	c^{V,L}_{\substack{eu\\e\mu}} = \left(\frac{1}{2} - \frac{4}{3} s_w^2\right) (c^{Hl(1)}_{e\mu} + c^{Hl(3)}_{e\mu}) + \frac{1}{2}\left(c^{lu}_{e\mu uu} + c^{lq(1)}_{e\mu 11} - c^{lq(3)}_{e\mu 11} \right) \, ,&\notag \\
&	c^{V,L}_{\substack{ed\\e\mu}} = \left(- \frac{1}{2} + \frac{2}{3} s_w^2\right) (c^{Hl(1)}_{e\mu} + c^{Hl(3)}_{e\mu}) + \frac{1}{2}\left(c^{ld}_{e\mu uu} + c^{lq(1)}_{e\mu 11} + c^{lq(3)}_{e\mu 11} \right) \, . &
\label{cvZeeConversion}
\end{align}
The analogous WCs involving heavier quarks are irrelevant, because $f^{q}_{Vi}=0$ for $q = c,s,b,t$. 
The tree-level contribution is relatively suppressed by a factor $m_{p,n}/v \simeq 0.004$, therefore the loop-induced pieces are comparable to it.
Moreover, in the limit $\epsilon_2 \to 0$ one finds $c^{eH} = 0$ and the tree-level WCs vanish. 
Finally, the dipole WC $c^{e\gamma}$ is given by the sum of the two-loop Barr-Zee contribution in Eq.~\eqref{cegBZ} with the one-loop contributions 
from Eqs.~\eqref{cegMatchTree}, \eqref{cegMatchLoop} and \eqref{cegRGE}.

Next, let us discuss the lepton decays into three charged leptons, introduced in section \ref{eto3e}. 
These are mediated at tree level by $\epsilon_2$, which induces $c^{le}$. 
In the limit $\epsilon_2 \to 0$, they are generated by $\epsilon_\delta$ only at one-loop order:
the scalar $\delta$ cannot induce any $\ell \to 3\ell$ decays at tree level, since  
it always couples to one neutrino and one charged lepton. 
Indeed, the combinations of $c^{ll}$ entries which enter into the decay rates vanish, in the Zee model at tree level, due to the antisymmetry of $\epsilon_\delta$.
In view of these considerations, to derive the bounds on $\epsilon_\delta$, 
we need to evaluate the relevant WCs at one-loop leading-log order. Let us consider the various decay channels in turn.
The constraints from the $\tau^-\to\ell^-_b\ell^-_b\ell^+_c$ decays are provided in table \ref{tab:WCbounds}. 
In the limit $\epsilon_2\to 0$,  we computed the four-lepton operator WCs at one-loop leading log using the RGEs of \cite{Alonso:2013hga,Jenkins:2017dyc}.
For the channels $\ell_a \to 3\ell_b$, the tree-level rate in the Zee model with $\epsilon_2$ only is 
\begin{align}
	\Gamma(\ell_a^- \to \ell_b^- \ell_b^+ \ell_b^-) = \frac{m_a^5}{384 \pi^3 v^4} \left( |c^{le}_{abbb}|^2 + |c^{le}_{bbab}|^2 \right) \, .
\end{align} 
On the other hand, the rate with $\epsilon_\delta$ only is given by
\begin{align}
	\Gamma(\ell_a^- \to \ell_b^- \ell_b^+ \ell_b^-) &= \frac{m_a^5}{384 \pi^3 v^4} \Bigg[ 2 \left|g_{\substack{4\\ab}}\right|^2 +  \left|g_{\substack{6\\ ab}}\right|^2 
	- \frac{8 ev}{\sqrt{2} m_a} \text{Re}\left[c^{e\gamma}_{ba} \left(2g_{\substack{4\\ ab}} + g_{\substack{6\\ ab}}\right) \right] \notag \\
	&+ \frac{16 e^2 v^2 }{m_a^2} \left(\log \frac{m_a^2}{m_b^2} - \frac{11}{4} \right) (|c^{e\gamma}_{ab}|^2 + |c^{e\gamma}_{ba}|^2)\Bigg]  \, ,
	\label{Fab}
\end{align}
where the form factors $g_{4,6}$ \cite{Kuno:1999jp} in the Zee model are
\begin{align}
	g_{\substack{4\\ab}} &= (1- 2 s_w^2) (c^{Hl(1)}_{ab} + c^{Hl(3)}_{ab}) + \frac{1}{24\pi^2} \left(
	[g_1^2 - g_2^2] \log \frac{M}{v} + 8 e^2\log \frac{v}{m_\tau} \right) c^{ll}_{abcc}  \, \label{g4} ,\\
	g_{\substack{6\\ab}} &= - 2 s_w^2 (c^{Hl(1)}_{ab} + c^{Hl(3)}_{ab}) + \frac{1}{12\pi^2} \left(g_1^2 \log \frac{M}{v} + 4 e^2 \log \frac{v}{m_\tau}\right) c^{ll}_{abcc} \label{g6} \, ,
\end{align}
with $c \neq a,b$, and where we used $c^{ll}_{abcc} = c^{ll}_{ccab} = - c^{ll}_{accb} = -c^{ll}_{cbac}$ because $\epsilon_\delta$ is antisymmetric and real. 
We have ignored terms relatively suppressed by additional loops or charged lepton Yukawas. 
Also, the contributions of the other form factors identified in \cite{Kuno:1999jp} are negligible.
The factors of $\log(M/v)$ account for the running from the Zee mass scale to the electroweak scale; 
the factors of $\log(v/m_\tau)$ account for the running down to the mass scale of the decaying $\tau$ or, in the case of $\mu \to 3e$, of the $\tau$ in the loop. 
Note $c^{Hl(1,3)}$ are also proportional to $\log(M/v)$, see table \ref{table-WCs1}. 
Finally, the rate for $\tau^- \to \ell_b^- \ell_c^+ \ell_c^-$ in the Zee model with only $\epsilon_2$ is
\begin{equation}
	\Gamma(\tau^- \to \ell_b^- \ell_c^+ \ell_c^-) = \frac{m_\tau^5}{384 \pi^3 v^4} \left(|c^{le}_{\tau b cc}|^2 + |c^{le}_{\tau ccb}|^2 + |c^{le}_{cc\tau b}|^2 + |c^{le}_{cb\tau c}|^2 \right) \, ,
\end{equation}
while with only $\epsilon_\delta$ the rate reads
\begin{align}
	\Gamma(\tau^- \to \ell_b^- \ell_c^+ \ell_c^-) &= \frac{m_\tau^5}{384 \pi^3 v^4} \Bigg[  \left|g_{\substack{4\\ \tau b}}\right|^2 
	+ \left|g_{\substack{6\\ \tau b}}\right|^2 - \frac{8ev}{\sqrt{2} m_\tau} \text{Re}\left[ c^{e\gamma}_{b\tau} \left(g_{\substack{4\\ \tau b}} + g_{\substack{6\\ \tau b}}\right) \right]  \notag\\
	&+ \frac{16e^2 v^2}{m_\tau^2} \left(\log \frac{m_\tau^2}{m_\mu^2} - 3 \right) 
	\left( |c^{e\gamma}_{b\tau}|^2 + |c^{e\gamma}_{\tau b}|^2 \right) \Bigg]  \, .
	\label{Gtbbc}
\end{align}
In summary, in the Zee model with $\epsilon_\delta= 0$ and $\epsilon_2\ne 0$, 
all $\ell\to3\ell$ decays are generated at tree level, proportionally to various combinations of $c^{le}$ entries.  
In the opposite limit, $\epsilon_\delta\ne 0$ and $\epsilon_2= 0$, all $\ell\to3\ell$ decays 
are generated at one loop, proportionally to various log-enhanced combinations of the WCs
$c^{ll}$, $c^{Hl(1,3)}$, and $c^{e\gamma}$.

Our results are summarised in Figs.~\ref{fig:ZeeEMu}, \ref{fig:ZeeETau} and \ref{fig:ZeeMuTau}. 
For the $\epsilon_\delta$ only case (left panels), we study constraints as a function of pairs of entries of the antisymmetric matrix $\epsilon_\delta$,  setting to zero the third independent entry. 
This completely fixes the phenomenology, since the entries of $\epsilon_\delta$ can be chosen real and positive with no loss of generality. 
Amusingly, if $(\epsilon_{\delta})_{bc}=0$ and $(\epsilon_{\delta})_{ab,ac}\ne 0$, where $a\ne b\ne c$ are the three lepton flavours, then dim-6 operators induce LFV only in the $b \leftrightarrow c$ channel. The plots display the bounds from all LFV observables we discussed, except for LFV Higgs decays, as they are too weak due to loop and $y_e$ suppression, and for  
$\tau^- \to \ell_b^- \ell_b^- \ell_c^+$ decays, since dim-6 operators induced by the scalar $\delta$ violate flavour by at most one unit.

For the $\epsilon_2$ only case (right panels), we study constraints as a function of pairs of diagonal entries of the matrix $\epsilon_2$, setting to zero the third diagonal entry and assuming $|(\epsilon_2)_{ab}| = \sqrt{|(\epsilon_2)_{aa} (\epsilon_2)_{bb}|}$. 
This benchmark allows us to compare flavour-conserving and violating observables on a level footing.
In general, the $\epsilon_2$ entries are all independent, therefore 
the relative strength of the flavour-conserving and flavour-violating bounds in our figures can be varied by 
rescaling the size of the off-diagonal entry $(\epsilon_2)_{ab}$. 
With the choice of parameters in the figures, the rates of $\tau\to \ell_a\ell_a\ell_b$ decays with $a\ne b$ vanish at tree level (the WCs $c^{\ell e}_{abcd}$ 
involving all three different flavours are zero).
Since the experimental constraints on these decays are comparable to those on $\tau\to 3\ell_a$, 
this is not consequential for our analysis.

Several of the observables we consider depend on the product $\epsilon_2 \epsilon_\lambda$, where $\epsilon_\lambda$ is related to the coupling between the two Higgs 
doublets, 
see our spurion analysis in section \ref{ssec:zee}. 
In particular, non-standard Higgs decays in the Zee model are directly proportional to $c^{eH} = \epsilon_\lambda^* \epsilon_2^\dag$. 
Furthermore, dipole moments and radiative charged lepton decays may be dominated by the Barr-Zee contribution of Eq.~\eqref{cegBZ}, which is also proportional to $\epsilon_\lambda^* \epsilon_2^\dag$. 
The $\mu \to e$ conversion is also driven by the Barr-Zee piece, except in the regions $|(\epsilon_2)_{ee}| \gg |(\epsilon_2)_{\mu \mu}|$ and $|(\epsilon_2)_{\mu \mu}| \gg 
|(\epsilon_2)_{ee}|$ (i.e. where the cyan line bends in the right panel of Fig. \ref{fig:ZeeEMu}). 
On the other hand, the Barr-Zee contribution is subleading for lepton-to-three-lepton decays. 
In the right panels of Figs. \ref{fig:ZeeEMu}-\ref{fig:ZeeMuTau}, as we vary the entries of the matrix $\epsilon_2$,  we fix $\epsilon_\lambda = 0.1$, which is roughly the
maximal value allowed by perturbativity when $M_2\sim10$ TeV. 
Reducing $\epsilon_\lambda$ would proportionally weaken the bounds from the set of observables just mentioned.

We see from the left (right) panels of Figs.~\ref{fig:ZeeEMu}-\ref{fig:ZeeMuTau} that $\epsilon_\delta$ ($\epsilon_2$) is predominantly bound by observables that conserve (violate) flavour. 
As explained in section \ref{EWPT}, many of the flavour-conserving constraints are due to EFT corrections to the muon and tau SM decays to three leptons. 
These are induced by $\epsilon_\delta$ at tree-level, since the decays can be mediated by the scalar $\delta$. 
In particular, the shift in the Fermi constant depends on $(\epsilon_{\delta})_{e\mu}$, thus this parameter is constrained by several observables. 
Of these, $m_W$ gives the 
strongest constraint since the model predicts a negative shift in the $W$-boson mass, in the opposite direction of the current anomaly. 
The best limits on $(\epsilon_{\delta})_{e \tau}$ and $(\epsilon_\delta)_{\mu \tau}$ come from $G_F$ universality. 
Only the very stringent experimental bounds on $\mu \to e$ transitions provide a stronger constraint, in the region $(\epsilon_\delta)_{e\tau}\sim(\epsilon_\delta)_{\mu\tau}$, see the left panel of Fig.~\ref{fig:ZeeEMu}. 
The best of these comes from $\mu \to e$ conversion in nuclei: assuming order-one Yukawa couplings, the current (future) limit probes scales as large as $M_\delta \simeq 90\, (2600)$ TeV.
By contrast, the flavour-conserving bounds are stronger than both current and expected future limits from $\tau \to e$ and $\tau \to \mu$ transitions, see the left panels of Figs.~\ref{fig:ZeeETau} and \ref{fig:ZeeMuTau}.

Flavour-conserving bounds are weaker in the limit $\epsilon_\delta \to 0$. 
In this case, the relevant shift to $G_F$ is quadratic in the dim-6 WCs, $\mathcal{O}(c^2)$.\footnote{More precisely, there exists an $\mathcal{O}(c)$ shift in $G_F$ in this limit, but it is suppressed both by a loop and two powers of the matrix $y_e$, so it is compatible with 
$\epsilon_2$ entries of order one.} 
Indeed, the operators $Q_{ll}$ and $Q_{Hl}^{(3)}$ (induced by $\epsilon_\delta$) generate a tree-level contribution to muon decay that interferes with the SM diagram, while $Q_{le}$ 
(induced by $\epsilon_2$) generates a decay that does not interfere. 
The measured Fermi constant in this case is 
\begin{equation}
	G_F \simeq G_{F,0} \Big(1 + \frac{1}{2} \sum \limits_{a,b = e,\mu,\tau} |c^{le}_{ab e\mu}|^2 \Big) \, ,
\end{equation} 
which corresponds, according \eq{gfShift}, to $c^G = (-1/2)\sum_{a,b = e,\mu,\tau} |c^{le}_{ab e\mu}|^2$. 
Note this should be taken only as an order of magnitude estimate, as we are neglecting the possible interference of dim-8 operators with the SM, which would be of the same order. 
Up to this caveat, the bounds on $c^G$ listed in table \ref{tab:WCbounds} can be applied, but are negligible, 
as one can see in the top right corner in the right panel of Fig. \ref{fig:ZeeEMu}.
The shift in $G_F$ vanishes in Figs.~\ref{fig:ZeeETau} and \ref{fig:ZeeMuTau}, as a result of the assumption $|(\epsilon_2)_{ab}| = \sqrt{|(\epsilon_2)_{aa}(\epsilon_2)_{bb}|}$: also in these cases 
the residual flavour-conserving bounds are negligible.

Coming to the LFV bounds in the $\epsilon_2$ case (with $\epsilon_\delta\to 0$), 
the current best limit  in the $\mu-e$ sector comes from $\mu \to e \gamma$ in most of the parameter space.
The particular shape of the $\mu\to e$ conversion constraint in the right panel of Fig.~\ref{fig:ZeeEMu} is due to the interference 
between the tree-level and one-loop contributions to the amplitude. 
For order-one Yukawa couplings, the current $\mu \to e \gamma$ (future $\mu \to e$ conversion) limits probe scales as large as $M_2 \simeq 400\, (3200)$ TeV. 
It is noticeable that, in the $\tau-e$ and $\tau-\mu$ sectors (right panels of Figs.~\ref{fig:ZeeETau} and \ref{fig:ZeeMuTau}), the LFV Higgs decays provide the best constraint
in the region $|(\epsilon_2)_{\tau \tau}|\gtrsim10^{-3}$, and they are superseded by $\tau\to 3\ell_a$ for $|(\epsilon_2)_{\tau \tau}|\lesssim 10^{-3}$. 
We note also that in the limit where $H_2$ has only off-diagonal couplings, the bounds from $\mu \to e$ processes do not apply,
since they are induced by a combination of flavour diagonal and off-diagonal couplings. In this limit, the best bounds on LFV may come from neutrino scattering experiments \cite{Babu:2019mfe} and muonium-antimuonium transitions \cite{Fukuyama:2021iyw}.

Let us now turn to dipole moments.
The real matrix $\epsilon_\delta$ contributes to the charged-lepton magnetic dipole moments but not the electric dipole moments since it induces a real $c^{e\gamma}_{aa}$. 
The aforementioned constraints on $\epsilon_\delta$ are sufficiently strong that the contributions to $a_e$ and $a_\mu$ are negligibly small compared to the present anomalies. 
On the other hand, $\epsilon_2$ induces a contribution to the dipole which has an unknown phase, thus it could affect both the magnetic and electric dipole moment.\footnote{Combining the two bounds $|\text{Re}(\epsilon_2)_{aa}| < X_1$ and $|\text{Im}(\epsilon_2)_{aa}| < X_2$ provides 
the limit $|(\epsilon_2)_{aa}| < \sqrt{X_1^2 + X_2^2}$, which is well approximated by the weaker of the two bounds.}
We therefore present bounds on $\epsilon_2$ from magnetic (electric) dipole moments as dotted lines, assuming the relevant $\epsilon_2$ entry to be 
real (imaginary). 
If $(\epsilon_2)_{ee}$ were purely imaginary, $d_e$ would give the strongest limit on the Zee model, probing $M_2 \simeq 8$ PeV for $(Y_2)_{ee}$ imaginary and of order one. 
Intriguingly, in the region $|(\epsilon_2)_{\mu \mu}| \gg |(\epsilon_2)_{ee,\tau\tau}|$, it appears possible to address the $(g-2)_\mu$ anomaly with a second Higgs doublets in the (slightly) decoupled regime. 
Indeed, the $2\sigma$ best fit region is lightly shaded in the right panels of Figs.~\ref{fig:ZeeEMu} and \ref{fig:ZeeMuTau}, and in both plots it overlaps with the allowed white region of parameter space.

Some additional 
bounds come from colliders. 
LEP sets an upper limit $\sim~0.1$ on $(\epsilon_\delta)_{e\mu}$, $(\epsilon_\delta)_{e\tau}$ and $|(\epsilon_2)_{ee}|$, as illustrated in Figs.~\ref{fig:ZeeEMu}-\ref{fig:ZeeMuTau}, from measurements of $e^+ e^- \to \ell^+ \ell^-$ \cite{Schael:2006wu}. 
However, this never provides the strongest constraint: note that, in the right panels of Figs.~\ref{fig:ZeeEMu} and \ref{fig:ZeeETau}, 
the bounds on ${\rm Re}(\epsilon_2)_{ee}$ and ${\rm Im}(\epsilon_2)_{ee}$,
coming from $a_e$ or $d_e$ respectively, always prevail since one cannot avoid both bounds at the same time.
A muon specific two-Higgs-doublet model was studied in Ref.~\cite{Abe:2017jqo}, 
which corresponds to $|(\epsilon_2)_{\mu \mu}|$ much larger than all other entries of $\epsilon_2$: it was found that such model can resolve 
the $(g-2)_\mu$ anomaly and avoid LHC searches as long as $M_2 \gtrsim 640$ GeV. For $|(Y_2)_{\mu \mu}| = 1$, this corresponds 
to $|(\epsilon_2)_{\mu \mu}| \lesssim 0.27$, shown in the right panel of Figs.~\ref{fig:ZeeEMu} and \ref{fig:ZeeMuTau}: of course, such upper bound changes linearly  with the Yukawa coupling. 
These findings are consistent with our allowed region for $a_\mu$, discussed in the previous paragraph.

We compared our results with previous literature on the phenomenology of the Zee model \cite{Herrero-Garcia:2017xdu,Babu:2019mfe}, 
the two-Higgs-doublet model \cite{Diaz:2000cm,Davidson:2016utf} (Zee in the limit $\epsilon_\delta \to 0$), and the SM plus $\delta$ only \cite{Crivellin:2020klg,Felkl:2021qdn}
(Zee in the limit $\epsilon_2 \to 0$), finding agreement where there is an intersection, except for a typo in the shift of $m_W$ in \cite{Felkl:2021qdn}.

\subsection{Leptoquark model}

\begin{figure}[t]
	\hspace{-1.0cm}
	\mbox{
		\includegraphics[width=0.52\textwidth]{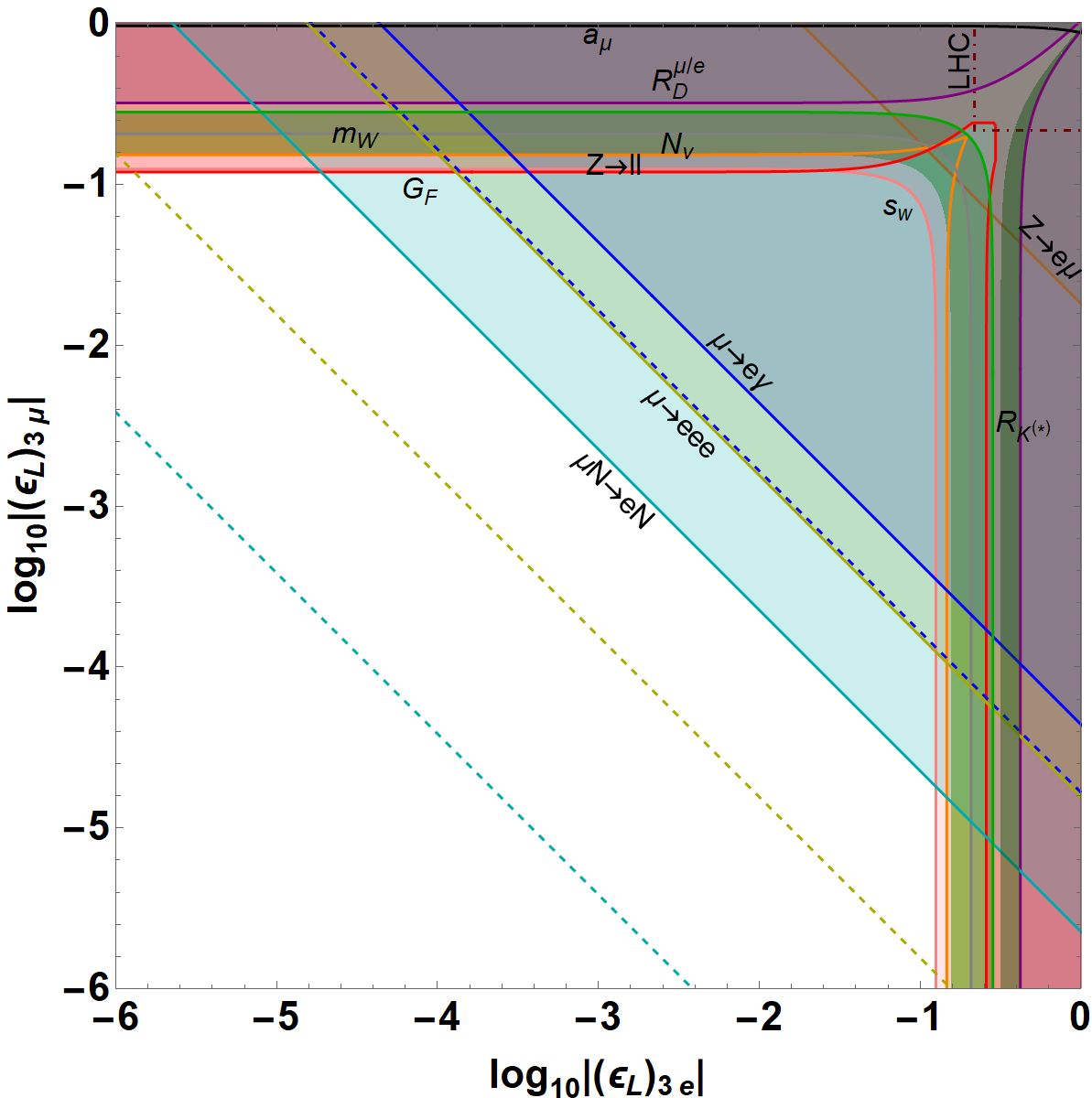}
		\quad
		\includegraphics[width=0.52\textwidth]{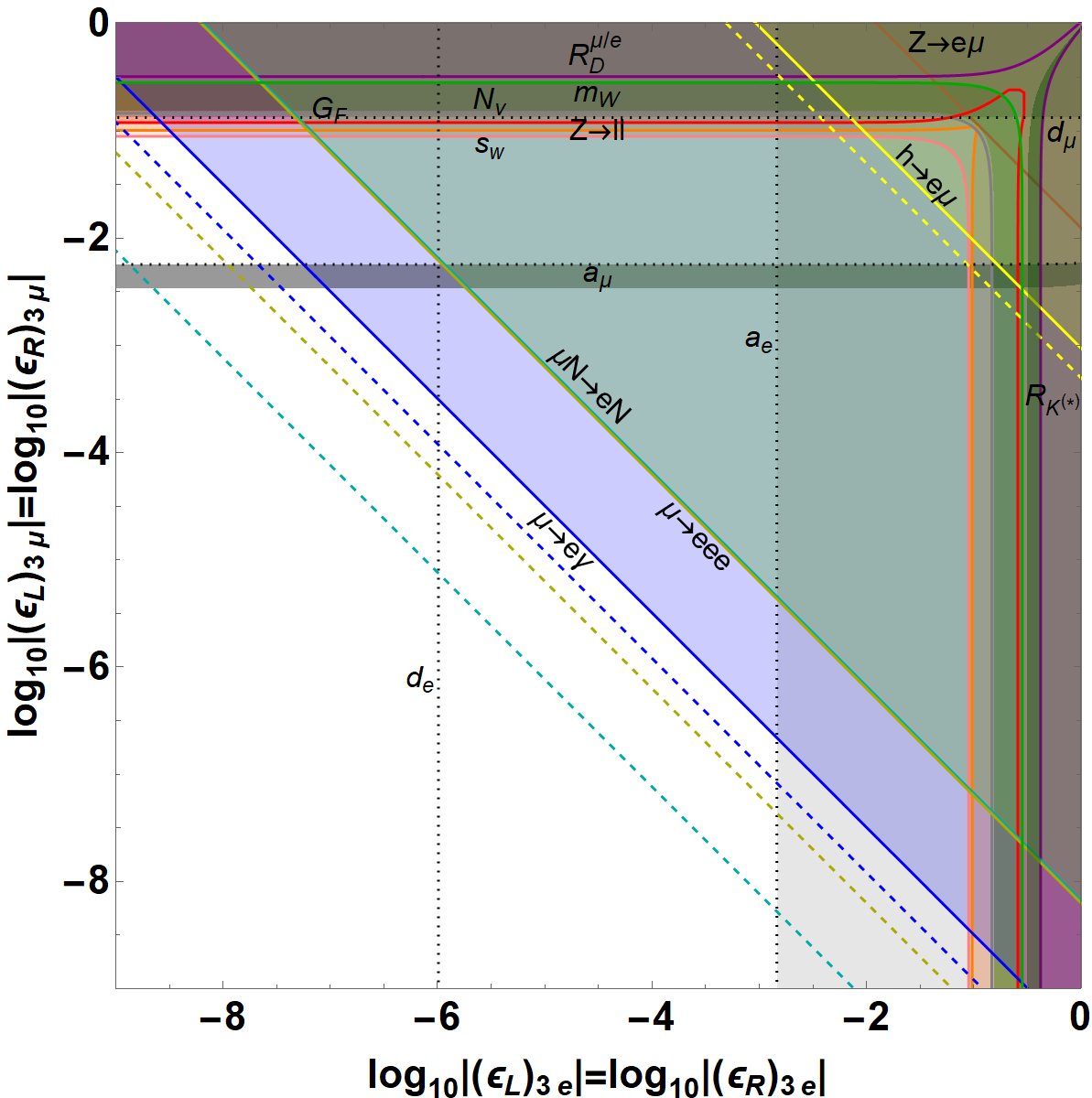}
	}
	\caption{
		Bounds on the LQ model, in the plane $|(\epsilon_L)_{3e}|$ versus $|(\epsilon_L)_{3\mu}|$ (left panel),
		and in the plane $|(\epsilon_L)_{3e}| = |(\epsilon_R)_{3e}|$ versus $|(\epsilon_L)_{3\mu}|=|(\epsilon_R)_{3\mu}|$ (right panel). 
		For each panel, all other entries of $\epsilon_L$ and $\epsilon_R$ (as well as $\epsilon_D$) are set to zero. 
		In the RGE-induced WCs proportional to $\log M_i$, we have taken  $M_i=10\,{\rm TeV}$. 
		The colour scheme for the constraints is the same as in Fig.~\ref{fig:SeesawEMU}, as indicated by the labels. 
		Additionally, the dark-red dot-dashed line in the left panel corresponds to the direct search for $S$ at the LHC, 
		the purple line and shading corresponds to the bound from $R_D^{\mu/e}$,
		and the dark green band corresponds to the $1\sigma$ best-fit region for the $R_{K^{(*)}}$ anomalies. 
		The magnetic (electric) dipole moment bounds in the right panel constrain only the real (imaginary) part of 
		$(\epsilon_L^\dag y_u^T \epsilon_R)_{aa}$ and are therefore dotted.
		The weaker of the two provides a robust bound on $|(\epsilon_L)_{3a}(\epsilon_R)_{3a}|$ (light grey shading). 
		For $a_\mu$/$m_W$ (right/both panels), we shaded in dark grey/light green the $2\sigma$ preferred region.}
	\label{fig:LQEMu}
\end{figure}

\begin{figure}[tb]
	\hspace{-1.0cm}
	\mbox{
		\includegraphics[width=0.52\textwidth]{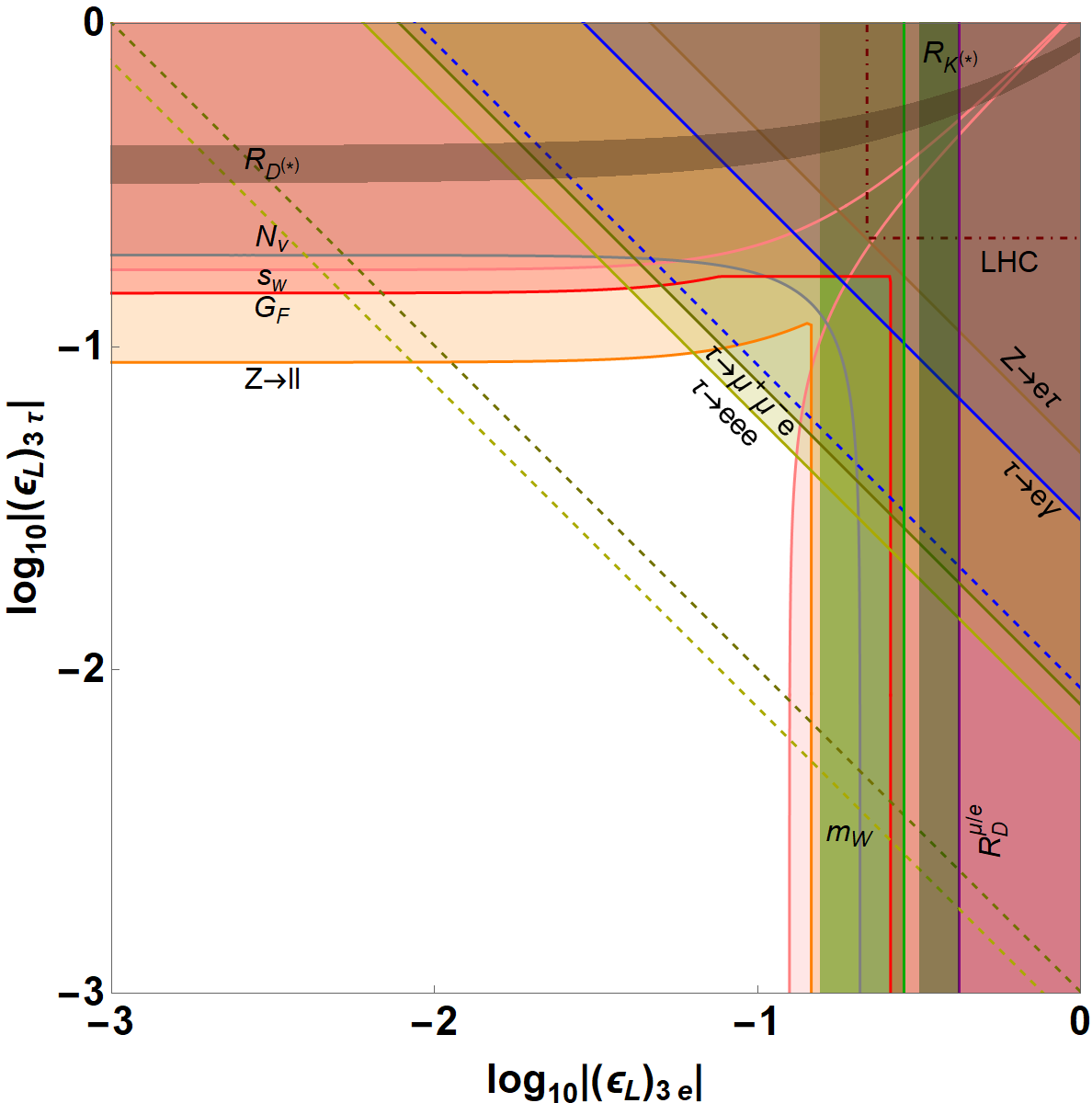}
		\quad
		\includegraphics[width=0.52\textwidth]{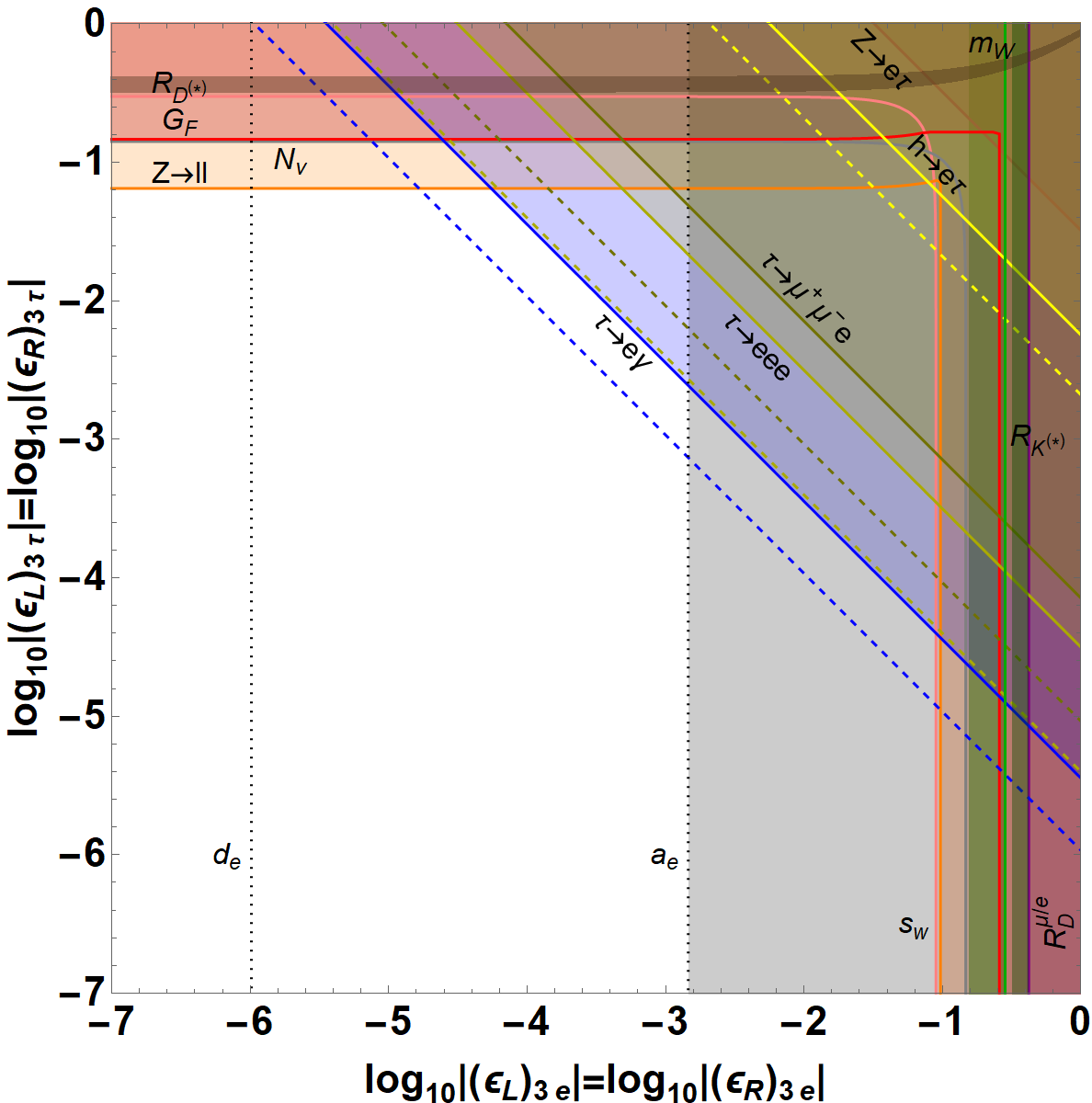}
	}
	\caption{Bounds on the LQ model, in the plane $|(\epsilon_L)_{3e}|$ versus $|(\epsilon_L)_{3\tau}|$ (left panel), and in the plane 
			$|(\epsilon_L)_{3e}| = |(\epsilon_R)_{3e}|$ versus $|(\epsilon_L)_{3\tau}| = |(\epsilon_R)_{3\tau}|$ (right panel).
			For each panel, all other entries of $\epsilon_L$ and $\epsilon_R$ (as well as $\epsilon_D$) are set to zero. 
			In the RGE-induced WCs proportional to $\log M_i$, we have taken  $M_i=10\,{\rm TeV}$. 
			The colour scheme for the constraints is the same as in Figs.~\ref{fig:SeesawETAU} and \ref{fig:LQEMu}, as indicated by the labels. 
			Additionally, the dark brown band corresponds to the $1\sigma$ best-fit region for the $R_{D^{(*)}}$ anomalies. 
			The magnetic (electric) dipole moment bounds in the right panel constrain only the real (imaginary) part of 
			$(\epsilon_L^\dag y_u^T \epsilon_R)_{ee}$ and are therefore dotted.
			The weaker of the two provide a robust bound on $|(\epsilon_L)_{3e}(\epsilon_R)_{3e}|$ (light grey shading). 
			For $m_W$, we shaded in light green the $2\sigma$ preferred region.
			}
	\label{fig:LQETau}
\end{figure}

\begin{figure}[tb]
	\hspace{-1.0cm}
	\mbox{
		\includegraphics[width=0.52\textwidth]{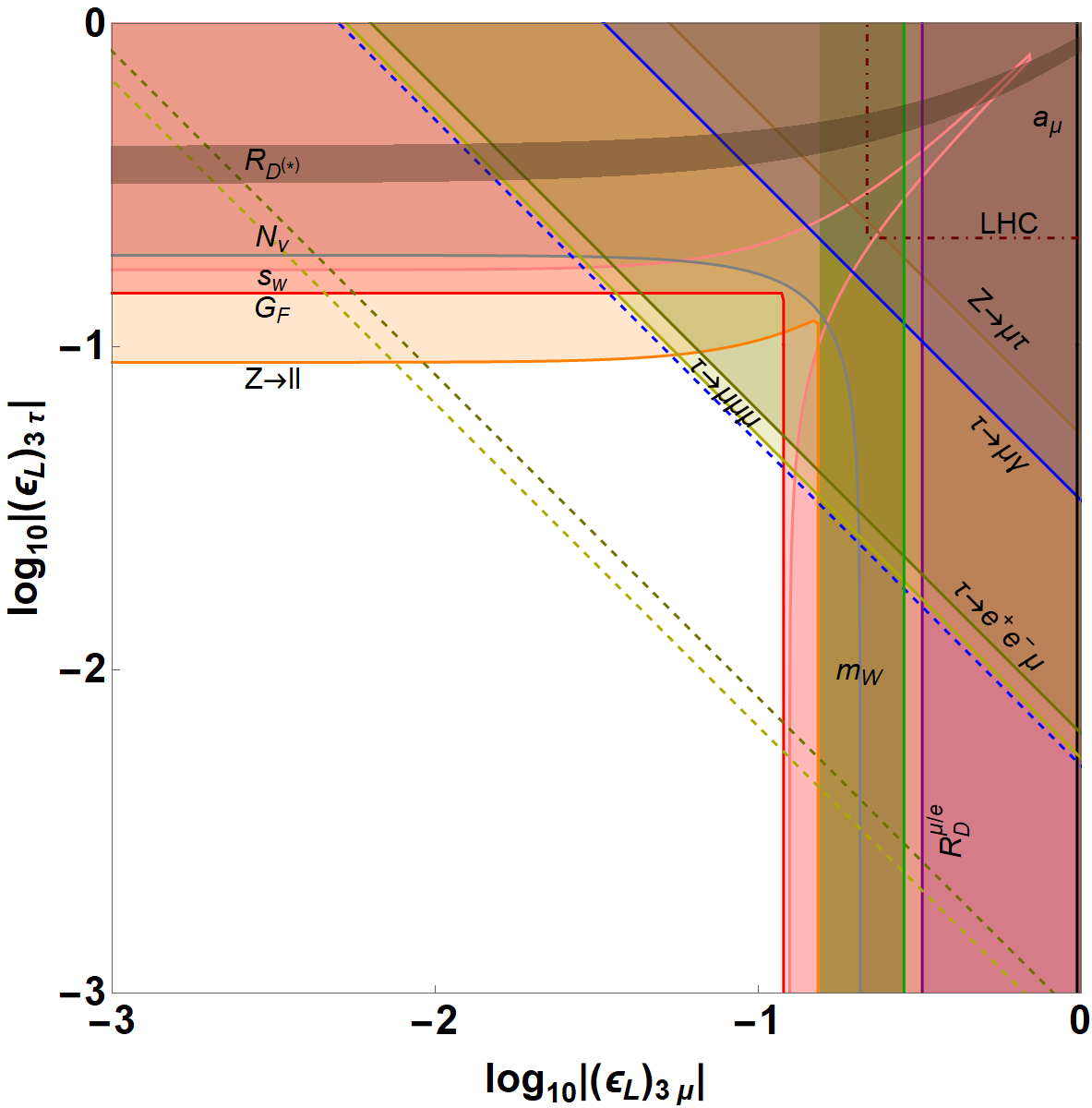}
		\quad
		\includegraphics[width=0.52\textwidth]{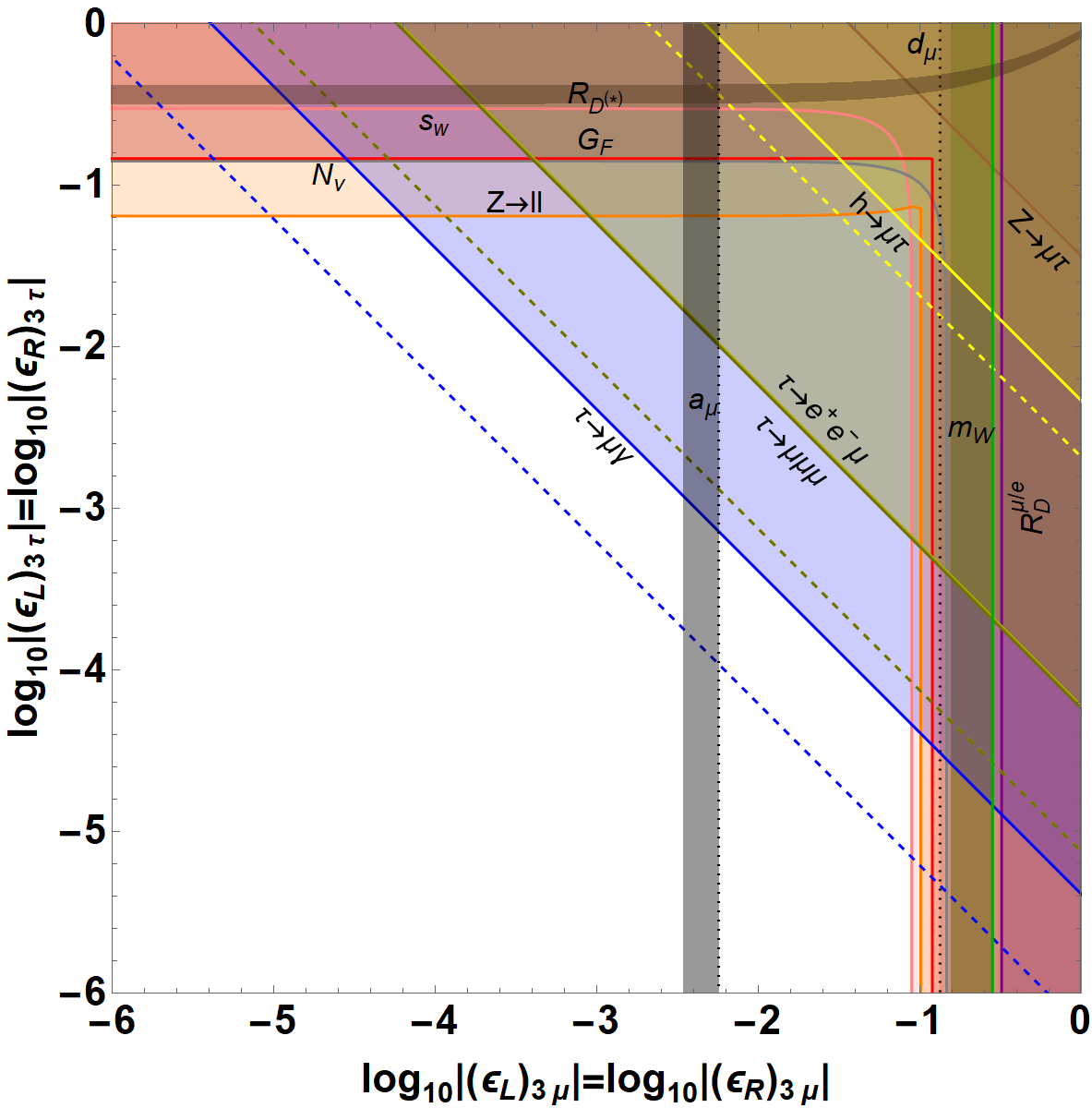}
	}
	\caption{Bounds on the LQ model, in the plane $|(\epsilon_L)_{3\mu}|$ versus $|(\epsilon_L)_{3\tau}|$ (left panel),
		and in the plane $|(\epsilon_L)_{3\mu}| = |(\epsilon_R)_{3\mu}|$ versus $|(\epsilon_L)_{3\tau}|=|(\epsilon_R)_{3\tau}|$ (right panel). 
		For each panel, all other entries of $\epsilon_L$ and $\epsilon_R$  (as well as $\epsilon_D$) are set to zero. 
		In the RGE-induced WCs proportional to $\log M_i$, we have taken  $M_i=10\,{\rm TeV}$. 
		The colour scheme for the constraints is the same as in Figs.~\ref{fig:SeesawMUTAU}, \ref{fig:LQEMu} and \ref{fig:LQETau}, as indicated by the labels. 
		The magnetic (electric) dipole moment bounds in the right panel constrain only the real (imaginary) part of 
		$(\epsilon_L^\dag y_u^T \epsilon_R)_{\mu\mu}$ and are therefore dotted.
		The weaker of the two provides a robust bound on $|(\epsilon_L)_{3\mu}(\epsilon_R)_{3\mu}|$ (light grey shading).
		For $a_\mu$/$m_W$ (right/both panels), we shaded in dark grey/light green the $2\sigma$ preferred region. 
	}
	\label{fig:LQMuTau}
\end{figure}

The minimal LQ model has an even larger parameter space than the Zee model, consisting of three 
flavour matrices, $\epsilon_L, \epsilon_R, \epsilon_D$. 
Neutrino masses are proportional to the combination $\mu_{DS} [(\epsilon_L^T y_d^\dag \epsilon_D) + (\dots)^T]$. There are four ways to take the limit $m_\nu\to 0$: 
three trivial solutions $\mu_{DS} \to 0$, $\epsilon_L \to 0$, or $\epsilon_D = 0$, and a tuned solution in which the entries of $\epsilon_D$ and $\epsilon_L$ are non-zero but conspire to suppress $m_\nu$. 
Considering the Lagrangian in \eq{eq:LQmodel}, one notices that the lepton number symmetry is restored only in the cases  $\mu_{DS} \to 0$ or $\epsilon_D \to 0$,  
while in the other two the smallness of $m_\nu$ is accidental. 
We will see that the phenomenological constraints on the doublet LQ, i.e. on $\epsilon_D$, are quite weak, therefore we will focus on the singlet LQ parameter space
by studying the bounds in the case where $\epsilon_L$ and $\epsilon_R$ entries are comparable (right panels of Figs.~\ref{fig:LQEMu}-\ref{fig:LQMuTau}), 
and in the case where one is much larger than the other (left panels of Figs.~\ref{fig:LQEMu}-\ref{fig:LQMuTau}).

Since we wish to focus specifically on lepton phenomenology, we make the simplifying assumption that, at the electroweak scale, the LQs only couple to third-generation quarks, in the basis where down quark masses are diagonal.\footnote{In particular, 
this implies that $S$ does couple to $u_L$ ($c_L$) and $l_{La}$, proportionally to $V_{ub}$ ($V_{cb}$), via the coupling $(Y_L)_{3a}$.}
There are two reasons for this choice. 
Firstly, there are relatively fewer bounds on processes involving the top and bottom quarks than on those involving lighter quarks. 
Secondly, certain loops involving the quarks are proportional to their Yukawa coupling (various WCs generated at one-loop leading log in Tables \ref{table-WCs1}-\ref{table-WCs4} depend on $y_u$ or $y_d$), thus our assumption accounts for the important loops involving the top quark, only ignoring those proportional to the much smaller Yukawa couplings of the lighter quarks. 
Below, we will briefly discuss 
the possibility of relaxing this ansatz in order to address the neutral and charged current $B$-meson anomalies.

Turning to the dipole operators, let us recall that their WCs are generated at one-loop leading-log order, according to table \ref{table-WCs1}. 
Since this contribution vanishes if either $\epsilon_L$ or $\epsilon_R$ vanishes, it is important to compute the one-loop finite contributions to the dipole WCs. 
As anticipated by \eq{dipoleLQ}, they are suppressed by a charged lepton Yukawa, however they are proportional to a single new physics coupling,  $\epsilon_L$, $\epsilon_R$, or $\epsilon_D$.
Computing the relevant loops, we find
\begin{align}
	&\delta c^{eB}_{ab} = \frac{3 g_1}{16\pi^2} \left[ \frac{(\epsilon_D^\dag \epsilon_D y_e^\dag)_{ab}}{48} + \frac{ (y_e^\dag \epsilon_R^\dag \epsilon_R)_{ab}}{24} \right] \, ,&
	&\delta c^{eW}_{ab} = \frac{3 g_2}{16\pi^2} \left[ \frac{(\epsilon_D^\dag \epsilon_D y_e^\dag)_{ab}}{48} - \frac{(\epsilon_L^\dag \epsilon_L y_e^\dag)_{ab}}{24} \right] \, ,& \label{cegLQ1}
\end{align}
which is in agreement with \cite{Dedes:2021abc}. 
According to \eq{cegMatchTree}, we find
the contribution to $c^{e \gamma}$ proportional to $\epsilon_D^\dag \epsilon_D$ vanishes at this order, as confirmed by
e.g. \cite{Dorsner:2016wpm,Zhang:2021dgl}. The pieces proportional to 
$\epsilon_L^\dag \epsilon_L$ and $\epsilon_R^\dag \epsilon_R$ do contribute to $c^{e \gamma}$, with an equal coefficient.

As for previous models, we will consider the phenomenological constraints collected in table \ref{tab:WCbounds} and particularise them to the LQ-induced WCs. The additional constraints that did not fit in that table are, as usual, 
$\mu\to e$ conversion in nuclei and additional lepton decays to three charged leptons. 
The $\mu \to e$ conversion can in principle be induced at tree level by LQs. 
In the third-generation ansatz, however, the tree-level contribution of vector operators in \eq{mueconversion} vanishes, since it is only summed over light quarks, while the scalar one is
\begin{equation}
	c^{S,R}_{\substack{et\\e \mu}} = - \frac{1}{2} c^{lequ(1)}_{e\mu tt} =\frac 14 (\epsilon^\dagger_L)_{e3}(\epsilon_R)_{3\mu}\, .
	\label{csrLQ}
\end{equation}
In fact, the rate is dominated by the contribution from the dipole WC since, although loop-suppressed, it is relatively enhanced by a factor 
$\sim m_t^2/(m_\mu m_p) \sim 10^5$ in the amplitude. 
In the limit of $\epsilon_L \to 0$ or $\epsilon_R \to 0$, the scalar-current WC vanishes, while $c^{e\gamma}$ remains non-zero
due to the one-loop finite part given in Eq.~\eqref{cegLQ1}. 
Since the latter is chirality-suppressed by $y_e$, the contribution from light-quark vector operators  becomes comparable, as they arise at one-loop leading-log order:
\begin{align}
	c_{\substack{eu\\ e\mu}}^{V,R} &= \left(\frac{1}{2} - \frac{4s_w^2}{3} \right) c^{He}_{e\mu} + \frac{1}{2}c^{qe}_{uue\mu} - \frac{ g_1^2}{18\pi^2} c^{eu}_{e\mu tt} \log \frac{M}{v} \, , \label{V1}\\
	c_{\substack{ed\\ e\mu}}^{V,R} &= \left( - \frac{1}{2} + \frac{2s_w^2}{3} \right) c^{He}_{e\mu} + \frac{1}{2}c^{qe}_{dde\mu} + \frac{1}{2}c^{ed}_{e\mu dd} \, , \\
	c_{\substack{eu\\ e\mu}}^{V,L} &= \left(\frac{1}{2} - \frac{4s_w^2}{3} \right) (c^{Hl(1)}_{e\mu} + c^{Hl(3)}_{e\mu}) + \frac{5g_1^2}{144\pi^2}(c^{ld}_{e\mu bb} - c^{lq(1)}_{e\mu bb}) \log \frac{M}{v} + \frac{g_2^2}{16\pi^2} c^{lq(3)}_{e\mu bb} \log \frac{M}{v} \, , \\
	c_{\substack{ed\\ e\mu}}^{V,L} &= \left( - \frac{1}{2} + \frac{2s_w^2}{3} \right) (c^{Hl(1)}_{e\mu} + c^{Hl(3)}_{e\mu}) + \frac{g_1^2}{144\pi^2}(c^{lq(1)}_{e\mu bb} - c^{ld}_{e\mu bb}) \log \frac{M}{v} - \frac{g_2^2}{16\pi^2} c^{lq(3)}_{e\mu bb} \log \frac{M}{v} \, . \label{V4}
\end{align}
These WCs have to be inserted in  Eq.~\eqref{mueconversion} and, for $\epsilon_L \to 0$ or $\epsilon_R \to 0$, they are relatively log-enhanced with respect to the dipole. 
Eqs.~(\ref{V1}-\ref{V4}) also give the leading contribution of $\epsilon_D$ to $\mu \to e$ conversion.

Turning to $\ell \to 3\ell$ decays, it can be seen that $\tau \to \ell_b^- \ell_c^+ \ell_b^-$ is not generated at one-loop leading log in the LQ model, 
by comparing the WCs in table \ref{tab:WCbounds} with those in table \ref{table-WCs3}. 
The reason  is that the semi-leptonic operators induced by the LQs at tree level violate flavour by at most one unit, and neither gauge nor Higgs loops can generate the necessary additional LFV 
for such decays to occur. 
When $\epsilon_L$ and $\epsilon_R$ are both sizeable, the other decay channels, $\ell_a^- \to \ell_b^- \ell_b^+ \ell_b^-$ and $\tau \to \ell_b^- \ell_c^+ \ell_c^-$, 
are dominated by the dipole contribution, with
\begin{align}
	\Gamma(\ell_a^- \to \ell_b^- \ell_b^+ \ell_b^-) &=  \frac{e^2 m_a^3}{24 \pi^3 v^2} \left(\log \frac{m_a^2}{m_b^2} - \frac{11}{4} \right) (|c^{e\gamma}_{ab}|^2 + |c^{e\gamma}_{ba}|^2) \, , \label{l3lLQ1} \\
	\Gamma(\tau^- \to \ell_b^- \ell_c^+ \ell_c^-) &= \frac{e^2m_\tau^3}{24 \pi^3 v^2} \left(\log \frac{m_\tau^2}{m_\mu^2} - 3 \right) 
	\left( |c^{e\gamma}_{b\tau}|^2 + |c^{e\gamma}_{\tau b}|^2 \right)  \, . \label{l3lLQ2}
\end{align}
When $\epsilon_L \to 0$ or $\epsilon_R \to 0$, again the dipole contribution is non-zero, however  
one also needs to include the comparable contributions from four-lepton operators \cite{Kuno:1999jp,Crivellin:2013hpa}
\begin{align}
	& \Gamma(\ell_a^- \to \ell_b^- \ell_b^+ \ell_b^-) = \frac{m_a^5}{384\pi^3 v^4} 
	\Bigg[ 2 |g_{\substack{3\\ ab}}|^2 + 2 |g_{\substack{4\\ ab}}|^2 + |g_{\substack{5\\ ab}}|^2 + |g_{\substack{6\\ ab}}|^2 \notag \\
	&- \frac{8ev}{\sqrt{2}m_a} \text{Re} \left[(2g_{\substack{3\\ ab}}^* + g_{\substack{5\\ ab}}^*) c^{e\gamma}_{ab} 
	+ (2 g_{\substack{4\\ ab}}^* + g_{\substack{6\\ ab}}^*) c^{e\gamma *}_{ba} \right] 
	+ \frac{16 e^2 v^2}{m_a^2} \left( \log \frac{m_a^2}{m_b^2} - \frac{11}{4} \right) (|c^{e\gamma}_{ab}|^2 + |c^{e\gamma}_{ab}|^2) \Bigg] \, , 
	\label{big3l}\\
	& \Gamma(\tau^- \to \ell_b^- \ell_c^+ \ell_c^-) = \frac{m_\tau^5}{384\pi^3 v^4} 
	\Bigg[ |g_{\substack{3\\ \tau b}}'|^2 + |g_{\substack{4\\ \tau b}}'|^2 + |g_{\substack{5\\ \tau b}}'|^2 
	+ |g_{\substack{6\\ \tau b}}'|^2 + |c^{le}_{\tau ccb}|^2 + |c^{le}_{cb\tau c}|^2 \notag \\
	&- \frac{8ev}{\sqrt{2}m_\tau} \text{Re} \left[(g_{\substack{3\\ ab}}'^* + g_{\substack{5\\ ab}}'^*) c^{e\gamma}_{ab} + (g_{\substack{4\\ ab}}'^* + g_{\substack{6\\ ab}}'^*) c^{e\gamma *}_{ba} \right] + \frac{16 e^2 v^2}{m_\tau^2} \left( \log \frac{m_\tau^2}{m_\mu^2} - 3 \right) (|c^{e\gamma}_{\tau b}|^2 + |c^{e\gamma}_{\tau b}|^2) \Bigg] \, , \label{big3lbis}
\end{align}
with form factors, in the present LQ model, given by
\begin{align}
	&g_{\substack{3\\ ab}} = - c^{ee}_{abbb} - c^{ee}_{bbab} - 2s_w^2 c^{He}_{ab} \, ,&
	&g_{\substack{4\\ ab}} = -c^{ll}_{abbb} - c^{ll}_{bbab} + (1 - 2s_w^2) (c^{Hl(1)}_{ab} + c^{Hl(3)}_{ab}) \,, & \\
	&g_{\substack{5\\ ab}} = -c^{le}_{bbab} + (1-2s_w^2) c^{He}_{ab} \,, &
	&g_{\substack{6\\ ab}} = -c^{le}_{abbb} - 2s_w^2 (c^{Hl(1)}_{ab} + c^{Hl(3)}_{ab}) \, , & \\
	&g_{\substack{3\\ ab}}' = - c^{ee}_{abcc} - c^{ee}_{ccab} - 2s_w^2 c^{He}_{ab} \, ,&
	&g_{\substack{4\\ ab}}' = -c^{ll}_{abcc} - c^{ll}_{ccab} + (1 - 2s_w^2) (c^{Hl(1)}_{ab} + c^{Hl(3)}_{ab}) \,, & \\
	&g_{\substack{5\\ ab}}' = -c^{le}_{ccab} + (1-2s_w^2) c^{He}_{ab} \,, &
	&g_{\substack{6\\ ab}}' = -c^{le}_{abcc} - 2s_w^2 (c^{Hl(1)}_{ab} + c^{Hl(3)}_{ab}) \, , & 
\end{align}
where $c \neq a,b$.

Since the LQs couple directly to quarks, there are several additional constraints on the model parameter space beyond the purely leptonic observables discussed in section \ref{sec:SMEFTconstraints}. 
Our pragmatic assumption that the LQs couple only to third-generation quarks evades many stringent bounds associated with $s \to d$, $b\to d$ and $b \to s$ transitions, which are necessarily  suppressed by loops and small CKM factors. Still, collider searches give some relevant  constraints. In particular,
LEP sets the bound $c^{ld}_{eebb} \in [-4.8, 2.5] \times 10^{-2}$ with $2\sigma$ confidence \cite{Schael:2006wu}. 
An analysis of LHC high-$p_T$ dilepton tails \cite{Greljo:2017vvb} found $c^{ld}_{eebb} \in [-8.3,7.5] \times 10^{-3}$ and $c^{ld}_{\mu \mu bb} \in [-3.4,6.4] \times 10^{-3}$, also at $2\sigma$. 
Note that neither experiment probes the $t\bar{t}\ell^+ \ell^-$ interaction, thus they do not constrain the other WCs generated at tree level by the LQs, that is,  $c^{eu}$, $c^{lequ(1,3)}$, and the linear combination $c^{lq(1)} - c^{lq(3)}$. 
There are also some notable LHC direct searches for LQs with only one or two decay channels, which 
provide a robust bound on their mass, independent of the size of their Yukawa couplings:
the charge $1/3$ component of $D$ decays only into $\bar{b}\nu$, for which case Ref.~\cite{Angelescu:2021lln} obtained $m_D \geq 1.1$ TeV. 
In the limit $\epsilon_R \to 0$, $S$ decays into $b \bar{\nu}$ with a branching ratio $50\%$, which implies $m_S \geq 0.8$ TeV \cite{Angelescu:2021lln}. 
This bound is plotted in the left panels of Figs. \ref{fig:LQEMu}. \ref{fig:LQETau} and \ref{fig:LQMuTau} in dot-dashed, dark-red lines, 
assuming $(Y_L)_{3a} = 1$. It becomes stronger linearly as the Yukawa coupling becomes smaller.

LQs which couple only to third generation quarks automatically affect $B$ mesons and therefore lead us to consider the $B$ physics anomalies and related observables. We begin with the neutral-current transitions and then consider charged-current processes. 
There are compelling indications of lepton-flavour-universality violation in $B \to K^{(*)}\ell^+ \ell^-$ decays at LHCb \cite{LHCb:2017avl,LHCb:2021trn} compared to the SM expectation \cite{Bordone:2016gaq}. 
The $S$ can modify this transition at one-loop via a box diagram with a $W$-boson \cite{Bauer:2015knc}, which generates
\begin{equation}
	C_9^a = - C_{10}^a = \frac{m_t^2 |(\epsilon_L)_{3a}|^2}{2 e^2 v^2}  \, ,
\end{equation}
in the basis of weak-effective-theory operators (see e.g. \cite{Buras:1998raa}). 
Since this is positive, it has the wrong sign to explain the anomaly when $a = \mu$, but the correct sign when $a = e$. 
Although new physics in the electron sector does not give as large a pull away from the SM as in the muon sector, it nonetheless provides a
good fit of the $R_{K^{(*)}}$ anomalies, 
with a $>4\sigma$ pull with respect to the SM \cite{Altmannshofer:2021qrr}. 
The $1\sigma$ best-fit region when the new physics couples only to electrons is $C_9^e = - C_{10}^e \in [0.27, 0.47]$ \cite{Altmannshofer:2021qrr}. 
In order to plot this region in Figs.~\ref{fig:LQEMu} and \ref{fig:LQETau}, we naively generalise this best fit region to $C_9^{e-\mu} =- C_{10}^{e-\mu}  \in [0.27, 0.47]$, where $C_{9,10}^{e-\mu} \equiv C_{9,10}^e - C_{9,10}^\mu$. 
As can be seen from those figures, this region is ruled out by various flavour-conserving constraints. 
We note that a very similar 
solution was identified in \cite{Coy:2019rfr}, although in that paper $S$ was assumed to couple to the bottom-quark mass eigenstate at the LQ mass scale, rather than at the electroweak scale. Remarkably, that solution remains viable 
due to significant RGE effects between the two scales.

Alternatively, one could depart from the third-generation ansatz, and permit LQ couplings to second-generation quarks,  opening additional possibilities to explain the neutral-current anomalies. 
For instance, adding the coupling $(\epsilon_{L})_{2\mu}$ permits further box diagrams which could allow for an explanation in the muon channel \cite{Bauer:2015knc}. 
However, limits from $B_s-\bar{B}_s$ mixing and $B \to K^{(*)}\nu \nu$ in particular have all but ruled out this solution, see e.g. \cite{Gherardi:2020qhc}. 
We also note that the solution suggested in a similar context by \cite{Cata:2019wbu}, involving $(\epsilon_D)_{2a}$ and 
$(\epsilon_D)_{3a}$, does not give a good fit to the data, because it induces $c^{ld}_{aa23}$, which involves fermions of the wrong chirality to explain the anomalies, for either $a=e$ or $\mu$.
In summary, $S$ cannot provide a fully satisfactory resolution of the neutral-current anomalies, and $D$ does not even provide
a suitable WC to begin with.

Since $S$ may induce $b \to c \ell \bar{\nu}$ transitions, it can also provide a possible explanation for lepton-flavour-universality violation in charged-current processes. Measurements of $R_{D^{(*)}}$ \cite{Aaij:2015yra,Aaij:2017deq,Abdesselam:2019dgh} have displayed a roughly $3\sigma$ tension with the SM \cite{HFLAV:2019otj}, although a recent form-factor calculation \cite{Martinelli:2021und} implies only a $1.4\sigma$ discrepancy. 
When the $S$ couples only to third-generation quarks (in the down-quark mass basis), it gives a tree-level, but CKM-suppressed modification to $R_{D^{(*)}}$ from its coupling $Y_L$. 
The $1\sigma$ best-fit region to explain these anomalies found by \cite{Angelescu:2021lln} is $c^{lq(3)}_{\tau \tau bb} \in [-0.045,-0.025]$. 
Allowing for couplings to electrons and muons, at first order this can be naively 
generalised to $c^{lq(3)}_{\tau \tau bb} - (c^{lq(3)}_{ee bb}+ c^{lq(3)}_{\mu \mu bb})/2 \in [-0.045,-0.025]$, which is the region plotted in Figs. \ref{fig:LQETau} and \ref{fig:LQMuTau}. 
From the figures one concludes that $S$ with only third-generation couplings is incapable of resolving the $R_{D^{(*)}}$ anomalies, with bounds such as those from $Z \to \ell_a^+ \ell_a^-$ and $s_w$ clearly forbidding this solution. 
Introducing couplings to second-generation quarks, however, opens additional possibilities to address the anomalies. 
One scenario which satisfies all current experimental constraints is a combination of $(\epsilon_L)_{3\tau}$ and $(\epsilon_L)_{2\tau}$ \cite{Bauer:2015knc}. 
In addition, the $b \to c \ell \bar{\nu}$ transitions provide a stringent limit on lepton-flavour universality for the case of light leptons,
from $R_D^{\mu/e} \equiv BR(B \to D \mu \nu)/BR(B \to D e \nu)$. 
Comparing the SM calculation \cite{PhysRevD92034506} 
with the experimental result \cite{BaBar:2008zui,Belle:2015pkj} gives $c^{lq(3)}_{eebb} - c^{lq(3)}_{\mu \mu bb} \in [-0.044, 0.026]$ at $2\sigma$. This bound is represented by the purple line and shading in Figs.~\ref{fig:LQEMu}-\ref{fig:LQMuTau}.

We summarise the various constraints in Figs.~\ref{fig:LQEMu}, \ref{fig:LQETau} and \ref{fig:LQMuTau}.
In the left panels, we switch on the couplings $(\epsilon_L)_{3a}$ only, taking the different pairs of lepton flavours in turn, while setting the third to zero,
analogously to our plots for the seesaw and Zee models.
In the right panels, we rather assumed $|(\epsilon_L)_{3a}|=|(\epsilon_R)_{3a}|$, again choosing a pair of flavours at a time. 
In both the left and right panels, flavour-conserving and violating observables are directly correlated, given this choice of parameters.
Of the three spurions, $\epsilon_L$ has the largest range of phenomenological consequences for a few reasons. 
Firstly, only via the coupling $Y_L$ does the $S$ interact with all four types of fermions: neutrinos, charged leptons, down-type and up-type quarks. 
Consequently, the mixing into $c^G$ at one-loop leading-log order comes only from $\epsilon_L$, and we have shown that the associated shift in $G_F$ modifies many important observables at the electroweak scale.
Secondly, the $S$ coupling to top quarks (via $Y_L$ as well as $Y_R$) is particularly relevant for various one-loop processes, since top-quark loops 
come with an anomalous dimension $\sim N_c y_t^2$, which is typically an order of magnitude larger than gauge loops $\sim g^2$. 
Thirdly, since $\epsilon_L$ corresponds to a coupling to quark doublets, it induces (CKM-suppressed) interactions with light quarks, even within our third-generation assumption, implying constraints from observables such as $R_{D^{(*)}}$ and $R_D^{\mu/e}$, as we just discussed. 
The case where both $\epsilon_L$ and $\epsilon_R$ are sizeable is relevant, of course, to generate dipole operators that are not chirality suppressed, 
largely increasing  the sensitivity of all dipole-driven observables.

By contrast, the doublet LQ $D$ (its associated spurion $\epsilon_D$) has a much milder phenomenological impact, 
therefore we do not display the allowed parameter space in a plot. The constraints on $\epsilon_D$ are generically weak for several reasons.
The $\epsilon_D$ contributions to observables such as $\mu \to e$ conversion, $\ell \to 3\ell$ decays, and $Z$-boson decays
are relatively smaller than those of $\epsilon_{L,R}$, since $D$ does not couple to the top quark, 
hence mixing effects are smaller than those from $S$. 
Moreover, $c^G$ is not generated at one-loop leading log, thus $\epsilon_D \sim 1$ remains consistent with the electroweak observables that we consider. 
Finally, as commented below Eq.~\eqref{cegLQ1}, the piece of the EM dipole WC proportional to $\epsilon_D^\dag \epsilon_D$ vanishes even at one-loop finite order, so bounds from dipole moments, radiative charged leptons decays, and other flavour-violating processes mediated by the dipole operator are much weaker for $\epsilon_D$, than for $\epsilon_L$ and $\epsilon_R$.

The left panels of Figs.~\ref{fig:LQEMu}, \ref{fig:LQETau} and \ref{fig:LQMuTau} show the complementarity of constraints on $\epsilon_L$ from flavour-conserving and flavour-violating observables. 
As usual, the best bounds overall come from $\mu \to e $ transitions, with the strongest being $\mu \to e$ conversion in nuclei, 
as long as $|(\epsilon_L)_{be}|$ and $|(\epsilon_L)_{b\mu}|$ are both larger than $\sim 10^{-4}$, 
while flavour-conserving observables take over when one of the two entries is smaller, see Fig. \ref{fig:LQEMu}. 
In contrast, the most stringent limits on $\tau$ flavour change, which come from $\tau \to eee$ and $\tau \to \mu\mu\mu$, are, at best,
comparable with those from flavour-conserving observables, the most constraining of which are 
$Z\to \ell_a^+ \ell_a^-$ and $G_F$ universality. 
The near future prospects for flavour-violating $\tau \to 3\ell$ decays indicate that these will become the most stringent probe as long as $|(\epsilon_L)_{3\tau}|$ and $|(\epsilon_L)_{3e,\mu}|$ are of similar size.

Turning to the right panels, the presence of $\epsilon_R = \epsilon_L$ 
permits a chirality flip that is not suppressed by $y_e$. As a consequence, 
the dipole WCs and the three-Higgs WC $c^{eH}$ can be much larger than in the $\epsilon_R = 0$ case. 
Constraints from radiative charged lepton decays, dipole moments, Higgs decays, and also $\mu \to e$ conversion (see the discussion below 
Eq.~\eqref{csrLQ}) and $\ell \to 3\ell$ decays (see Eqs.~\eqref{l3lLQ1} and \eqref{l3lLQ2}) are thus orders of magnitude stronger. 
On the other hand, observables involving vector-current processes, for which the chirality flip is irrelevant, are barely affected. 
The $G_F$ universality, $Z$ decays (to both charged leptons and neutrinos), $s_w$, and $m_W$ all fit into this category. 
The small shifts in the corresponding bounds from the left to the right panels in Figs.~\ref{fig:LQEMu}-\ref{fig:LQMuTau}
are due to additional contributions from $\epsilon_R$, which are of the same order as those from $\epsilon_L$. 
In this $\epsilon_L = \epsilon_R$ scenario, flavour-violating observables give the best bounds, even in the $\tau$ sector, 
as long as $|(\epsilon_{L,R})_{3\tau}|$ is comparable to $|(\epsilon_{L,R})_{3e,\mu}|$. 
We note finally that both the $\epsilon_L$ only and $\epsilon_R = \epsilon_L$ cases lead to a positive shift in $m_W$, however the region in parameter space which ameliorates the present $m_W$ tension (given by the green bands in the figures) is ruled out by other flavour-conserving observables.

In summary, as one moves from $\epsilon_L=\epsilon_R$ to $\epsilon_L \gg \epsilon_R$, the constraints in the right panels of 
Figs.~\ref{fig:LQEMu}-\ref{fig:LQMuTau} shift progressively to those in the left panels. The opposite limit, $\epsilon_L \ll \epsilon_R$, behaves in
a qualitatively similar way, although there are fewer bounds in this case, in particular the $m_W$, $R_D^{\mu/e}$ and LHC constraints do not apply.

The dipoles play an important role in this LQ model with $\epsilon_L=\epsilon_R$. 
Note that the dipole WC is proportional to $(\epsilon_L^\dag y_u^T \epsilon_R)$, thus its phase is arbitrary. 
Consequently, the bounds from dipole moments in the right panels of the figures are represented by dotted lines, 
as they do not directly constrain $|(\epsilon_{L,R})_{3a}|$. 
In the electron sector, there is an extremely strong limit on the imaginary part of $(\epsilon_L^\dag y_u^T \epsilon_R)_{ee}$ from $d_e$. 
Furthermore, there is also a powerful limit on the real part of this combination from $a_e$. 
Combining the two bounds gives $|(\epsilon_{L,R})_{3e}|\lesssim 10^{-3}$, which excludes a large portion of parameter space.
In the muon sector, this model has the capacity to explain the $a_\mu$ anomaly 
for $(\epsilon_L^*)_{3\mu} (\epsilon_R)_{3\mu} \simeq -10^{-5}$. 
The region which reduces the discrepancy to $<2\sigma$ is plotted in Figs.~\ref{fig:LQEMu} and \ref{fig:LQMuTau}, 
assuming that $(\epsilon_L^\dag y_u^T \epsilon_R)_{\mu\mu}$ is real and negative. 
As can be seen, this solution of the $a_\mu$ anomaly  is compatible with all constraints as long as 
$|(\epsilon_{L})_{3e}| = |(\epsilon_{R})_{3e}| \lesssim 10^{-7}$ and $|(\epsilon_{L})_{3\tau}| = |(\epsilon_{R})_{3\tau}| \lesssim 10^{-3}$,
in order to avoid the $\mu \to e \gamma$ and $\tau \to \mu \gamma$ bounds, respectively.

We compared our results with various previous studies in the literature \cite{Dorsner:2016wpm,Arnan:2019olv,Mandal:2019gff,Gherardi:2020qhc,Zhang:2021dgl,Dedes:2021abc}, 
finding agreement when there is intersection, except where already noted above.

\section{Models' comparison and conclusions}\label{sec:Comp}

Each of the neutrino-mass models induces a different set of WCs, and thus has unique experimental implications, as detailed in section \ref{sec:Pheno}. 
In particular, observables are correlated differently in each model. 
If future experiments display one or several indications of new physics, 
this could shed light on the mechanism of neutrino mass generation, even if the energy scale of the experiments is far below the new physics scale $M$. 
The ability to discriminate between the models is already apparent in Figs.~\ref{fig:SeesawEMU}-\ref{fig:LQMuTau}. 
The most constraining bounds on a given model highlight through which observables that model, if realised in Nature, is most likely to first be discovered. Vice versa, if a deviation from the SM emerges in some other observables first, the model may be excluded.

\begin{figure}[tb]
\centering		\includegraphics[width=0.8\textwidth]{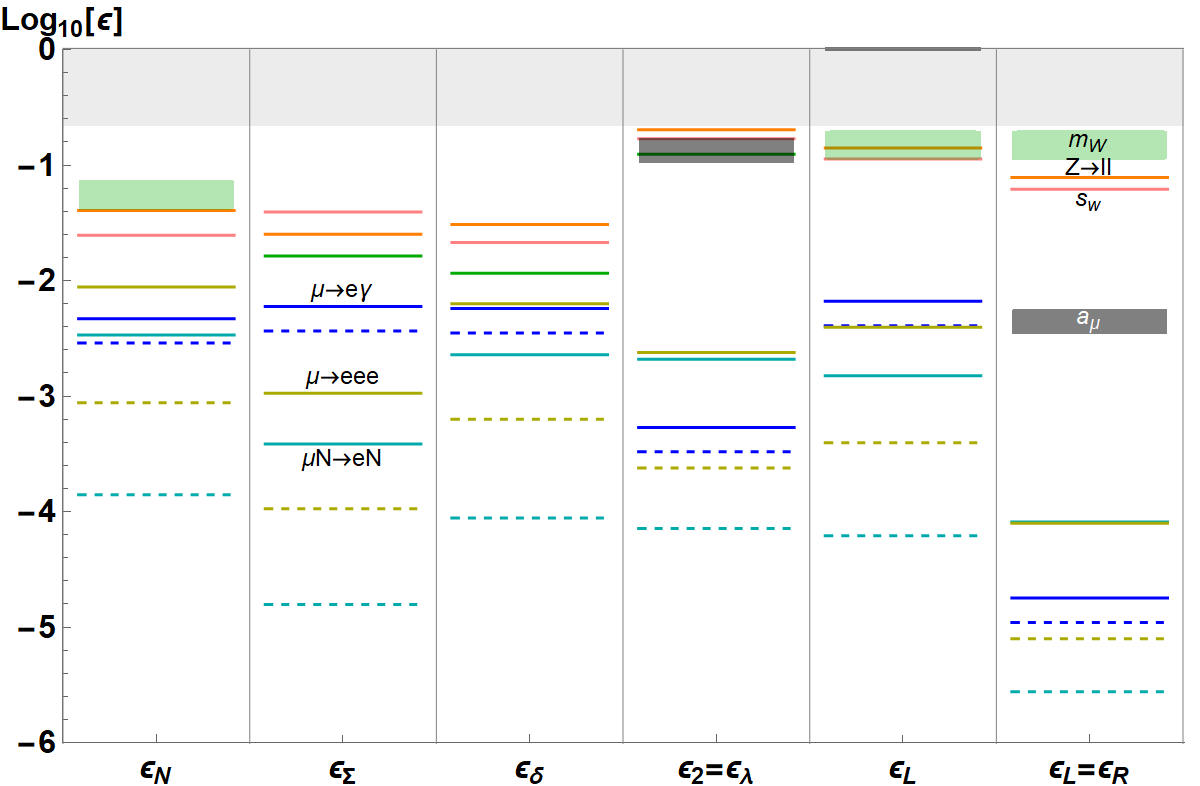}
	\caption{Comparison of bounds on the spurions, assuming they take an equal value $\epsilon$ for each choice of the lepton flavours $e,\mu,\tau$. 
	We fixed $M_i=10\,{\rm TeV}$ in order to specify (i) the RGE-induced WCs, proportional to $\log M_i$, and (ii) the perturbativity bound 
	$\epsilon< 4\pi v/(\sqrt{2}M_i)$ from the requirement $Y<4\pi$, which excludes the light-grey shaded region. 
	We display only the constraints from a subset of observables: 
	the most sensitive $e-\mu$ flavour-violating ones, namely $\mu \to e$ conversion, $\mu \to eee$ and $\mu \to e\gamma$, and a few 
	flavour-conserving 	ones, $m_W$, $s_w$ and $Z\to \ell_a^+ \ell_a^-$. 
	We also shade in dark grey (light green) the regions which reduce the $a_\mu$ ($m_W$) anomaly to within $2\sigma$, for the cases where this is possible. 
	The colour conventions are those of previous figures, as also indicated by the labels. }
	\label{fig:Comparison}
\end{figure}

\begin{figure}[tb]
\centering		\includegraphics[width=0.8\textwidth]{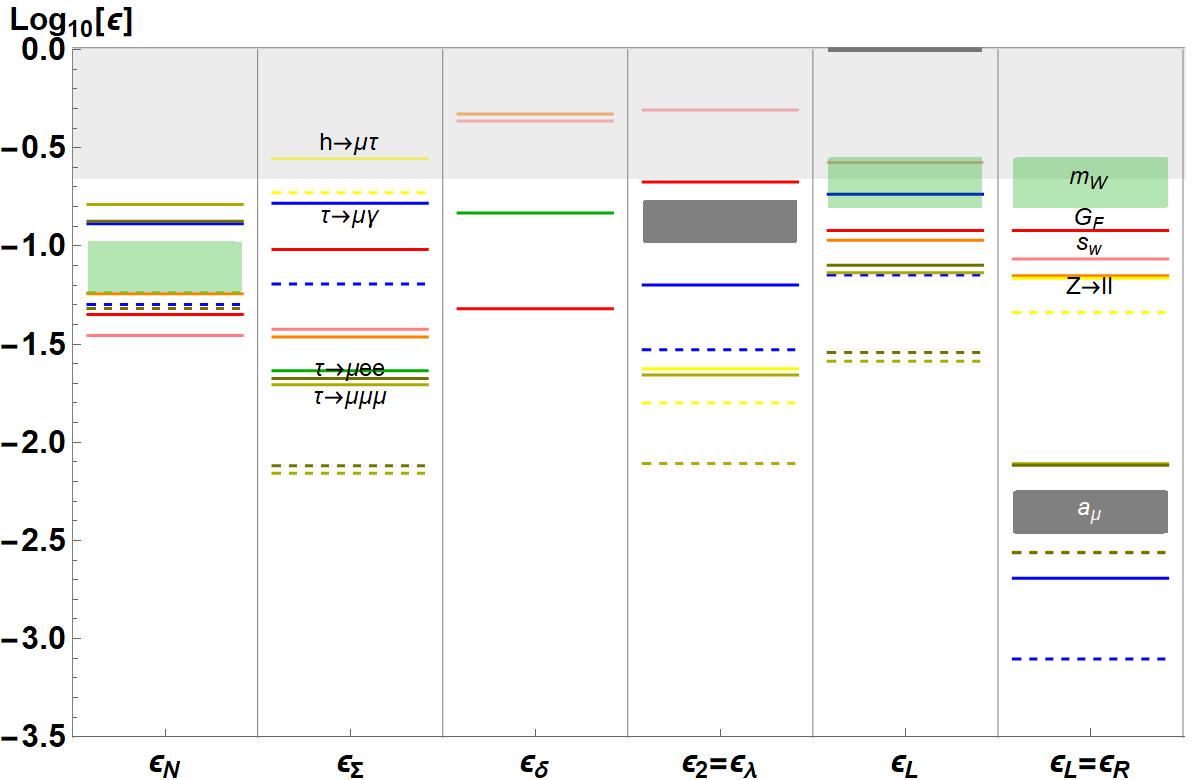}
	\caption{Comparison of bounds on the spurions, assuming that all the entries involving the $\mu$ and $\tau$ flavours take an equal value $\epsilon$, 
	while all the entries involving the $e$ flavour are set to zero.
	We display only the constraints from a relevant subset of observables: 
the $\mu-\tau$ flavour-violating decays of the $\tau$ lepton and of the Higgs, 
	the few flavour-conserving bounds already shown in Fig.~\ref{fig:Comparison}, plus the $G_F$ universality bound in red. 
	The other conventions are as in Fig.~\ref{fig:Comparison}.}
	\label{fig:ComparisonNoE}
\end{figure}

\begin{figure}[tb]
\centering		\includegraphics[width=0.8\textwidth]{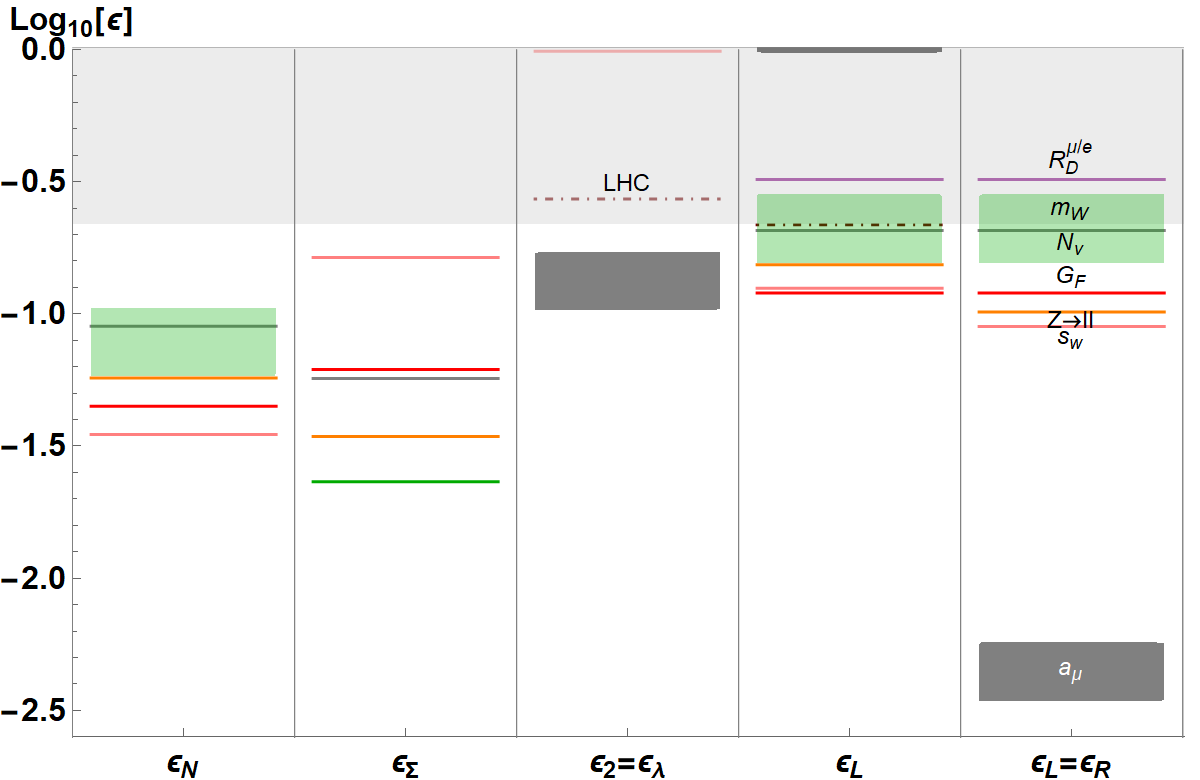}
	\caption{Comparison of bounds on the spurions, assuming that only the $\mu\mu$ entry does not vanish and is equal to $\epsilon$.
	Note we dropped the $\epsilon_\delta$ column as, under this assumption, the $\epsilon_\delta$ matrix identically vanishes, being antisymmetric. 
	We display the constraints from all 
	flavour-conserving observables considered in our analysis. 
	The other conventions are as in Fig.~\ref{fig:Comparison}.}
	\label{fig:ComparisonMuOnly}
\end{figure}

A direct comparison between neutrino-mass models is made in Figs.~\ref{fig:Comparison}-\ref{fig:ComparisonMuOnly}.
We plot the upper bounds on the spurions $\epsilon$ of each model, from a relevant subset of flavour-conserving and flavour-violating observables.
More precisely, for the type-I and type-III seesaw, we consider the spurions $|(\epsilon_N^\dag\epsilon_N)_{ab}|^{1/2}$ 
and $|(\epsilon_\Sigma^\dag\epsilon_\Sigma)_{ab}|^{1/2}$ respectively (corresponding to the left and right panels of Figs.~\ref{fig:SeesawEMU}-\ref{fig:SeesawMUTAU}), 
for the Zee model either the spurion $(\epsilon_\delta)_{ab}$ or $|(\epsilon_2)_{ab}|$ combined with $\epsilon_\lambda$
(left and right panels of Figs. \ref{fig:ZeeEMu}-\ref{fig:ZeeMuTau}),
and for the LQ model either $|(\epsilon_L)_{3a}|$ or $|(\epsilon_L)_{3a}|= |(\epsilon_R)_{3a}|$
(left and right panels of Figs. \ref{fig:LQEMu}-\ref{fig:LQMuTau}). 
In each of the corresponding columns of Figs.~\ref{fig:Comparison}-\ref{fig:ComparisonMuOnly}, all other spurions are set to zero.

Since $\epsilon\equiv Yv/(\sqrt{2}M)$, for any given value of the new physics scale $M$, the perturbativity 
bound on the Yukawa couplings, say $Y\lesssim 4\pi$, translates into an upper bound on $\epsilon$. Since in Figs.~\ref{fig:Comparison}-\ref{fig:ComparisonMuOnly} we picked $M=10$ TeV, this excludes the grey-shaded region at $\epsilon \gtrsim 0.3$.
Independently of this perturbativity argument, the EFT power counting is consistent only for $\epsilon$ significantly smaller than one, as 
we neglected higher orders in $\epsilon$ in our analysis of the constraints. 
Figs.~\ref{fig:Comparison}-\ref{fig:ComparisonMuOnly} show that the validity of the EFT approach is ensured in all models,
for most of the flavour-violating observables and also for the most sensitive flavour-conserving ones.

As a first illustration, in Fig.~\ref{fig:Comparison} we assume that the spurion entries are equal for all values of the lepton flavour indices, $a,b=e,\mu,\tau$.\footnote{In the $\epsilon_2$ model, 
we also fix the flavour-blind spurion $\epsilon_\lambda$, according to $\epsilon_\lambda=(\epsilon_2)_{ab}$ for any $a,b$, which departs from the choice $\epsilon_\lambda = 0.1$ adopted in Figs. \ref{fig:ZeeEMu}-\ref{fig:ZeeMuTau}.}
While this flavour-democratic assumption tends to turn on all possible constraints, there are exceptions,
e.g. the $G_F$-universality constraint vanishes under this condition.  
Fig.~\ref{fig:Comparison} shows that, in every model, the strongest future constraints are from $\mu \to e$ conversion in titanium, $\mu \to eee$ and $\mu \to e\gamma$, in that order. 
However, the gap between sensitivities varies for each model: in the type-III seesaw, for instance, the $\mu \to 3e$ decay is induced at tree-level, thus this observable is considerably more sensitive than $\mu \to e \gamma$. By contrast, the Zee model with the $\epsilon_2$ spurion 
and the LQ model with $\epsilon_L = \epsilon_R$ are the two scenarios in which the chirality flip in the dipole is generated without 
$y_e$-suppression, thus the future sensitivity of the dipole-induced radiative muon decay is comparable to that of $\mu \to eee$. 
Overall, the LQ model with $\epsilon_L = \epsilon_R$ is the most stringently constrained, with the current $\mu \to e \gamma$ bound probing 
$\epsilon \sim 10^{-4.5}$ and the future $\mu \to e$ conversion probing $\epsilon \sim 10^{-5.5}$. 
This is mainly due to the dipole arising at one loop without $y_e$ suppression (whereas in the $\epsilon_2$ case it arises at two loops via the Barr-Zee diagrams). 
The least constrained spurions are $\epsilon_N$ in the type-I seesaw and $\epsilon_\delta$ in the Zee model. 
For both, (i) the dipole operators are $y_e$-suppressed, and (ii) the corrections to $Z$-boson couplings to charged leptons (which mediates $\mu \to e$ conversion and $\ell \to 3\ell$ decays) are suppressed by a gauge loop. 
These spurions are presently bound at the level $\epsilon \sim 10^{-2.5}$, with expected future bounds of $\epsilon \sim 10^{-4}$. 
In this flavour-democratic setting, the constraints from all flavour-conserving observables are negligible compared to the flavour-violating ones: a few examples of the former are shown in Fig.~\ref{fig:Comparison} for reference.

The situation can be radically different if the spurion entries in lepton-flavour space are hierarchical.
Indeed, in each model there
are limits in which flavour violation is suppressed in some or even all channels. 
To illustrate this possibility, in Fig.~\ref{fig:ComparisonNoE} we switched off all spurions with electron flavour indices, while keeping spurions with $\mu$ and/or $\tau$ indices 
equal to each other.
In this case, $e-\mu$ and $e-\tau$ transitions are forbidden. Thus, in this figure we show the most sensitive $\mu-\tau$ flavour-violating observables, together with a few flavour-conserving ones.
One can observe that the most stringent constraints come from very different observables depending on the model: $\tau$ decays to three leptons are the most sensitive in the cases of $\epsilon_\Sigma$, $\epsilon_2$ and $\epsilon_L$, while $\tau\to\mu\gamma$ prevails in the case $\epsilon_L=\epsilon_R$, with the strong constraint $\epsilon\lesssim 10^{-3}$. 
In the case of $\epsilon_N$, the most stringent bounds, $\epsilon\lesssim 10^{-1.5}$, come from flavour-conserving observables, and the same is true in the $\epsilon_\delta$ case, where $G_F$ universality is by far the dominant constraint.

In  Fig.~\ref{fig:ComparisonMuOnly} we switched off all the spurions with $e$ and/or $\tau$ flavour indices, retaining only the $\mu\mu$ entry of each spurion matrix, as motivated by the muon anomalies. This limit removes all bounds on $\epsilon_\delta$, which vanishes by antisymmetry, but in all other models the pattern of constraints remains 
non-trivial. We include in the figure all flavour-conserving observables that we analysed in the previous sections. One can observe that the ratios of sensitivities to the various observables are strongly model-dependent, allowing for powerful model discrimination in the case of a positive future signal. 
This is due to the different combinations of $c^G$, $c^{Hl(1,3)}$ and $c^{He}$ which enter into these observables.
Further discriminating power could be obtained e.g. by considering individually $Z$ decays to each flavour of leptons: in the figure
we rather selected the strongest of the three to set the best bound. 
In the $\epsilon_2$ case, the leading correction to electroweak precision observables is suppressed by a weak gauge loop, consequently the associated constraints are negligible;
the only significant bound comes from the LHC direct searches for a charged scalar.
Finally, in the cases of $\epsilon_2$ and $\epsilon_L=\epsilon_R$, the most sensitive observable is the muon magnetic dipole moment, meaning that these two models can address the $a_\mu$ anomaly consistently. 
On the other hand, regions which explain the $m_W$ anomaly are always constrained by other flavour-conserving constraints, and we therefore conclude that none of the models considered can resolve this discrepancy.

\subsection{Summary and perspective}

It is difficult to directly attack
the question of the origin of neutrino masses: 
there is a large number of viable models, and typically all lepton-number-violating observables 
are highly suppressed by the tininess of the neutrino masses. 
Still, a crucial handle may come from 
lepton-number-conserving precision observables which are available in the lepton sector.
To this end, we have investigated four neutrino-mass models through the lens of the EFT. 

The four models cover a range of possibilities: new fermion singlets, new charged fermions, new colourless scalars, or new coloured scalars. 
For each case, we firstly performed a spurion analysis of the symmetries of the model,
in order to efficiently identify the few combinations of parameters relevant for the low-energy phenomenology.  
We then computed the leptonic WCs at tree-level,
one-loop leading log, as well as one-loop finite order when necessary. 
Our results are summarised in Tables \ref{table-WCs1}-\ref{table-WCs4}. 
The set of models was chosen, in particular, to cover different possibilities for the origin of the all-important Weinberg operator (tree-level or one-loop) and dipole operators (one-loop, leading log or finite).

As precision observables are available at or below the electroweak scale $v$, our EFT approach is suitable for any new physics model, as long as its degrees of freedom are all heavier than $v$. 
For example, traditional scenarios addressing the naturalness of the  electroweak scale, such as low-energy supersymmetry or Higgs compositeness, 
can easily incorporate a mechanism for neutrino mass generation, and correct other lepton observables as well. Since supersymmetric and composite states have been pushed by LHC data beyond the TeV frontier, even in these scenarios the EFT approach -- to fully integrate out the supersymmetric or composite spectrum -- is the most convenient and efficient way to compare with low-energy lepton observables.

Let us also note that, for a given model, an observable may arise only at some high order in the EFT expansion, which is worth computing if the experimental sensitivity for that observable is sufficiently good. 
For instance, in some of the models we studied, it was necessary to consider the dipole WCs at one-loop finite or even two-loop order.
In different models, also other WCs may be needed to higher order,
in view of the extreme sensitivities of lepton observables such as $\mu\to e$ transitions or lepton EDMs. 
This strongly motivates the current efforts of the community to implement one-loop matching and two-loop running systematically.

In section \ref{sec:SMEFTconstraints},
we derived the model-independent bounds on the WCs of the SMEFT leptonic and Higgs operators 
by collecting, combining, and extending results from the literature. 
Most of these bounds are summarised in table \ref{tab:WCbounds}. 
We restricted ourselves to the set of operators listed in Tables \ref{table-WCs1}-\ref{table-WCs4}: for most lepton observables,
these are the only relevant SMEFT operators (exceptions are specified in section \ref{sec:SMEFTconstraints}). 
These model-independent bounds are suitable for a comprehensive phenomenological analysis of any UV model, once its associated set of WCs has been computed. 

The phenomenological study of our four models is presented in section \ref{sec:Pheno}, 
with the main results illustrated in Figs.~\ref{fig:SeesawEMU}-\ref{fig:LQMuTau}. 
The WCs of each model are fully determined by one or two spurion matrices in lepton-flavour space, 
thus implying strong correlations between the various observables. 
In both the type-I and type-III seesaw (Figs.~\ref{fig:SeesawEMU}-\ref{fig:SeesawMUTAU}), there is only one spurion matrix 
and, if the number of heavy fermions is $\le 3$, than the flavour-violating observables are fully correlated to the flavour-conserving ones.
In the Zee model, there are two spurion matrices $\epsilon_\delta$ and $\epsilon_2$, corresponding to the Yukawa couplings of the charged singlet scalar and of the second Higgs doublet, respectively. While they combine to determine neutrino masses, their contributions to other observables are independent and can be studied separately
(Figs.~\ref{fig:ZeeEMu}-\ref{fig:ZeeMuTau}).
In the case of $\epsilon_\delta$, there is an enforced correlation between flavour-conserving and violating observables, emerging from the antisymmetry of the singlet Yukawa couplings.  
Instead, in the case of $\epsilon_2$, the flavour structure is arbitrary in general.
In the LQ model, it turns out that the doublet LQ is subject only to weak constraints, which makes it an interesting target for future direct searches. On the other hand, the rich phenomenology of the singlet LQ is characterised by two spurions,
$\epsilon_L$ and $\epsilon_R$, which correspond to coupling to weak-doublet and singlet SM fermions, respectively. 
We restricted ourselves to couplings to the third family of quarks only, and we studied in detail the phenomenology as a function of the
 size of 
$\epsilon_L$ and $\epsilon_R$ (Figs.~\ref{fig:LQEMu}-\ref{fig:LQMuTau}), also discussing the interplay with the anomalies in $B$-meson semileptonic decays.

Our EFT-educated analysis 
of the phenomenology 
elucidates the similarities and differences in the experimental status and prospects for each model. 
Although the electroweak precision observables have more limited sensitivity than LFV ones, they have a strong discriminating power, in the sense that the pattern of potential deviations from the SM predictions is highly model-dependent. 
On the other hand, only $\mu\to e$ flavour-violating transitions can access very heavy or very weakly-coupled new physics. 
In this regime, in order to distinguish between the models one would need to compare the rates of $\mu \to e$ conversion, $\mu \to eee$ and $\mu \to e \gamma$ decays precisely. 
Both in the Zee model (by means of a second Higgs doublet coupling mostly to the muon) and in the minimal LQ model 
(by means of a singlet LQ coupled both to the left-handed and right-handed muon), 
there is the possibility of addressing the $a_\mu$ anomaly while evading all other constraints. To this end,  
the flavour-violating couplings of these putative scalar particles should be strongly suppressed.

\section*{Acknowledgements}
We thank Sacha Davidson for a number of enlightening discussions, Tobias Felkl and Olcyr Sumensari for discussions on $B$ anomalies, 
Peter Stoffer for clarifying an issue in \cite{Aebischer:2021uvt}, David Marzocca and Ennio Salvioni for discussions on one-loop matching.
This project has received support from the European Union Horizon 2020 research and innovation programme under the Marie Sk\l odowska-Curie grant agreement No 860881-HIDDeN, and from the IISN convention 4.4503.15. 
R.C. thanks the UNSW School of Physics for their hospitality during part of this project.

\bibliographystyle{JHEP}
\bibliography{neutrinomassEFT.bib}

\end{document}